\newcommand{\be}{\begin{equation}}
\newcommand{\ee}{\end{equation}}
\newcommand{\bea}{\begin{eqnarray}}
\newcommand{\eea}{\end{eqnarray}}
\newcommand{\nn}{\nonumber}
\newcommand{\Appendix}[1]%
    {\renewcommand{\thesection}{Appendix~\Alph{section}:}%
         \section{#1}}%

\catcode`@=11
\long\def\@makecaption#1#2{
   \vskip 10pt
   \setbox\@tempboxa\hbox{{\small\bf #1.} \ {\small #2}}
   \ifdim \wd\@tempboxa >\hsize       
   {\small\bf #1.} \ {\small #2}\par  
   \else                              
        \hbox to\hsize{\hfil\box\@tempboxa\hfil}
   \fi}
\catcode`@=12

\catcode`@=11
\def\secteqno{\@addtoreset{equation}{section}%
\def\theequation{\thesection.\arabic{equation}}}
\def\endsecteqno{\def\theequation{\@ifundefined{chapter}%
{\arabic{equation}}{\thechapter.\arabic{equation}}}}
\newcounter{subequation}
\def\thesubequation{\alph{subequation}}
\def\sneqnarray{\stepcounter{equation}\let\@currentlabel=\theequation
\setcounter{subequation}{1}
\def\@eqnnum{{\rm (\theequation\thesubequation)}}
\global\@eqcnt\z@\tabskip\@centering\let\\=\@eqncr\let\@@eqncr=\@@sneqncr
$$\halign to \displaywidth\bgroup\@eqnsel\hskip\@centering
 $\displaystyle\tabskip\z@{##}$&\global\@eqcnt\@ne
 \hskip 2\arraycolsep \hfil${##}$\hfil
 &\global\@eqcnt\tw@ \hskip 2\arraycolsep
$\displaystyle\tabskip\z@{##}$\hfil
tabskip\@centering&\llap{##}\tabskip\z@\cr}
\def\endsneqnarray{\@@sneqncr\egroup $$\global\@ignoretrue}
\def\@@sneqncr{\let\@tempa\relax
   \ifcase\@eqcnt \def\@tempa{& & &}\or \def\@tempa{& &}
   \else \def\@tempa{&}\fi
     \@tempa \if@eqnsw\@eqnnum\stepcounter{subequation}\fi
     \global\@eqnswtrue\global\@eqcnt\z@\cr}
\def\nobiblabels{\def\@lbibitem[##1]##2{\@bibitem{##2}}}
\catcode`@=12

\def\beq{\begin{equation}}
\def\eeq{\end{equation}}
\def\bea{\begin{eqnarray}}
\def\eea{\end{eqnarray}}
\def\nn{\nonumber}

  \def\pa{\partial}

\def\pda{\overleftrightarrow{\partial}}
\def\lQ{\Lambda_{\rm QCD}}

\documentclass[aps,10pt,prd,showpacs,amsmath,amssymb,preprintnumbers,superscriptaddress,nofootinbib,showkeys]{revtex4-1}

\usepackage[german,english]{babel}
\usepackage{hyperref}
\usepackage{amssymb}
\usepackage{amsmath}
\usepackage{bm}
\usepackage{braket}
\usepackage{slashed}
\usepackage{multirow}
\usepackage{epsfig}
\usepackage{epstopdf}
\usepackage{bbm}
\usepackage{bbold}
\usepackage{soul}

\begin{document}

\title{Effective QCD string and doubly heavy baryons}
\author{Joan Soto}
\email{joan.soto@ub.edu}
\affiliation{Departament de F\'\i sica Qu\`antica i Astrof\'isica and Institut de Ci\`encies del Cosmos, Universitat de Barcelona (IEEC-UB), Mart\'\i$\,$ i Franqu\`es 1, 08028 Barcelona, Catalonia, Spain}

\author{Jaume Tarr\'us Castell\`a}
\email{jtarrus@ifae.es}
\affiliation{Grup de F\'\i sica Te\`orica, Departament de F\'\i sica and IFAE-BIST, Universitat Aut\`onoma de Barcelona,\\ 
E-08193 Bellaterra (Barcelona), Catalonia, Spain}

\date{\today}

\begin{abstract}
Expressions for the potentials appearing in the nonrelativistic effective field theory description of doubly heavy baryons are known in terms of operator insertions in the Wilson loop. However, their evaluation requires nonperturbative techniques, such as lattice QCD, and the relevant calculations are often not available. We propose a parametrization of these potentials with a minimal model dependence based on an interpolation of the short- and long-distance descriptions. The short-distance description is obtained from weakly-coupled potential NRQCD and the long-distance one is computed using an effective string theory. The effective string theory coincides with the one for pure gluodynamics with the addition of a fermion field constrained to move on the string. We compute the hyperfine contributions to the doubly heavy baryon spectrum. The unknown parameters are obtained from heavy quark-diquark symmetry or fitted to the available lattice-QCD determinations of the hyperfine splittings. Using these parameters we compute the double charm and bottom baryon spectrum including the hyperfine contributions. We compare our results with those of other approaches and find that our results are closer to lattice-QCD determinations, in particular for the excited states. Furthermore, we compute the vacuum energy in the effective string theory and show that the fermion field contribution produces the running of the string tension and a change of sign in the L\"uscher term.
\end{abstract}

\maketitle

\section{Introduction}

The discovery of more than two dozen exotic quarkonium states, as well as the more recent measurements of pentaquarks and double charm baryons, has increased interest in the wider class of hadrons containing two heavy quarks. All doubly heavy hadrons have in common that the constituent heavy quarks are nonrelativistic and that the dynamics of the heavy quarks and the light degrees of freedom, light quarks and gluons, can be factorized in an adiabatic expansion. An effective field theory (EFT) for doubly heavy hadrons built upon these two expansions was presented in Ref.~\cite{Soto:2020xpm}. Since the EFT reproduces the Born-Oppenheimer (BO) approximation at leading order we will refer to it as BOEFT. In the construction of the EFT no assumption is made about the heavy-quark distance and hence the EFT is valid both for short and long distances with respect to $\Lambda^{-1}_{\rm QCD}$, the inverse of the intrinsic scale of the nonperturbative effects in QCD. Therefore, the EFT can be seen as a generalization of strongly coupled potential NRQCD (pNRQCD)~\cite{Brambilla:2000gk,Pineda:2000sz} for quarkonium states to any heavy-quark-pair state with nontrivial light degrees of freedom. The matching coefficients of BOEFT depend on the heavy-quark-pair distance and, therefore, correspond to potential interactions. Expressions for these potentials in terms of operator insertions in the Wilson loop can be obtained by matching BOEFT to NRQCD~\cite{Caswell:1985ui,Bodwin:1994jh,Manohar:1997qy}, which can also be found in Ref.~\cite{Soto:2020xpm}. Since the Wilson loops involve nonperturbative dynamics, in principle they should be evaluated with lattice QCD.

BOEFT has been applied to doubly heavy baryons in Ref.~\cite{Soto:2020pfa}. In this case, the Wilson loop with light quark operator insertions corresponding to the static potential, has been obtained in the lattice~\cite{Najjar:2009da,Najjarthesis} including several excited states. This lattice data was used in Ref.~\cite{Soto:2020pfa} to obtain the double charm and bottom baryon spectrum at leading order in BOEFT. The leading-order spectrum is formed by spin-symmetry multiplets of states with total angular momentum $j$ and parity ${\eta_p}$. The degeneracy of the states in the multiplets is broken by $1/m_Q$ suppressed operators in BOEFT, where $m_Q$ is the heavy-quark mass. These operators correspond to different couplings of the heavy-quark spin and angular momentum to the light-quark spin. Unfortunately, at the moment there is no lattice data available for the potentials of these heavy-quark spin and angular-momentum dependent operators.

The main aim of this paper is to develop a parametrization for the subleading potentials for doubly heavy baryons. The short-distance regime is defined as $r\ll 1/\Lambda_{\rm QCD}$, which is equivalent to assuming that there is an energy gap between the relative momentum of the heavy quarks $m_Q v$, with $v$ the relative velocity, and $\Lambda_{\rm QCD}$. Therefore, in the short-distance regime one can build BOEFT in two steps. First, the relative momentum is integrated out perturbatively in order to build weakly-coupled pNRQCD~\cite{Pineda:1997bj,Brambilla:1999xf,Brambilla:2005yk}, and second, one integrates out the $\Lambda_{\rm QCD}$ modes. This procedure results in multipole expanded expressions of the potentials in BOEFT where the dependence on the heavy-quark-pair distance is explicit and the nonperturbative dynamics is encoded in some unknown constants. Examples of this two-step matching can be found in Refs.~\cite{Brambilla:2018pyn,Brambilla:2019jfi} for the heavy-quark spin dependent potentials of quarkonium hybrids and in Refs.~\cite{Pineda:2019mhw,Castella:2021hul} for the hybrid to standard quarkonium transitions. 

In the long-distance regime, $r\gg 1/\Lambda_{\rm QCD}$, it is known that the heavy-quark-antiquark static potential obtained from lattice QCD is well described in terms of an Effective String Theory (EST)~\cite{Nambu:1978bd} modeling the flux tube formed between the heavy-quark-antiquark pair at large separations. Corrections to the long-distance linear behavior of the static potential can be calculated in a systematic manner in the EST~\cite{Luscher:1980fr,Luscher:2002qv} (see also \cite{Polchinski:1991ax,Aharony:2013ipa}), including the contribution from the vacuum energy of the string which has also been confirmed by lattice QCD~\cite{Luscher:2002qv,Juge:2002br,Brandt:2017yzw}. The long-distance behavior of the subleading potentials for quarkonium can be computed in the EST given a mapping of the Wilson loop with operator insertions into EST correlation functions. This mapping was worked out in Ref.~\cite{PerezNadal:2008vm} and some of the subleading potentials were computed. This computation was later extended up to next-to-leading order in the EST in Refs.~\cite{Brambilla:2014eaa,Hwang:2018rju}. The parametrization given by these computations agree well with the lattice determinations of Refs.~\cite{Koma:2006si,Koma:2006fw}. The excitations of the string produce a spectrum of excited states, corresponding to quarkonium hybrid static potentials, which accurately describe the lattice determinations at long distances~\cite{Juge:2002br}. The mapping of operators to the EST to compute subleading potentials for hybrid quarkonium was introduced in Ref.~\cite{Oncala:2017hop}.

In this paper we present an EST for two static heavy quarks and one valence light quark, which is suitable to compute the long-distance part of the potentials of BOEFT for doubly heavy baryons. We obtain the mapping between different operator insertions in the Wilson loop and correlators in the EST and use it to compute the static potential and the heavy-quark spin and angular-momentum dependent potentials in the long-distance regime. A parametrization of the potentials for any distance between the heavy-quark pair is built by interpolating between the short- and long-distance descriptions. The free parameters of the short- and long-distance descriptions of the potentials are then fitted to a broad set of lattice data on the hyperfine splittings of doubly heavy baryons~\cite{Briceno:2012wt,Namekawa:2013vu,Brown:2014ena,Alexandrou:2014sha,Bali:2015lka,Padmanath:2015jea,Alexandrou:2017xwd,Lewis:2008fu,Brown:2014ena,Mohanta:2019mxo,Bahtiyar:2020uuj}. Using this parametrization of the potentials we compute the hyperfine contributions to the double charm and bottom baryons states of Ref.~\cite{Soto:2020pfa} corresponding to spin $1/2$ light-quark states. These include all the states below threshold of double bottom baryons, for which no lattice determination exists beyond the ground state spin doublet. Finally, we compare our results with previous model based determinations of the masses of doubly heavy baryons.

We present the paper as follows. In Sec.~\ref{sec:dhbp}, we review the general structure of the doubly heavy baryon potentials at next-to-leading order in the $1/m_Q$ expansion. We also discuss the short-distance constraints for those potentials. The leading-order parameters can be extracted from the heavy-light meson spectrum using heavy quark-diquark symmetry. In Sec.~\ref{sec:est}, we propose an EST with fermionic degrees of freedom in order to describe the long-distance behavior of the potentials. Based on the $D_{\infty h}$ group, we put forward a mapping from the NRQCD operator insertions in the Wilson loop to EST operators, and use it to compute the potentials. At leading order, they turn out to depend on two parameters only. In Sec.~\ref{hysc} we review the expressions of the doubly heavy baryon hyperfine splittings. In Sec.~\ref{sec:intfp} we model the spin-dependent potentials using suitable interpolations between the known short-distance behavior and the just calculated long-distance one. Then, the remaining unknown parameters of the parametrization of the potentials are fitted to lattice data on the hyperfine splittings of doubly heavy baryons. Using these parameters, we predict the spectrum of doubly heavy baryons including hyperfine contributions in Sec.~\ref{sec:dhbs}. We compare our results with other approaches in Sec.~\ref{sec:models}. We close the paper with some conclusions in Sec~\ref{sec:con}. In Appendix~\ref{app:ce} we calculate the Casimir energy (L\"uscher term) in the EST. Finally, in Appendix~\ref{app:sdec} we give expressions for the short-distance regime constants as correlators in weakly-coupled pNRQCD.

\section{Doubly heavy baryon potentials}\label{sec:dhbp}

\subsection{General expressions}

A general EFT framework to describe any doubly heavy hadron has been presented in Ref.~\cite{Soto:2020xpm}. The EFT was worked out up to $1/m_Q$ including the terms that depend on the heavy-quark spin and angular momentum. The matching expressions of the potentials in terms of operator insertions in the Wilson loop can also be found in Ref.~\cite{Soto:2020xpm}. This EFT framework has been applied to doubly heavy baryons in Ref.~\cite{Soto:2020pfa} where the spectrum associated to the four lowest lying static energies was obtained. These static energies are characterized by the representation of $D_{\infty h}$ and the quantum numbers of the light-quark operator that interpolates them, in particular the spin $\kappa$ and parity $p$. In the present work we will only consider the cases with light-quark interpolating operator with $\kappa^p=(1/2)^{\pm}$. These spin-$1/2$ cases only have one possible projection into the heavy-quark axis and therefore each correspond to a single $D_{\infty h}$ representation. These are $(1/2)_g$ and $(1/2)_u'$ for the $\kappa^p=(1/2)^{+}$ and $\kappa^p=(1/2)^{-}$ operators, respectively.

The Hamiltonian densities associated to the $\kappa^p=(1/2)^{\pm}$ light-quark states~\cite{Soto:2020pfa} have the following expansion up to $1/m_Q$
\begin{align}
h_{(1/2)^{\pm}}=\frac{\bm{p}^2}{m_Q}+\frac{\bm{P}^2}{4m_Q}+V_{(1/2)^{\pm}}^{(0)}(\bm{r})+\frac{1}{m_Q}V_{(1/2)^{\pm}}^{(1)}(\bm{r},\,\bm{p})\,.\label{hamden}
\end{align}
 At leading order we have just the static potential
\begin{align}
V_{(1/2)^{\pm}}^{(0)}(\bm{r})=&V_{(1/2)^{\pm}}^{(0)}(r)\,.\label{lopot}
\end{align}
The heavy-quark spin and angular-momentum dependent operators appear at next-to-leading order and read as
\begin{align}
V_{(1/2)^{\pm}{\rm SD}}^{(1)}(\bm{r})=&V^{s1}_{(1/2)^{\pm}}(r)\bm{S}_{QQ}\cdot\bm{S}_{1/2}+V^{s2}_{(1/2)^{\pm}}(r)\bm{S}_{QQ}\cdot\left(\bm{{\cal T}}_{2}\cdot\bm{S}_{1/2}\right)+V^{l}_{(1/2)^{\pm}}(r)\left(\bm{L}_{QQ}\cdot\bm{S}_{1/2}\right)\,,\label{nlopot}
\end{align}
with ${\cal T}^{ij}_2=\hat{\bm{r}}^i\hat{\bm{r}}^j-\delta^{ij}/3$, $\bm{S}_{1/2}=\bm{\sigma}/2$ and $2\bm{S}_{QQ}=\bm{\sigma}_{QQ}=\bm{\sigma}_{Q_1}\mathbb{1}_{2\,Q_2}+\mathbb{1}_{2\,Q_1}\bm{\sigma}_{Q_2}$, where $\bm{\sigma}$ are the standard Pauli matrices and $\mathbb{1}_2$ is an identity matrix in the heavy-quark spin space for the heavy quark labeled in the subindex.

The matching expressions of the potentials in terms of operator insertions in the Wilson loop can be found in Ref.~\cite{Soto:2020xpm}. For the potentials in Eqs.~\eqref{lopot} and \eqref{nlopot} the expressions in Ref.~\cite{Soto:2020xpm} reduce to
\begin{align}
V_{(1/2)^{\pm}}^{(0)}(\bm{r})&=\lim_{t\to\infty}\frac{i}{t}\log\left({\rm Tr}\left[\langle1\rangle^{(1/2)^{\pm}}_{\Box}\right]\right)\,,\label{lopotst}
\end{align}
and
\begin{align}
V^{s1}_{(1/2)^{\pm}}(r)&=-c_F\lim_{t\to\infty}\frac{4}{3t}\int^{t/2}_{-t/2} dt^{\prime}\frac{{\rm Tr}\left[\bm{S}_{1/2}\cdot\langle g\bm{B}(t^{\prime},\bm{x}_1)\rangle^{(1/2)^{\pm}}_{\Box}\right]}{{\rm Tr}\left[\langle1\rangle^{(1/2)^{\pm}}_{\Box} \right]}\,,\label{nlopots1}\\
V^{s2}_{(1/2)^{\pm}}(r)&=-c_F\lim_{t\to\infty}\frac{6}{t}\int^{t/2}_{-t/2} dt^{\prime}\frac{{\rm Tr}\left[\left(\bm{S}_{1/2}\cdot \bm{{\cal T}}_{2}\right)\cdot\langle g\bm{B}(t^{\prime},\bm{x}_1)\rangle^{(1/2)^{\pm}}_{\Box}\right]}{{\rm Tr}\left[\langle1\rangle^{(1/2)^{\pm}}_{\Box} \right]}\,,\label{nlopots2}\\
V^{l}_{(1/2)^{\pm}}=&-\lim_{t\to\infty}2\int^{1}_{0} ds\,s\frac{{\rm Tr}\left[\bm{S}_{1/2}\cdot\left(\frac{2}{3}\mathbb{1}_2 -\bm{{\cal T}}_{2}\right)\cdot\langle g\bm{B}(t/2,\bm{z}(s))\rangle^{(1/2)^{\pm}}_{\Box}\right]}{{\rm Tr}\left[\langle1\rangle^{(1/2)^{\pm}}_{\Box} \right]}\,,\label{nlopotl}
\end{align}
where $\bm{z}(s)=\bm{x}_1+s(\bm{R}-\bm{x}_1)$ and we use the following notation for the Wilson loop averages
\begin{align}
&\langle \dots\rangle^{(1/2)^{\pm}}_{\Box}=\langle {\cal Q}_{(1/2)^{\pm}}(t/2,\,\bm{R})\dots {\cal Q}_{(1/2)^{\pm}}^{\dagger}(-t/2,\,\bm{R})P\left\{e^{-ig\int_{{\cal C}_1+{\cal C}_2}dz^{\mu}A^{\mu}(z)}\right\}\rangle\,,
\end{align}
with ${\cal C}_1$ and ${\cal C}_2$ the upper and lower paths of a rectangular Wilson loop. Note that, unlike the quark-antiquark case, the flow is in the same direction for both paths. The interpolating operators are
\begin{align}
{\cal Q}^{\alpha}_{(1/2)^+}(t,\bm{x})&=\left[P_+q^{l}(t,\bm{x})\right]^\alpha\underline{T}^l \,,\\ 
{\cal Q}^{\alpha}_{(1/2)^-}(t,\bm{x})&=\left[P_+\gamma^5q^l(t,\bm{x})\right]^\alpha\underline{T}^l \,,
\end{align}
where $\alpha=-1/2,1/2$, and we have used the following $\bar{3}$ tensor invariants 
\begin{align}
&\underline{T}^l_{ij}  = \frac{1}{\sqrt{2}} \epsilon_{lij},\quad i,\,j,\,l=1,2,3\,.
\end{align}

\subsection{Short-distance potentials}\label{sec:sdp}

The short-distance regime is characterized by $r\ll \Lambda^{-1}_{\rm QCD}$. Since, in this regime the heavy-quark-pair distance and $\Lambda^{-1}_{\rm QCD}$ are well-separated scales the matching of NRQCD to the BOEFT for doubly heavy baryons can be done in two steps. First, one integrates out the heavy-quark-pair distance, which can be done in perturbation theory. This produces weakly-coupled potential NRQCD (pNRQCD) for doubly heavy systems presented in Ref.~\cite{Brambilla:2005yk}. Then, integrating out the $\Lambda_{\rm QCD}$ modes one recovers BOEFT. This procedure delivers expressions of the potentials in Eqs.~\eqref{lopotst}-\eqref{nlopotl} as an expansion in the heavy-quark-pair distance. An analogous approach was used in Refs.~\cite{Brambilla:2018pyn,Brambilla:2019jfi} to determine short distance expansion of the hybrid quarkonium potentials. All the potentials follow the same general structure in the short-distance regime; a possible nonanalytic term in $r$ produced by integrating out the heavy-quark-pair distance and an expansion in powers of $r^2$ with nonperturbative coefficients. These nonperturbatice coefficients only depend on the $\Lambda_{\rm QCD}$ scale and can be expressed as weakly-coupled pNRQCD correlators of light quark and gluon operators.

\begin{figure}[ht!]
   \centerline{\includegraphics[width=.9\textwidth]{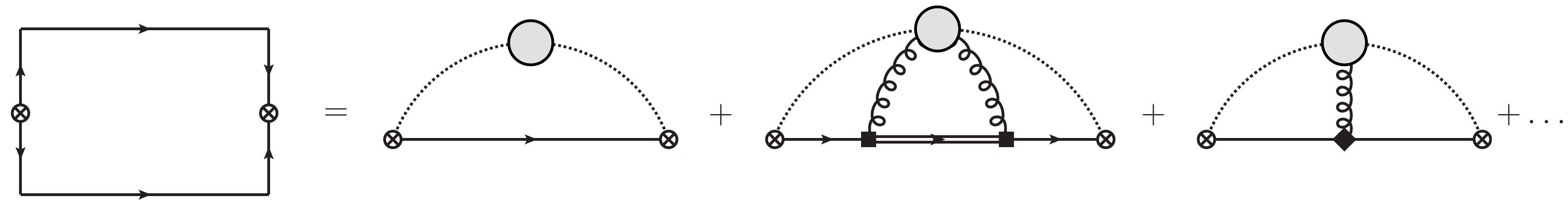}}
	\caption{Matching of the Wilson loop for the static potential for doubly heavy baryons, with the expansion in weakly-coupled pNRQCD up to next-to-leading order. The single lines represent the antitriplet fields, the double lines the sextet field, the dotted and the curly lines the light-quark and transverse gluon fields respectively (emissions of longitudinal gluon fields from the triplet and sextet fields and from the vertices are omitted). The crossed circles indicate the insertion of a ${\cal Q}$ operator and the square or diamond the insertion of a chromoelectric dipole or quadrupole operator, respectively.}
	\label{sp_matching}
 \end{figure}

The expansion of the static potential in Eq.~\eqref{lopotst} is given diagrammatically in Fig.~\ref{sp_matching} and corresponds to the following form
\begin{align}
V^{(0)}_{(1/2)^{\pm}}(r)&=-\frac{2}{3}\frac{\alpha_s}{r}+\overline{\Lambda}_{(1/2)^{\pm}}+\overline{\Lambda}^{(1)}_{(1/2)^{\pm}}r^2+\dots\,,\label{sdstp}
\end{align}
with the nonperturbative constants given as pNRQCD correlators in Appendix~\ref{app:sdec}.

For the heavy-quark spin and angular-momentum dependent potentials the short-distance expansion of the potentials is given diagrammatically in Fig.~\ref{sdp_matching} and are as follows:

\begin{figure}[ht!]
   \centerline{\includegraphics[width=.99\textwidth]{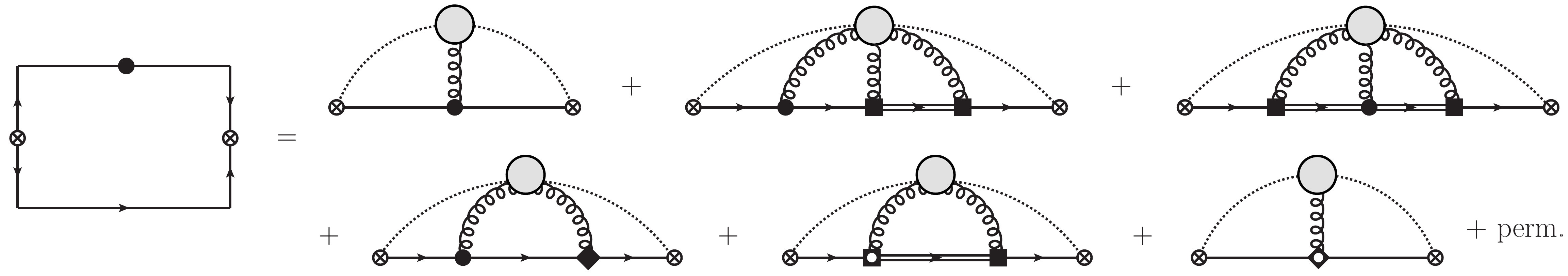}}
	\caption{Matching of the heavy-quark spin dependent potentials up to next-to-leading order in weakly-coupled pNRQCD. The legend is as in Fig.~\ref{sp_matching} with the addition of the solid dot, a white-dotted square, and a white-dotted diamond representing the insertion of a leading-order, dipole, and quadrupole heavy-quark spin chromomagnetic couplings, respectively. Further next to leading diagrams can be generated by changing the order of the different internal vertices, and by adding an extra transverse gluon emission to the heavy-quark spin chromomagnetic couplings. The potential of the heavy-quark angular-momentum dependent operator is matched to an analogous expansion.}
	\label{sdp_matching}
 \end{figure}

\begin{align}
V^{s1}_{(1/2)^{\pm}}(r)&=c_F\left(\Delta^{(0)}_{(1/2)^{\pm}}+\Delta^{(1,0)}_{(1/2)^{\pm}}r^2+\dots\right)\,,\label{sdps1}\\
V^{s2}_{(1/2)^{\pm}}(r)&=c_F\Delta^{(1,2)}_{(1/2)^{\pm}}r^2+\dots\,,\label{sdps2}\\
V^{l}_{(1/2)^{\pm}}=&\frac{1}{2}\left[\Delta^{(0)}_{(1/2)^{\pm}}+\left(\Delta^{(1,0)}_{(1/2)^{\pm}}-\frac{1}{3}\Delta^{(1,2)}_{(1/2)^{\pm}}\right)r^2\right]+\dots\,,\label{sdpl}
\end{align}
with the nonperturbative constants given in Appendix~\ref{app:sdec}. At leading order both Eqs.~\eqref{sdps1} and \eqref{sdpl} depend on the same correlator and the difference in the contribution to the potential stems from different factors in the coupling of the heavy-quark spin and angular momentum to the chromomagnetic field in the Lagrangian of Eq.~(9) in Ref.~\cite{Brambilla:2005yk}. The potential of the spin-tensor coupling in the Lagrangian of Eq.~\eqref{nlopot} vanishes at leading order since, to appear, it requires the insertion in the pNRQCD correlator of operators carrying the dependence on $\bm{r}$ which are suppressed in the multipole expansion. This type of correlator is also responsible for the next-to-leading order contributions to Eqs.~\eqref{sdps1} and \eqref{sdpl}. It is interesting to note, that the next-to-leading coefficient of Eq.~\eqref{sdpl} can be written as a combination of the next-to-leading coefficients of Eqs.~\eqref{sdps1} and \eqref{sdps2}.

In the static and $r\to 0$ limits the heavy-quark pair becomes indistinguishable from a single heavy antiquark. This is the so-called heavy quark-diquark duality~\cite{Savage:1990di,Hu:2005gf,Mehen:2017nrh,Mehen:2019cxn}. One can use this duality to relate the leading-order coefficients of the expansions of the potentials in Eqs.~\eqref{sdstp}-\eqref{sdpl} to the heavy meson masses. The value of $\overline{\Lambda}_{(1/2)^{+}}$ is equal to the leading-nonperturbative contribution to the lowest laying D- or B-meson masses, usually referred to as just $\overline{\Lambda}$.

The value of $\overline{\Lambda}$ has been obtained in Refs.~\cite{Bazavov:2018omf,Ayala:2019hkn} combining lattice determinations of the heavy-meson masses and perturbative computations of the heavy-quark masses. It is therefore necessary to use values of $\overline{\Lambda}$ and the heavy-quark masses computed in the same scheme. We use the values of Ref.~\cite{Bazavov:2018omf} in the MRS scheme
\begin{align}
&m_c=1.392(11)~{\rm GeV}\,, \label{mc}\\
&m_b=4.749(18)~{\rm GeV}\,, \label{mb}\\
&\overline{\Lambda}_{(1/2)^+}=0.555(31)~{\rm GeV} \,.\label{lbar12}
\end{align}
Following from the heavy quark-diquark duality, the difference $\overline{\Lambda}_{(1/2)^{-}}-\overline{\Lambda}_{(1/2)^{+}}$ is equal to the mass gap between the ground and first excited heavy-light mesons, up to corrections of order $\Lambda^2_{\rm QCD}/m_Q$. The values for this difference are collected in Table~\ref{ldt}. The values are compatible with the short-distance energy gaps between the static energies $(1/2)_g$ and $(1/2)_u'$ of Refs.~\cite{Najjar:2009da,Najjarthesis} associated to the light-quark operators $(1/2)^{+}$ and $(1/2)^{-}$, respectively.

\begin{table}[ht!]
\begin{tabular}{||cc||}\hline\hline
{\rm Heavy Mesons} & $(\overline{\Lambda}_{(1/2)^{-}}-\overline{\Lambda}_{(1/2)^{+}})[{\rm GeV}]$ \\ \hline
$m_{D^*_0(2300)^0}-m_{D^0}$              & $0.435(19)$ \\
$m_{D^*_0(2300)^\pm}-m_{D^\pm}$          & $0.465(7)$ \\
$m_{D_1(2420)^0}-m_{D^{*}(2007)^0}$      & $0.41375(7)$ \\
$m_{D_1(2420)^\pm}-m_{D^{*}(2010)^\pm}$  & $0.4129(24)$ \\
\hline\hline
\end{tabular}
\caption{Determination of $\overline{\Lambda}_{(1/2)^{-}}-\overline{\Lambda}_{(1/2)^{+}}$ from D meson mass differences. The masses are taken from the PDG~\cite{Zyla:2020zbs}. The uncertainty corresponds only to the experimental uncertainty of the meson masses. The uncertainty in the determination of $(\overline{\Lambda}_{(1/2)^{-}}-\overline{\Lambda}_{(1/2)^{+}})$ due to neglected higher-order terms is expected to be about $30\%$.}
\label{ldt} 
\end{table}

Finally, the value of $\Delta^{(0)}_{(1/2)^{\pm}}$ can be related to the hyperfine splittings in D or B mesons~\cite{Brambilla:2005yk}
\begin{align}
&m_{P^*_{\bar{Q}q}}-m_{P_{\bar{Q}q}}=\frac{2c_F(m_Q)}{m_Q}\Delta^{(0)}_{(1/2)^{\pm}}\,,\label{hmhf}
\end{align}
with corrections expected to be of order $\Lambda^3_{\rm QCD}/m^2_Q$. The values of $\Delta^{(0)}_{(1/2)^{\pm}}$ from Eq.~\eqref{hmhf} for various heavy-meson masses are found in Table~\ref{delahmm}.

\begin{table}[ht!]
\begin{tabular}{||cc||}\hline\hline
{\rm Heavy Mesons} & $\Delta^{(0)}_{(1/2)^{+}}~[{\rm GeV}^2]$ \\ \hline

$m_{D^*(2007)^{0}}-m_{D^0}$ & $0.08819(2)$ \\

$m_{D^*(2010)^{\pm}}-m_{D^{\pm}}$ & $0.087317(9)$ \\

$m_{B^{0*}}-m_{B^0}$ & $0.1222(6)$ \\

$m_{B^{\pm*}}-m_{B^{\pm}}$ & $0.1226(6)$ \\ \hline\hline
{\rm Heavy Mesons} &  $\Delta^{(0)}_{(1/2)^{-}}~[{\rm GeV}^2]$  \\ \hline
$m_{D_1(2420)^0}-m_{D^*_0(2300)^0}$ & 0.075(11) \\

$m_{D_1(2420)^\pm}-m_{D^*_0(2300)^\pm}$ & 0.0461(3)\\
\hline\hline
\end{tabular}
\caption{Determination of $\Delta^{(0)}_{(1/2)^{\pm}}$ from the heavy meson masses, taken from the PDG~\cite{Zyla:2020zbs}, using Eq.~\eqref{hmhf}. We take the renormalization group improved expression for $c_F$ at $1$~GeV. The uncertainty corresponds only to the experimental uncertainty of the meson masses. The uncertainty in the determination of $\Delta^{(0)}_{(1/2)^{\pm}}$ due to neglected higher order terms is expected to be of $\sim 30\%$ for the charm mesons and $\sim 10\%$ for the bottom mesons.}
\label{delahmm}
\end{table}

\section{Effective string theory}\label{sec:est}                                                                                                                                                
\subsection{Motivation}\label{est:m}

The QCD potentials for heavy quarks can be calculated assuming the heavy quarks to be static color sources. For a heavy-quark-antiquark system, the leading-order (static) potential is the energy of a source in the fundamental representation and a source in the complex conjugate representation separated at a distance $r$. Since the system must be a color singlet object, a certain gluon configuration must exist between the two sources in order to achieve so. When the distance is larger than the typical QCD scale $r\lQ \gg 1$, a flux tube emerges \cite{Bali:1994de}, with a typical radius $\sim \lQ^{-1}$. Assuming a constant energy per unit length in the flux tube leads to a linear potential. The flux-tube dynamics can be described by an EST, which matches the lattice QCD calculations very well for the static potential at long distances in the absence of light quarks \cite{Luscher:2002qv,Juge:2002br,Brandt:2017yzw}. When light quarks are present, the flux-tube configuration is still observed \cite{Ichie:2002dy} even though it may break due to light quark-antiquark pair creation, a phenomenon known as string breaking \cite{Bali:2005fu,Bulava:2019iut}. Nevertheless a flux-tube like configuration leading to a linear potential remains as an excited state for $r$ beyond the string-breaking scale.

For a baryon with two heavy quarks, we have an analogous situation. The two sources are now in the fundamental representation, and the gluon configuration linking them must also contain a valence light quark. When $r\lQ \gg 1$ we expect a flux tube to emerge from each source and to joint at the point between them where the valence light quark is at each time. Hence, the naive expectation would be to have a potential with the same string tension as in the quark-antiquark system plus a constant contribution $\sim \lQ$ due to the extra energy provided by the link to the valence light quark. Lattice QCD simulations indeed observe a linear potential \cite{Yamamoto:2008jz,Najjar:2009da}. Hence, we expect an EST to account for the long-distance behavior of the potential as well. Locally, the EST should be the same as the one for the quark-antiquark system, but it must contain some additional degrees of freedom describing the link to the valence light quark. In particular it must keep its transformation properties under $D_{\infty h}$ and flavor. We propose to add a fermion to the usual EST which transforms like the light quark under flavor and the Lorentz group. We write down a reparametrization invariant Lagrangian, and expand it at the desired order in the effective theory expansion.

\subsection{Construction}\label{est:c}

A string has one spatial dimension and its motion through spacetime defines a world sheet. The world sheet can be parametrized with two variables, which we will denote by $x=(\tau,\,\lambda)$. The embedding of the world-sheet in Minkowsky space is given by
\begin{align}
\bm{\xi}=\left(\xi^0(\tau,\lambda),\,\xi^1(\tau,\lambda),\,\xi^2(\tau,\lambda),\,\xi^3(\tau,\lambda)\right)\,.
\end{align}
The metric $g_{ab}$ induced on the string reads
\begin{align}
g_{ab}=\eta_{\alpha\beta}e^{\alpha}_ae^{\beta}_b\,,
\end{align}
with $\eta_{\alpha\beta}$ the Minkowsky metric, and $e^{\alpha}_a\equiv\partial \xi^{\alpha}/\partial x^a$ the Zweibein. The action of the gluonic string is just proportional to the area of the string world sheet
\begin{align}
S_{\rm g}=-\sigma\int d^2 x\sqrt{g}\,,\label{gst}
\end{align}
with $\sigma$ the string tension and $g=|$det$g_{ab}|$.

The action of a four-dimensional Dirac field constrained on a string is given by
\begin{align}
S_{\rm l.q}=\int d^2 x\sqrt{g}\bar{\psi}(x)\left(i\rho^a\pda_a-m_{\rm l.q.}\right)\psi(x)\quad,\quad \bar{\psi}\rho^a\pda_a\psi\equiv\left(\bar{\psi}(\rho^a\pa_a\psi)-(\pa_a\bar{\psi})\rho^a\psi\right)/2
\label{ngaplq}
\end{align}
with $\rho^a\equiv \gamma^{\mu}e^a_{\mu}$. The antisymmetrization of the partial derivative is required by Hermiticity. Note that the action in Eq.~\eqref{ngaplq} is  invariant under reparametrizations of the string if we choose $\psi (x)$ to transform like a scalar, and under Lorentz symmetry if we choose $\psi (x)$ to transform like a four-dimensional Dirac field but keeping $x$ invariant.

Let us choose the Gauge or string parametrization
\begin{align}
&\xi^0=\tau= t\,,\\
&\xi^3=\lambda= z\,.
\end{align}
Expanding the action in Eq.~\eqref{gst} for small string fluctuations we arrive at
\begin{align}
S_{\rm g}=-\sigma \int dt dz\left(1-\frac{1}{2}\pa^a\xi^l\pa_a\xi^l+\dots\right)\,,\label{gstex}
\end{align}
and for the case of the fermionic action in Eq.~\eqref{ngaplq} we find
\begin{align}
S_{\rm l.q}&=\int  dt dz\left(\bar{\psi}(t,z)i\gamma^a\pda_a\psi(t,z)-m_{\rm l.q.}\bar{\psi}(t,z)\psi(t,z)-\pa^a\xi^l\bar{\psi}(t,z)i\gamma^l\pda_a\psi(t,z)+\dots\right)\,,\label{ngaplqexp}
\end{align}
with $l=1,2,$ and $a=0,3$.

The fermion field mode expansion is
\begin{align}
\psi(t,\,z)=\sum_{n=-\infty}^\infty\sum_s\frac{1}{\sqrt{2rE_n}}\left(u^s_+(n)a^s_n e^{ip_nz}e^{-iE_nt}+u^s_-(n)b_n^{s\,\dagger} e^{-ip_n z}e^{iE_nt}\right)\label{fme1}
\end{align}
where $E_n=\sqrt{p^2_n+m_{\rm l.q.}^2}$. If we consider both periodic and antiperiodic solutions $p_n=n\pi/$r, $n\in\mathbb{Z}$. The spinors are defined as
\begin{align}
&u^s_{+}(E,\,p)=\frac{1}{\sqrt{E+m_{\rm l.q.}}}\left(\begin{array}{c} E+m_{\rm l.q.} \\ p\sigma_3 \end{array}\right)\chi_s\,,\\
&u^s_{-}(E,\,p)=\frac{1}{\sqrt{E+m_{\rm l.q.}}}\left(\begin{array}{c} p\sigma_3 \\ E+m_{\rm l.q.} \end{array}\right){\tilde\chi}_s,
\end{align}
with $\chi_{+1/2}=(1,\,0)$, $\chi_{-1/2}=(0,\,1)$ and ${\tilde\chi}_s=-i\sigma_2{\chi}_s^\ast$. The commutation relations for the creation an annihilation operators are
\begin{align}
\{a^s_n,\,a_{n^\prime}^{s^\prime\,\dagger}\}&=\delta^{sr}\delta_{nn^\prime}\,,\\
\{b^s_n,\,b_{n^\prime}^{s^\prime\,\dagger}\}&=\delta^{sr}\delta_{nn^\prime}\,,
\end{align}
all the other anticommutators vanish.

The field mode expansion in Eq.~\eqref{fme1} contains both positive- and negative-parity modes. Since the spinors fulfill the relation $u_\pm(E,-p)=\pm\gamma^0 u_\pm(E,p)$ a convenient choice for the transformation of the creation and annihilation operators under parity is
\begin{align}
Pa^s_nP= a^s_{-n}\,,\quad Pb^s_nP=-b^s_{-n}\label{partt}\,.
\end{align}
One can split the field mode expansion into two components of well-defined parity with the following definitions
\begin{align}
\psi_{n\eta_P}(t,z)=\sum_s\frac{1}{\sqrt{2E_n}}&\left[\varphi^s_{\eta_P\,+}(z,\,n)a^s_{n\eta_P} e^{-iE_nt}+\varphi^s_{\eta_P\,-}(z,\,n)b^{s\,\dagger}_{n\eta_P} e^{iE_nt}\right]\,,\label{fme2}
\end{align}
with
\begin{align}
a^s_{n\eta_P}&=\frac{a^s_{n}+\eta_P a^s_{-n}}{\sqrt{2}}\,,\quad b^s_{n\eta_P}=\frac{b^s_{n}+\eta_P b^s_{-n}}{\sqrt{2}}\,,\\
\varphi^s_{\eta_P\pm}(z,\,n)&=\frac{1}{\sqrt{2r}}\left(u_\pm^s(n)e^{i\frac{n\pi}{r}z}+\eta_P u_\pm^s(-n)e^{-i\frac{n\pi}{r}z}\right)\,,
\end{align}
with $\eta_P$ the parity eigenvalue
\begin{align}
P\psi_{n\eta_P}(t,z)P&=\eta_P\gamma^0\psi_{n\eta_P}(t,-z)\,.
\end{align}
The field mode expansion in Eq.~\eqref{fme1} can be rewritten in term of the two components of well-defined parity as
\begin{align}
\psi(t,z)=\sum^{\infty}_{n=1}\left(\psi_{n+}(t,z)+\psi_{n-}(t,z)\right)\,.
\end{align}

\subsection{Mapping}

Our aim is to use the EST introduced in Sec.~\ref{est:c} to compute the Wilson loops with operator insertions in Eqs.~\eqref{lopotst}-\eqref{nlopotl} which correspond to the potentials in the BOEFT. In order to do so we need a correspondence between NRQCD and EST correlators. This correspondence is defined by a mapping of NRQCD operators to the EST ones with matching symmetry properties. The symmetry transformations which leave a system of two static particles invariant form the group $D_{\infty h}$, which is the symmetry group of a cylinder. The basic transformations are rotations around the cylinder axis, reflections across a plane including the cylinder axis and parity. The conventional notation for the representations of $D_{\infty h}$ is $\Lambda_{\eta}^\sigma$. $\Lambda$ is the rotational quantum number, which for integer values is customarily labeled with capital Greek letters, $\Sigma,\,\Pi,\,\Delta\dots$ for $0\,,1\,,2\dots$. The parity eigenvalue is given as the index $\eta$ which is labeled as $g$ or $u$ for positive and negative parity, respectively. Finally, $\sigma$ gives the sign under reflections as $+$ or $-$; however, it is only written explicitly for the $\Sigma$ states, because for $\Lambda > 0$ rotations around the cylinder axis mix states in this quantum number. An operator belonging to $SO(3)\otimes P$ representation $\kappa^{p}$ can be projected into $D_{\infty h}$ representations: the rotational quantum number can take values corresponding to the absolute value of the projections of the spin of the operator into the heavy-quark axis $0\leq \Lambda\leq|\kappa|$ and the reflection eigenvalue corresponds to $\sigma=\eta(-1)^\kappa$. To simplify, we align the heavy-quark-pair axis with the $z$-axis, i.e., $\bm{r}=(0\,,0\,,z)$, set the heavy-quark positions at $z=\pm r/2$ and the center of mass at $\bm{R}=\bm{0}$.

Both Dirac and string fermions are spin-$1/2$ fields and have the same properties under rotations and reflections. Moreover they can only be projected to $\Lambda=1/2$. Therefore, to find the mapping of NRQCD to the EST operators we just need to make sure that the parities coincide
\begin{align}
&{\cal Q}_{(1/2)^+}(t,\,\bm{0})\mapsto\, P_+\psi_{1+}(t,\,0)\,,\label{map:e1}\\
&{\cal Q}_{(1/2)^-}(t,\,\bm{0})\mapsto\, P_-\psi_{1-}(t,\,0)\,,\label{map:e2}
\end{align}
with $P_{\pm}=(1\pm\gamma_0)/2$. Now, let us focus on the mapping for the chromomagnetic field $\bm{B}$, which can be projected into $\Sigma^-_u$ and $\Pi_u$ representations. Since we have chosen to align the heavy-quark-pair axis with the $z$-axis, then  $\bm{B}^l$, $l=1,2$ and $\bm{B}^3$ correspond to the $\Pi_u$ and $\Sigma^-_u$ representations, respectively. The mapping of the chromomagnetic field into string fluctuations can be found in Ref.~\cite{PerezNadal:2008vm}. 
\bea
\bm{B}^l(t,z)&\mapsto &\Lambda'\epsilon^{lm}\partial_t\partial_z\xi^m(t,z)\,,\label{bos}\\
\bm{B}^3(t,z)&\mapsto &{\Lambda'''}\epsilon^{lm}\partial_t\partial_z\xi^l(t,z)\partial_z\xi^m(t,z)\,.\label{bos2}
\eea
This implies that $\bm{B}^l$, $l=1,2$ is $\mathcal{O}(1/r^2)$ and $\bm{B}^3$ is $\mathcal{O}(1/\lQ r^3)$. However, mappings into string fermion operators are now possible and in fact provide the leading order contribution to the potentials in Eqs.~\eqref{nlopots1}-\eqref{nlopotl}. This mapping is as follows:
\begin{align}
\bm{B}^l(t,\,z)&\mapsto\Lambda_f\bar{\psi}(t,\,z)\frac{\bm{\Sigma}^l}{2}\psi(t,\,z)\,,
\label{map:e3}\\
\bm{B}^3(t,\,z)&\mapsto\Lambda^{\prime}_f\bar{\psi}(t,\,z)\frac{\bm{\Sigma}^3}{2}\psi(t,\,z)\,,\label{map:e4}
\end{align}
with $\bm{\Sigma}={\rm diag}(\bm{\sigma},\,\bm{\sigma})$. Note that here both $\bm{B}^l$, $l=1,2$ and $\bm{B}^3$ are $\mathcal{O}(\lQ/ r)$, and hence are more important than the corresponding bosonic operators in Eqs.~\eqref{bos} and \eqref{bos2}. Finally, to convert the two-dimensional spin operators in Eqs.~\eqref{nlopots1} and \eqref{nlopots2} into four-dimensional spin operators, we will use the following prescription 
\begin{align}
\bm{S}_{1/2}\mapsto\,\frac{1}{2}\bm{\Sigma}\,.\label{map:e5}
\end{align}

\subsection{Long-distance potentials}

Using the mapping of NRQCD operators in the Wilson loop to EST operators defined by Eqs.~\eqref{map:e1}-\eqref{map:e5} we compute the potentials in Eqs.~\eqref{lopotst}-\eqref{nlopotl} as correlators in the EST. For example, let us apply the mapping to the Wilson loop with the insertion of just the light-quark operators in the spatial sides of the loop
\begin{align}
\langle 1\rangle^{(1/2)^{\pm}}_{\Box}\,\mapsto \,P_\pm\langle \psi_{1\pm}(t/2,\,0)\psi^{\dagger}_{1\pm}(-t/2,\,0)\rangle P_\pm&=\frac{e^{-i(\sigma r+E_1)t}}{rE_1}(E_1\pm m)P_{\pm}
\end{align}
then the static potential is just
\begin{align}
V_{(1/2)^{\pm}}^{(0)}(\bm{r})&=\sigma r +E_1\,.
\end{align}
Similarly, one can apply the mapping to compute the heavy-quark spin and angular-momentum dependent potentials
\begin{align}
V^{s1}_{(1/2)^{\pm}}(r)&=\frac{c_F}{3r}\left(1\mp\frac{m_{\rm l.q.}}{E_1}\right)\left(\Lambda^{\prime}_f-2\Lambda_f\right)\,,\label{ldps1}\\
V^{s2}_{(1/2)^{\pm}}(r)&=\frac{c_F}{r}\left(1\mp\frac{m_{\rm l.q.}}{E_1}\right)\left(\Lambda^{\prime}_f+\Lambda_f\right)\,,\label{ldps2}\\
V^{l}_{(1/2)^{\pm}}(r)=&-\frac{1}{2r}\left(1\mp\frac{4}{\pi^2}\frac{m_{\rm l.q.}}{E_1}\right)\Lambda_f\,.\label{ldpl}
\end{align}

\section{Doubly heavy baryon hyperfine splittings}\label{sec:hysp}
\subsection{Hyperfine contributions}\label{hysc}

The hyperfine contributions to the masses of doubly heavy baryons have been computed in Ref.~\cite{Soto:2020pfa} for the states associated to the static energies $(1/2)_g$ and $(1/2)_u'$. These two static energies are interpolated by $(1/2)^+$ and $(1/2)^-$ light-quark operators, respectively. We summarize the quantum numbers available for the states associated to these static energies in Table~\ref{quantumnumbers}. Since the results of this section are equivalent for both $\kappa^p=(1/2)^\pm$ we will not display these labels.
\begin{table}[ht!]
\begin{tabular}{||c|c|c|c|c|c|c||}
 \hline\hline
$\kappa^p$                     & $\Lambda_\eta$                                  & $l$             & $\ell$        & $s_{QQ}$ & $j$                                       & $\eta_P$   \\ \hline
\multirow{4}{*}{$(1/2)^{\pm}$} & \multirow{4}{*}{$(1/2)_{g/u'}$}                 & $0$             & $1/2$         & $1$      & $(1/2,\,3/2)$                             & $\pm$ \\ 
                               &                                                 & $1$             & $(1/2,\,3/2)$ & $0$      & $(1/2,\,3/2)$                             & $\mp$ \\ 
                               &                                                 & $2$             & $(3/2,\,5/2)$ & $1$      & $((1/2,\,3/2,\,5/2)\,,(3/2,\,5/2,\,7/2))$ & $\pm$ \\ 
															 &                                                 & $3$             & $(5/2,\,7/2)$ & $0$      & $(5/2,\,7/2)$                             & $\mp$ \\ \hline\hline
\end{tabular}
\caption{Quantum numbers of doubly heavy baryons associated with the $(1/2)_{g}$ and $(1/2)_{u}'$ static energies. The quantum numbers are as follows: $l(l+1)$ is the eigenvalue of $\bm{L}^2_{QQ}$, $\ell(\ell+1)$ is the eigenvalue of $\bm{L}^2=(\bm{L}_{QQ}+\bm{S}_{1/2})^2$, $s_{QQ}(s_{QQ}+1)$ is the eigenvalue of $\bm{S}^2_{QQ}$. Note that the Pauli exclusion principle constrains $s_{QQ}=0$ for odd $l$ and $s_{QQ}=1$ for even $l$. The total angular momentum $\bm{J}^2=(\bm{L}+\bm{S}_{QQ})^2$ has eigenvalue $j(j+1)$. Finally, $\eta_P$ stands for the parity eigenvalue. Numbers in parentheses correspond to degenerate multiplets at leading order. Notice that $\pm$ in the parity column does not indicate degeneracy in that quantum number but correlates to the $\pm$ parity of the light-quark operator in the first column.}
\label{quantumnumbers}
\end{table}
Let us label the mass of the states as $M_{njl\ell}=M^{(0)}_{nl}+M^{(1)}_{njl\ell}+\dots$ with $M^{(0)}_{nl}$ the mass solution of the Schr\"odinger equation with the static potential and $M^{(1)}_{njl\ell}$ the hyperfine contribution. Recall that due to the Pauli principle the heavy-quark spin is $s_{QQ}=0$ for $l$ odd and $s_{QQ}=1$ for $l$ even. Let us denote the expectation values of the potentials between the radial wave functions as
\begin{align}
{\cal V}^{i}_{nl}=\int_0^{\infty}dr\,r^2\,\psi^{nl\,\dagger}(r)V^{i}(r)\psi^{nl}(r)\,,\quad i=s1,\,s2,\,l.
\end{align}

The hyperfine contributions for $l=0$ are given by
\begin{align}
M^{(1)}_{nj0\frac{1}{2}}=\frac{1}{2}\left(j(j+1)-\frac{11}{4}\right)\frac{{\cal V}^{s1}_{n0}}{m_Q}\,,
\end{align}
and the splitting is
\begin{align}
M_{n\frac{3}{2}0\frac{1}{2}}-M_{n\frac{1}{2}0\frac{1}{2}}=\frac{3}{2}\frac{{\cal V}^{s1}_{n0}}{m_Q}\,.\label{l0hfs}
\end{align}
In the case $l$ is an odd number the hyperfine contribution is as follows:
\begin{align}
M^{(1)}_{njlj}=\frac{1}{2}\left(j(j+1)-\frac{3}{4}-l(l+1)\right)\frac{{\cal V}^{l}_{nl}}{m_Q}\,,
\end{align}
which for the cases $l=1,3$ leads to the following splittings
\begin{align}
M_{n\frac{3}{2}1\frac{3}{2}}-M_{n\frac{1}{2}1\frac{1}{2}}&=\frac{3}{2}\frac{{\cal V}^{l}_{n1}}{m_Q}\,,\label{l1hfs}\\
M_{n\frac{7}{2}3\frac{7}{2}}-M_{n\frac{5}{2}3\frac{5}{2}}&=\frac{7}{2}\frac{{\cal V}^{l}_{n3}}{m_Q}\,\label{l3hfs}
\end{align}
For $l=2$ the hyperfine contributions are more complicated since they depend on all three potentials in Eq.~\eqref{nlopot} and the states $j=3/2,5/2$ with $\ell=3/2$ and $\ell=5/2$ are mixed. For $j=1/2$ and $7/2$ the contributions are
\begin{align}
&M^{(1)}_{n\frac{1}{2}2\frac{3}{2}}=\frac{1}{2}\frac{{\cal V}^{s1}_{n2}}{m_Q}-\frac{1}{3}\frac{{\cal V}^{s2}_{n2}}{m_Q}-\frac{3}{2}\frac{{\cal V}^{l}_{n2}}{m_Q}\,,\label{ml21}\\
&M^{(1)}_{n\frac{7}{2}2\frac{5}{2}}=\frac{1}{2}\frac{{\cal V}^{s1}_{n2}}{m_Q}-\frac{2}{21}\frac{{\cal V}^{s2}_{n2}}{m_Q}+\frac{{\cal V}^{l}_{n2}}{m_Q}\,.\label{ml26}
\end{align}
For $j=3/2,5/2$ we have the mixing matrices for $\ell=3/2$ and $\ell=5/2$ states\footnote{The off-diagonal terms were initially overlooked in Ref.~\cite{Soto:2020pfa}. They have been included in an Erratum.} 
\begin{align}
&M^{(1)}_{n\frac{3}{2}2}=\frac{1}{m_Q}\left(\begin{array}{cc} \frac{1}{5}{\cal V}^{s1}_{n2}-\frac{2}{15}{\cal V}^{s2}_{n2}-\frac{3}{2}{\cal V}^{l}_{n2} & \frac{3}{5}{\cal V}^{s1}_{n2}+\frac{1}{10}{\cal V}^{s2}_{n2} \\ \frac{3}{5}{\cal V}^{s1}_{n2}+\frac{1}{10}{\cal V}^{s2}_{n2} & -\frac{7}{10}{\cal V}^{s1}_{n2}+\frac{2}{15}{\cal V}^{s2}_{n2}+{\cal V}^{l}_{n2}\end{array}\right)\,,\\
&M^{(1)}_{n\frac{5}{2}2}=\frac{1}{m_Q}\left(\begin{array}{cc} -\frac{3}{10}{\cal V}^{s1}_{n2}+\frac{1}{5}{\cal V}^{s2}_{n2}-\frac{3}{2}{\cal V}^{l}_{n2} & \frac{\sqrt{14}}{5}{\cal V}^{s1}_{n2}+\frac{1}{15}\sqrt{\frac{7}{2}}{\cal V}^{s2}_{n2} \\ \frac{\sqrt{14}}{5}{\cal V}^{s1}_{n2}+\frac{1}{15}\sqrt{\frac{7}{2}}{\cal V}^{s2}_{n2} & -\frac{1}{5}{\cal V}^{s1}_{n2}+\frac{4}{105}{\cal V}^{s2}_{n2}+{\cal V}^{l}_{n2}\end{array}\right)\,.
\end{align}
We diagonalize to obtain the physical states
\begin{align}
M^{(1)}_{n\frac{3}{2}2\pm}=&-\frac{1}{4m_Q}\left\{{\cal V}^{s1}_{n2}+{\cal V}^{l}_{n2}\pm\frac{1}{3}\left[81\left({\cal V}^{s1}_{n2}\right)^2+4\left({\cal V}^{s2}_{n2}\right)^2+225\left({\cal V}^{l}_{n2}\right)^2-6{\cal V}^{l}_{n2}\left(27{\cal V}^{s1}_{n2}-8{\cal V}^{s2}_{n2}\right)\right]^{1/2}\right\}\,,\\
M^{(1)}_{n\frac{5}{2}2\pm}=&-\frac{1}{84m_Q}\left\{21{\cal V}^{s1}_{n2}-10{\cal V}^{s2}_{n2}+21{\cal V}^{l}_{n2}\pm\left[3969\left({\cal V}^{s1}_{n2}\right)^2+156\left({\cal V}^{s2}_{n2}\right)^2+11025\left({\cal V}^{l}_{n2}\right)^2+126{\cal V}^{s1}_{n2}\left(10{\cal V}^{s2}_{n2}+7{\cal V}^{l}_{n2}\right)\right.\right.\nn\\
&\left.\left.-1428{\cal V}^{s2}_{n2}{\cal V}^{l}_{n2}\right]^{1/2}\right\}\,.
\end{align}
For simplicity we consider the following hyperfine splittings among $l=2$ which are linear in the expectation values of the potentials
\begin{align}
M_{n\frac{5}{2}2+}+M_{n\frac{5}{2}2-}-M_{n\frac{3}{2}2+}-M_{n\frac{3}{2}2-}=&\frac{5}{21 m_Q}{\cal V}^{s2}_{n2}\,,\label{l2hfs1}\\
M_{n\frac{1}{2}2\frac{3}{2}}-\frac{1}{2}\left(M_{n\frac{3}{2}2+}+M_{n\frac{3}{2}2-}\right)=&\frac{1}{12m_Q}\left(9{\cal V}^{s1}_{n2}-4{\cal V}^{s2}_{n2}-15{\cal V}^{l}_{n2}\right)\,,\\
M_{n\frac{7}{2}2\frac{5}{2}}-\frac{1}{2}\left(M_{n\frac{3}{2}2+}+M_{n\frac{3}{2}2-}\right)=&\frac{1}{m_Q}\left(\frac{3}{4}{\cal V}^{s1}_{n2}-\frac{2}{21}{\cal V}^{s2}_{n2}+\frac{5}{4}{\cal V}^{l}_{n2}\right)\,.\label{l2hfs3}
\end{align}
These formulas fix ${\cal V}^{s1}_{n2}$, ${\cal V}^{s2}_{n2}$ and ${\cal V}^{l}_{n2}$ in terms of physical masses. Then, we have the following model-independent predictions
\begin{align}
M^{(1)}_{n\frac{3}{2}2+}-M^{(1)}_{n\frac{3}{2}2-}=&-\frac{1}{6m_Q}\left[81\left({\cal V}^{s1}_{n2}\right)^2+4\left({\cal V}^{s2}_{n2}\right)^2+225\left({\cal V}^{l}_{n2}\right)^2-6{\cal V}^{l}_{n2}\left(27{\cal V}^{s1}_{n2}-8{\cal V}^{s2}_{n2}\right)\right]^{1/2}\,,\label{l2hfs4}\\
M^{(1)}_{n\frac{5}{2}2+}-M^{(1)}_{n\frac{5}{2}2-}=&-\frac{1}{42m_Q}\left[3969\left({\cal V}^{s1}_{n2}\right)^2+156\left({\cal V}^{s2}_{n2}\right)^2+11025\left({\cal V}^{l}_{n2}\right)^2+126{\cal V}^{s1}_{n2}\left(10{\cal V}^{s2}_{n2}+7{\cal V}^{l}_{n2}\right)-1428{\cal V}^{s2}_{n2}{\cal V}^{l}_{n2}\right]^{1/2}\,.\label{l2hfs5}
\end{align}

\subsection{Interpolation of the full potentials}\label{sec:intfp}

We have obtained descriptions of the potentials of the spin and angular-momentum dependent operators in the short- and long-distance regimes in Eqs.~\eqref{sdps1}-\eqref{sdpl} and \eqref{ldps1}-\eqref{ldpl}, respectively. In this section we propose an interpolation between the descriptions of the potentials in these two regions to model the potential in the intermediate distance regime $r\sim1/\Lambda_{\rm QCD}$. Using this interpolation and the wave functions obtained in Ref.~\cite{Soto:2020pfa}, we compute the hyperfine splittings of Sec.~\ref{hysc} in terms of the parameters of the short- and long-distance descriptions. These parameters are then determined by fitting the hyperfine splittings of lattice determinations~\cite{Briceno:2012wt,Namekawa:2013vu,Brown:2014ena,Alexandrou:2014sha,Bali:2015lka,Padmanath:2015jea,Alexandrou:2017xwd,Lewis:2008fu,Brown:2014ena,Mohanta:2019mxo} of the double charm and bottom baryon spectrum and in the case of the short-distance parameters using heavy quark-diquark symmetry.

The interpolation we propose is constructed by summing the short- and long-distance descriptions multiplied by weight functions depending of $r$ and a new $r_0$ parameter. The weight functions are $w_s=r_0^n/(r^n+r_0^n)$ and $w_l=r^n/(r^n+r_0^n)$ for the short- and long-distance pieces, respectively. The sum of the weight functions is $w_s+w_l=1$ and the $r_0$ parameter determines the value of $r$ where both weights are equal. The value of the exponent $n$  is chosen as the minimal value that ensures that the product of the short- and long-distance potentials and the respective weight functions vanishes in the long- and short-distance limits, respectively. For the short-distance potentials we consider the contributions up to next-to-leading order. The resulting interpolated potentials are as follows:
\begin{align}
&V^{s1\,{\rm int}}_{(1/2)^\pm}=c_F\frac{\left(\Delta^{(0)}_{(1/2)^{\pm}}+\Delta^{(1,0)}_{(1/2)^{\pm}}r^2 \right)r^6_0+\frac{\left(\Lambda^{\prime}_f-2\Lambda_f\right)}{3}\left(1\mp\frac{m_{\rm l.q.}}{E_1}\right)r^5}{r^6+r^6_0}\,,\label{s1int}\\
&V^{s2\,{\rm int}}_{(1/2)^\pm}=c_F\frac{\Delta^{(1,2)}_{(1/2)^{\pm}}r^2 r^6_0+\left(\Lambda^{\prime}_f+\Lambda_f\right)\left(1\mp\frac{m_{\rm l.q.}}{E_1}\right)r^5}{r^6+r^6_0}\,,\\
&V^{l\,{\rm int}}_{(1/2)^\pm}=\frac{1}{2}\frac{\left[\Delta^{(0)}_{(1/2)^{\pm}}+\left(\Delta^{(1,0)}_{(1/2)^{\pm}}-\frac{1}{3}\Delta^{(1,2)}_{(1/2)^{\pm}}\right)r^2\right]r^6_0-\Lambda_f\left(1\mp\frac{4}{\pi^2}\frac{m_{\rm l.q.}}{E_1}\right)r^5}{r^6+r^6_0}\,,\label{lint}
\end{align}
with $E_1=\sqrt{(\pi/r)^2+m^2_{\rm l.q.}}$. Note that for $r_0=0$ we recover the long-distance potentials and for $r_0\to\infty$ we recover the short-distance potentials.

An accurate determination of $m_{\rm l.q.}$ would require lattice data for the static energies at longer distances than the one currently available. Nevertheless, fitting the long-distance part of the static potential to the lattice data of Refs.~\cite{Najjar:2009da,Najjarthesis}, we find the value 
\begin{align}
m_{\rm l.q.}=0.226~{\rm GeV}\,,\label{mlq}
\end{align}
for which the contribution to the potentials of the terms proportional to $m_{\rm l.q.}$ is small.

To obtain the unknown parameters in the interpolated potentials in Eqs.~\eqref{s1int}-\eqref{lint} for the case $\kappa^p=(1/2)^+$ we minimize $\chi^2$ function constructed as the sum of the hyperfine splittings of Sec.~\ref{hysc} taking the masses of the doubly heavy baryons from lattice determinations. The list of contributions to the $\chi^2$ function is as follows: For the double charm baryons $1S$ splitting in Eq.~\eqref{l0hfs}, there are six data points corresponding to Refs.~\cite{Briceno:2012wt,Namekawa:2013vu,Brown:2014ena,Alexandrou:2014sha,Bali:2015lka,Padmanath:2015jea,Alexandrou:2017xwd}. For the double bottom baryons $1S$ splitting, there are three data points corresponding to Refs.~\cite{Lewis:2008fu,Brown:2014ena,Mohanta:2019mxo}. The rest are single data points for double charm baryons from Ref.~\cite{Padmanath:2015jea} corresponding to the splittings for $2S$ and $3S$ from Eq.~\eqref{l0hfs}, $1P$ and $2P$ from Eq.~\eqref{l1hfs}, $1D$ and $2D$ from Eqs.~\eqref{l2hfs1}-\eqref{l2hfs3}, and finally, $1F$ from Eq.~\eqref{l3hfs}. The concrete assignments of quantum numbers to the states of Ref.~\cite{Padmanath:2015jea} that we have used are specified in Table~\ref{latstates}.  We performed several sets of fits varying the value of $r_0$; in Table~\ref{fits4p} we present the results with all parameters free, in Table~\ref{fits3p} we fix $\Delta^{(0)}_{(1/2)^{+}}=0.122$~{GeV}$^2$ from the $B$-meson splittings in Table~\ref{delahmm} and in Table~\ref{fits2p} we fix $\Delta^{(0)}_{(1/2)^{+}}=0.122$~{GeV}$^2$ and set $\Delta^{(1,0)}_{(1/2)^{+}}=\Delta^{(1,2)}_{(1/2)^{+}}=0$~{GeV}$^4$.

\begin{table}
\begin{tabular}{||c|c|c|c|c||} \hline \hline
                $l$     & $n$                  & $j$      & $M-M_{\eta_c}[{\rm GeV}]$ \\ \hline                      
\multirow{6}{*}{$0$}    & \multirow{2}{*}{$1$} & $1/2$    & $0.6532(80)$   \\ \cline{3-4}
                        &                      & $3/2$    & $0.7474(88)$   \\ \cline{2-4}
                        & \multirow{2}{*}{$2$} & $1/2$    & $1.3163(216)$  \\ \cline{3-4}
                        &                      & $3/2$    & $1.3297(332)$  \\ \cline{2-4}
                        & \multirow{2}{*}{$3$} & $1/2$    & $1.5427(142)$  \\ \cline{3-4}
                        &                      & $3/2$    & $1.5435(291)$  \\ \hline
\multirow{4}{*}{$1$}  & \multirow{2}{*}{$1$} & $1/2$    & $1.0243(114)$  \\ \cline{3-4}
                        &                      & $3/2$    & $1.0733(113)$  \\ \cline{2-4}
                        & \multirow{2}{*}{$2$} & $1/2$    & $1.5829(296)$  \\ \cline{3-4}
                        &                      & $3/2$    & $1.6315(353)$  \\ \hline          
\multirow{12}{*}{$2$} & \multirow{7}{*}{$1$} & $1/2$    & $1.3114(213)$  \\ \cline{3-4}
                        &                      & $3/2\,+$ & $1.2653(232)$  \\ \cline{3-4}
 &                      & $3/2\,-$ & $1.3697(131)$  \\ \cline{3-4}
                        &                      & $5/2\,+$ & $1.3075(130)$  \\ \cline{3-4}
                        &                      & $5/2\,-$ & $1.3542(141)$  \\ \cline{3-4}
 &                      & $7/2$    & $1.3715(97)$   \\ \cline{2-4}
                        & \multirow{7}{*}{$2$} & $1/2$    & $1.5044(181)$  \\ \cline{3-4}
                        &                      & $3/2\,+$ & $1.4243(296)$  \\ \cline{3-4}
 &                      & $3/2\,-$ & $1.5331(222)$  \\ \cline{3-4}
                        &                      & $5/2\,+$ & $1.5017(193)$  \\ \cline{3-4}
                        &                      & $5/2\,-$ & $1.5127(157)$  \\ \cline{3-4}
 &                      & $7/2$    & $1.5366(154)$  \\ \hline
\multirow{2}{*}{$3$}    & \multirow{2}{*}{$1$} & $5/2$    & $1.5502(221)$  \\ \cline{3-4}
                        &                      & $7/2$    & $1.5618(678)$  \\ \hline\hline
\end{tabular}
\caption{Assignments of quantum numbers of the lattice states of Ref.~\cite{Padmanath:2015jea} used in the fits of Sec.~\ref{sec:intfp}.}
\label{latstates}
\end{table}

\begin{table}
\begin{tabular}{||ccccccc||}\hline\hline
 $r_0~[{\rm fm}]$ & $\Delta^{(0)}_{(1/2)^{+}}~[{\rm GeV}^2]$  & $\Delta^{(1,0)}_{(1/2)^{+}}~[{\rm GeV}^4]$  & $\Delta^{(1,2)}_{(1/2)^{+}}~[{\rm GeV}^4]$ & $\Lambda_f~[{\rm GeV}]$ & $\Lambda_f'~[{\rm GeV}]$ & $\chi^2_{\rm d.o.f}$\\\hline
$0.0$    & n/a          & n/a         & n/a           & $-0.341(8)$  & $-0.268(16)$  & $0.62$ \\   
$0.1$    & $-3.13(12)$  & $15.17(37)$ & $19(77)$      & $-0.231(10)$ & $-0.282(19)$  & $0.67$ \\
$0.2$    & $-0.076(22)$ & $0.514(22)$ & $0.35(1.76)$  & $-0.196(13)$ & $-0.274(23)$  & $0.64$ \\
$0.3$    & $0.135(10)$  & $0.047(5)$  & $-0.045(203)$ & $-0.169(18)$ & $-0.264(32)$  & $0.63$ \\
$0.4$    & $0.163(6)$   & $-0.006(2)$ & $-0.041(64)$  & $-0.154(27)$ & $-0.272(45)$  & $0.63$ \\
$0.5$    & $0.165(5)$   & $-0.016(1)$ & $-0.023(26)$  & $-0.176(43)$ & $-0.322(64)$  & $0.64$ \\
$0.6$    & $0.159(4)$   & $-0.016(1)$ & $-0.012(14)$  & $-0.256(67)$ & $-0.427(90)$  & $0.66$ \\
$\infty$ & $0.086(3)$   & $-0.002(1)$ & $-0.002(2)$   & n/a          & n/a           & $0.94$ \\\hline\hline
\end{tabular}
\caption{Global fit of $\kappa^p=(1/2)^+$ $l=0,1,2,3$ multiplets hyperfine splittings for all the lattice data available for various values of $r_0$.}
\label{fits4p}
\end{table}

\begin{table}
\begin{tabular}{||cccccc||}\hline\hline
$r_0~[{\rm fm}]$  & $\Delta^{(1,0)}_{(1/2)^{+}}~[{\rm GeV}^4]$ & $\Delta^{(1,2)}_{(1/2)^{+}}~[{\rm GeV}^4]$ & $\Lambda_f~[{\rm GeV}]$ & $\Lambda_f'~[{\rm GeV}]$ & $\chi^2_{\rm d.o.f}$\\\hline
$0.1$    & $1.89(37)$    & $-8.5(77.2)$  & $-0.308(10)$ & $-0.267(19)$ & $0.66$ \\
$0.2$    & $0.231(22)$   & $0.39(1.72)$  & $-0.249(12)$ & $-0.283(23)$ & $0.62$ \\
$0.3$    & $0.056(5)$    & $-0.055(226)$ & $-0.158(18)$ & $-0.258(31)$ & $0.59$ \\
$0.4$    & $0.013(2)$    & $-0.061(63)$  & $-0.086(27)$ & $-0.223(44)$ & $0.61$ \\
$0.5$    & $-0.0006(14)$ & $-0.036(26)$  & $-0.048(43)$ & $-0.216(64)$ & $0.64$ \\
$0.6$    & $-0.005(1)$   & $-0.019(14)$  & $-0.061(67)$ & $-0.262(91)$ & $0.66$ \\ 
$\infty$ & $-0.0054(5)$  & $-0.0089(29)$ & n/a          & n/a          & $2.67$ \\ \hline\hline
\end{tabular}
\caption{Global fit of $\kappa^p=(1/2)^+$ $l=0,1,2,3$ multiplets hyperfine splittings for all the lattice data available for various values of $r_0$ with $\Delta^{(0)}_{(1/2)^{+}}=0.122$~{GeV}$^2$ from the $B$-meson splittings in Table~\ref{delahmm}.}
\label{fits3p}
\end{table}

\begin{table}
\begin{tabular}{||cccc||}\hline\hline
$r_0~[{\rm fm}]$ & $\Lambda_f~[{\rm GeV}]$ & $\Lambda_f'~[{\rm GeV}]$ & $\chi^2_{\rm d.o.f}$\\\hline
$0.1$    & $-0.355(10)$ & $-0.265(19)$  & $0.66$ \\
$0.2$    & $-0.368(13)$ & $-0.264(25)$  & $0.72$ \\
$0.3$    & $-0.348(19)$ & $-0.270(33)$  & $0.69$ \\
$0.4$    & $-0.266(27)$ & $-0.286(44)$  & $0.60$ \\
$0.5$    & $-0.085(41)$ & $-0.314(61)$  & $0.58$ \\
$0.6$    & $ 0.224(75)$ & $-0.353(102)$ & $0.83$ \\  \hline\hline
\end{tabular}
\caption{Global fit of $\kappa^p=(1/2)^+$ $l=0,1,2,3$ multiplets hyperfine splittings for all the lattice data available for various values of $r_0$ with $\Delta^{(0)}_{(1/2)^{+}}=0.122$~{GeV}$^2$ from the $B$-meson splittings in Table~\ref{delahmm} and $\Delta^{(1,0)}_{(1/2)^{+}}=\Delta^{(1,2)}_{(1/2)^{+}}=0$.}
\label{fits2p}
\end{table}

\begin{figure}
\includegraphics[width=0.4\linewidth]{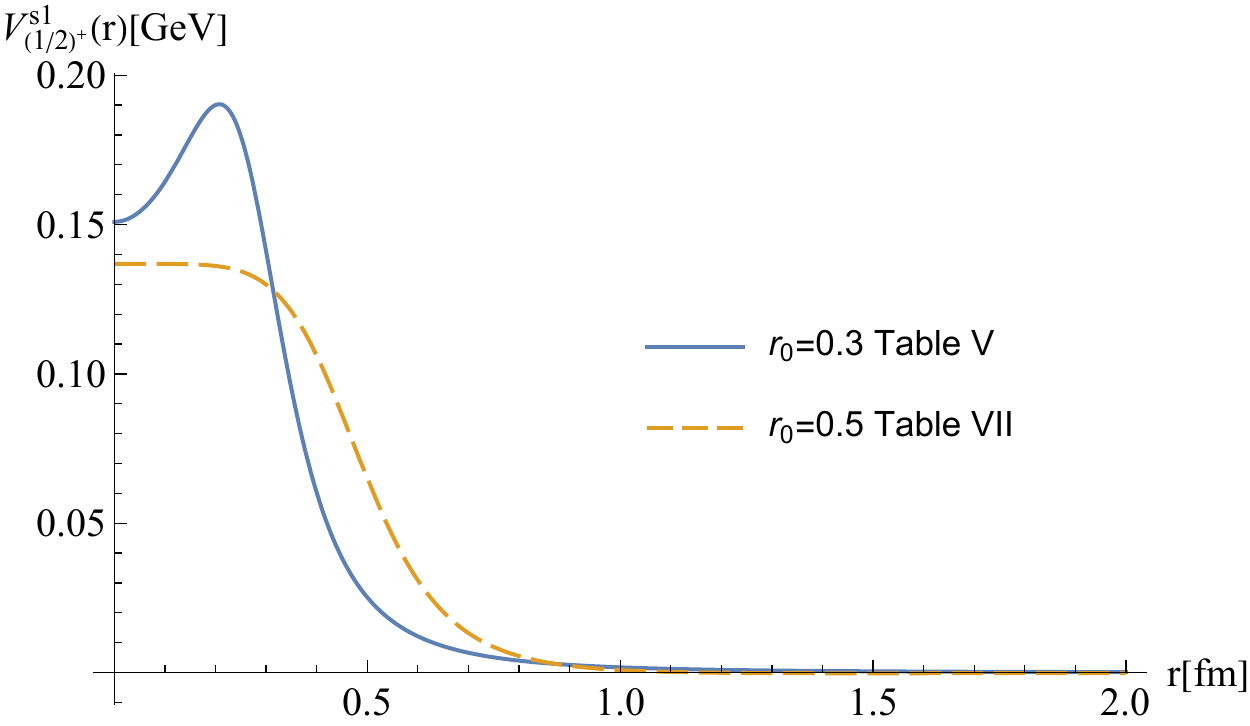}\includegraphics[width=0.4\linewidth]{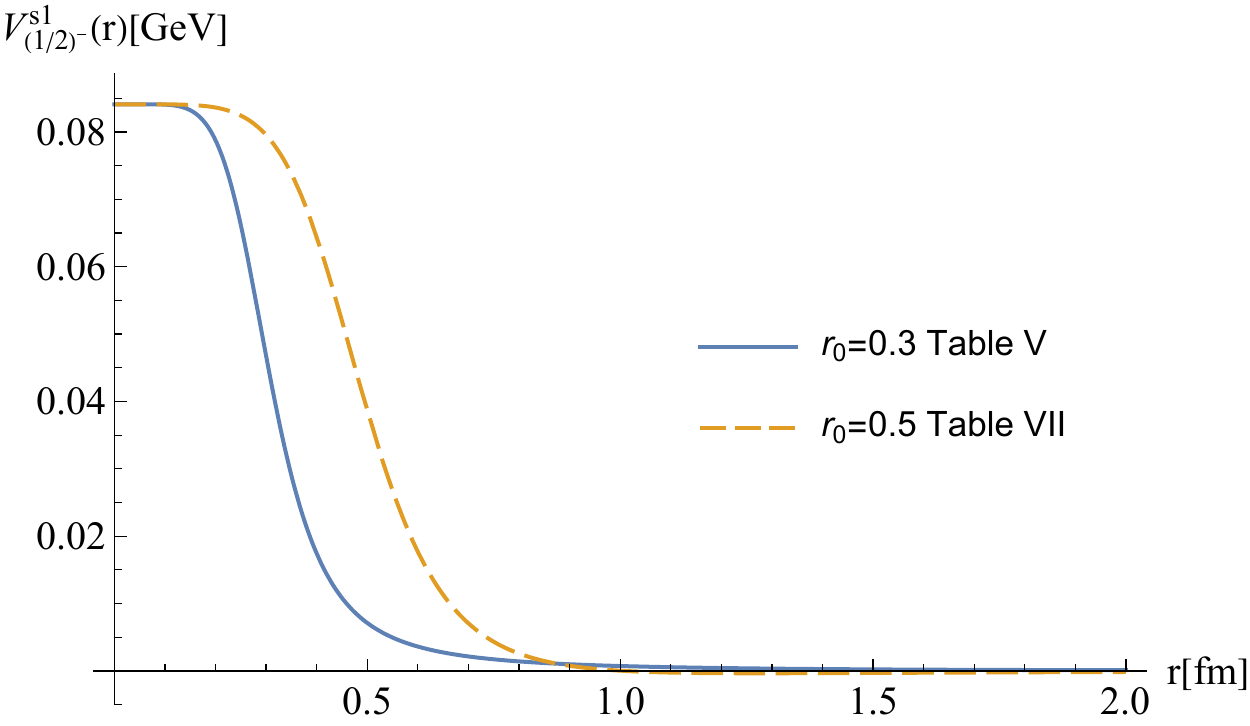} \\
\includegraphics[width=0.4\linewidth]{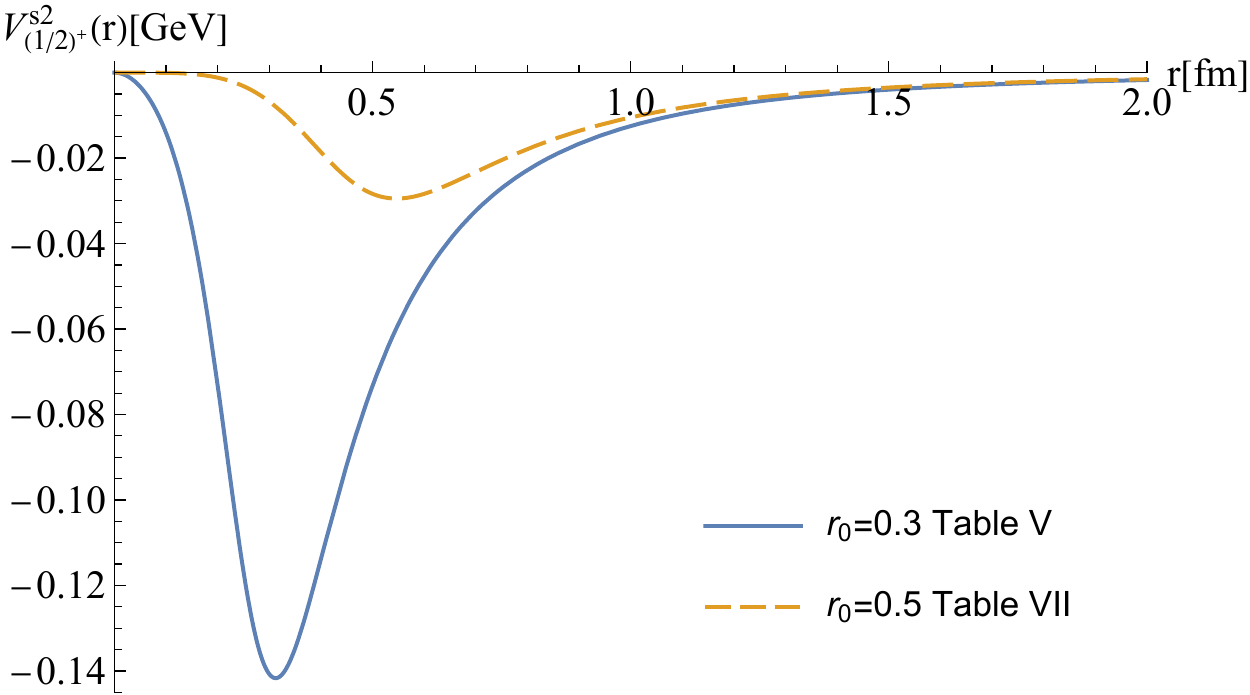}\includegraphics[width=0.4\linewidth]{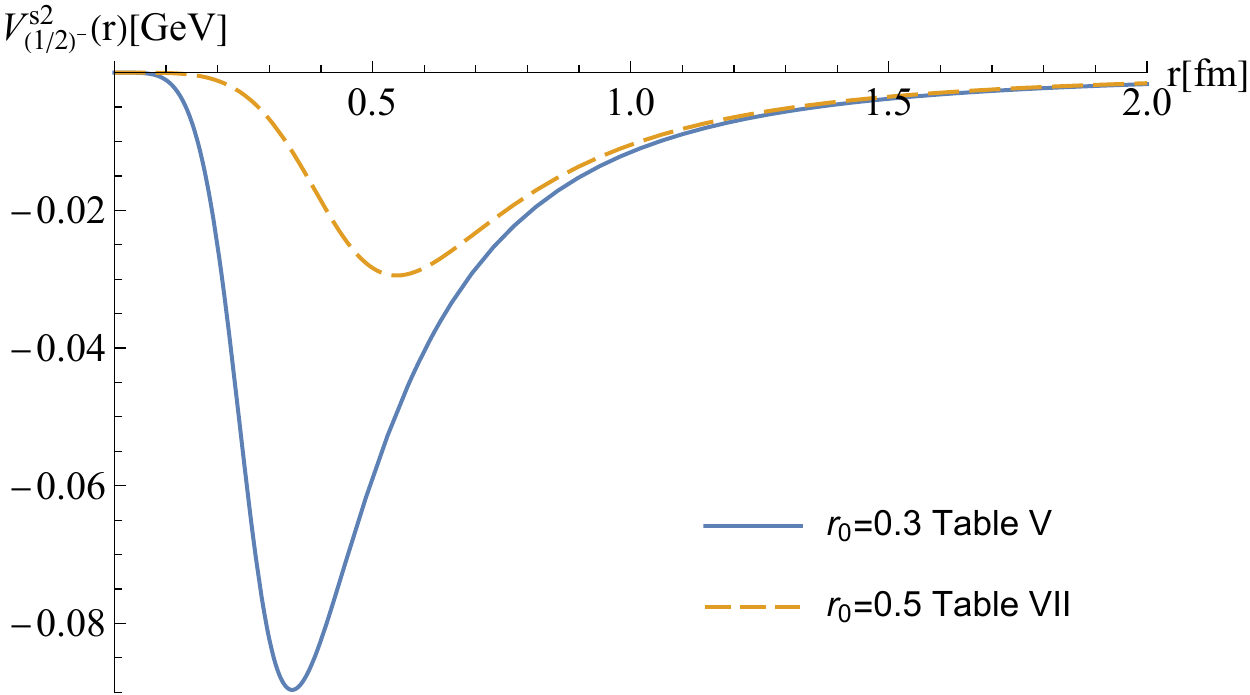} \\
\includegraphics[width=0.4\linewidth]{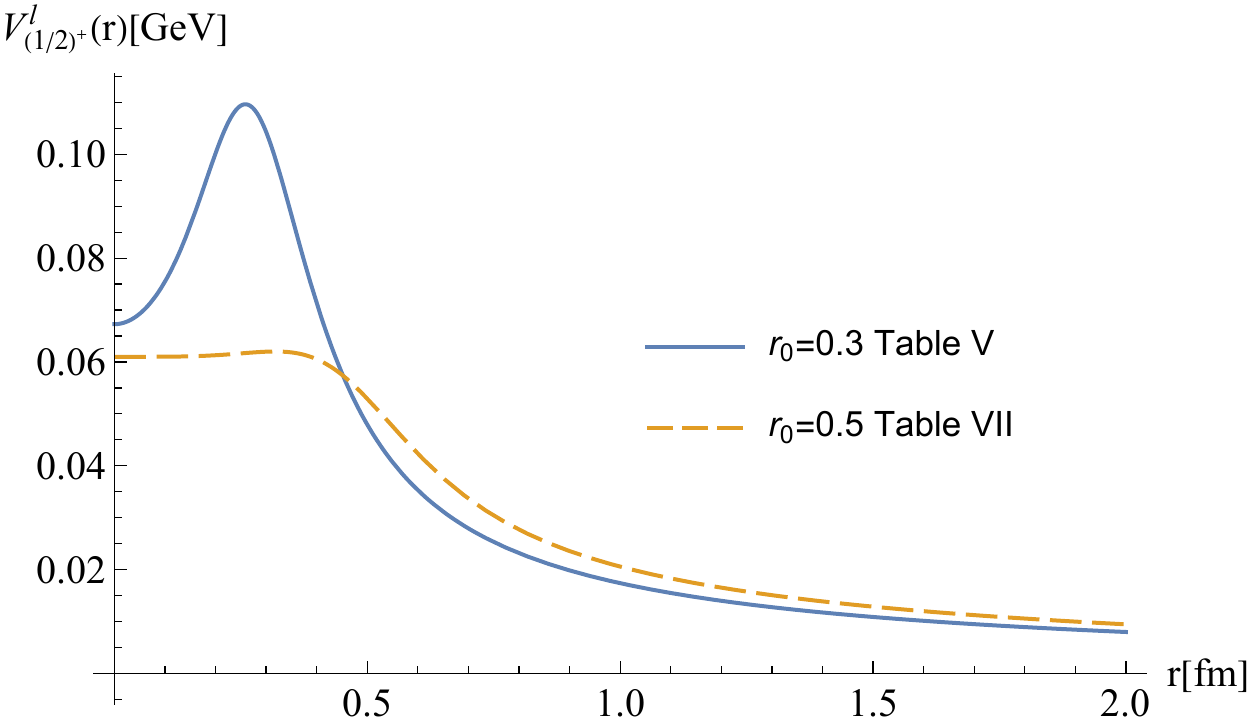}\includegraphics[width=0.4\linewidth]{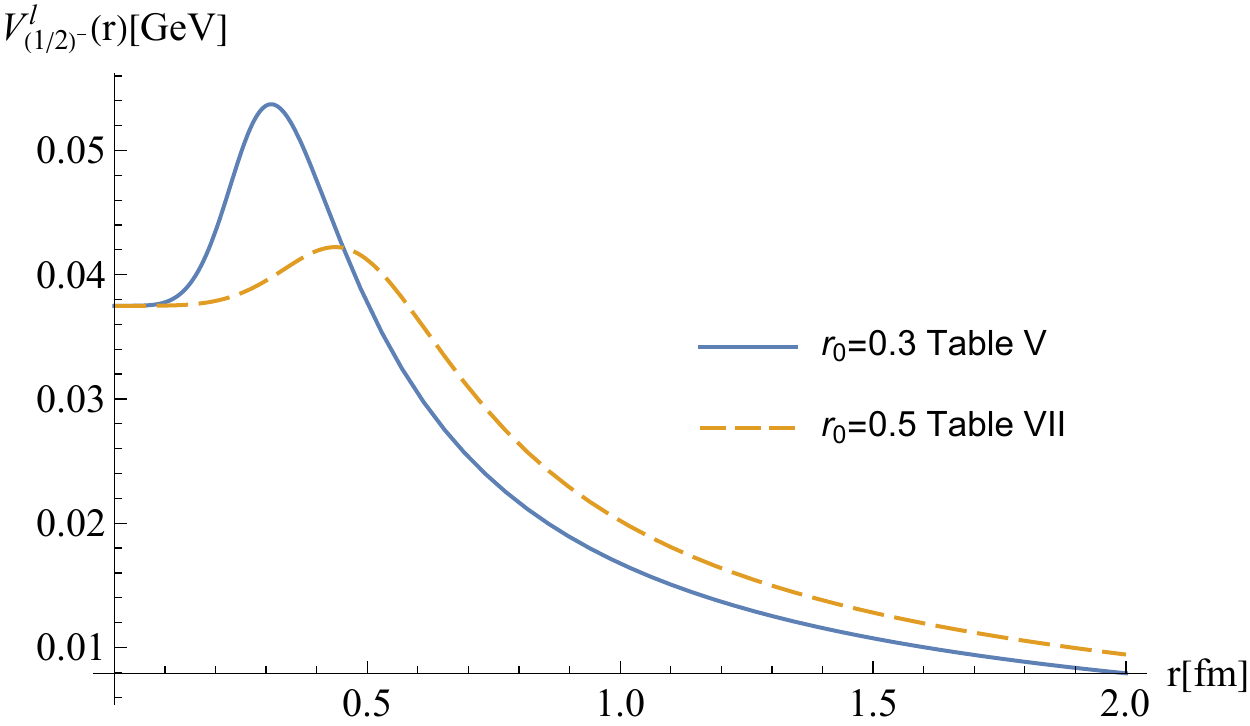} \\
\caption{Plot of the potentials in Eqs.~\eqref{s1int}-\eqref{lint} for the values of the parameters of $r_0=0.5$~fm in Table~\ref{fits2p} and  $r_0=0.3$~fm in Table~\ref{fits4p}. In the case of $\kappa^p=(1/2)^-$ we take $\Delta^{(0)}_{(1/2)^{-}}=0.075$~GeV$^2$ and $\Delta^{(1,2)}_{(1/2)^{-}}=0$~GeV$^4$ and the values of $\Lambda_f$ and $\Lambda_f'$ indicated in the legend. In the potentials $V^{s1}$ and $V^{s2}$ we use the two-loop, RG improved expression of $c_F=c_F(1~{\rm GeV},\,m_c)$.}
\label{plot_pot}
\end{figure}

Several conclusions can be extracted from the fits. First of all, when we restrict the fit to either the short-distance form of potential ($r_0=\infty$) or the long-distance form of it ($r_0=0$), we see from Table \ref{fits4p} that the latter produces a much better fit than the former. This indicates both that the long distance form is important and that the EST provides a good description of it. We observe that the value of $\Delta^{(1,2)}_{(1/2)^{+}}$ changes significantly, carries large uncertainty, and in the best fits it is compatible with $0$. In the case of $\Delta^{(1,0)}_{(1/2)^{+}}$ its value also shows variation, however it is significantly different from zero. Nevertheless, the inclusion of the next-to-leading order terms in the short-distance potentials does not improve the overall quality of the fits. Therefore, it seems that with the current lattice data it is not possible to constrain these next-to-leading order terms in the multipole expansion. The values of $\Lambda_f$ and $\Lambda_f'$ stay consistent across the different sets of fits, with $\Lambda_f'$ being very stable while $\Lambda_f$ decreasing in absolute value as $r_0$ gets larger and even changing sign. Our preferred fit is the one with minimal $\chi^2_{\rm d.o.f}$ in Table~\ref{fits2p} corresponding to $r_0=0.5$~fm. In Fig.~\ref{plot_pot} we plot the interpolated expressions of the potentials in Eqs.~\eqref{s1int}-\eqref{lint} for the parameter set in the entry for $r_0=0.5$~fm in Table~\ref{fits2p} and  $r_0=0.3$~fm in Table~\ref{fits4p}. In the case of the potentials for $\kappa^p=(1/2)^-$, we plot the potentials with $\Delta^{(0)}_{(1/2)^{-}}=0.075$~GeV$^2$ from the neutral $D$-meson entry in Table~\ref{delahmm}, $\Delta^{(1,0)}_{(1/2)^{-}}=\Delta^{(1,2)}_{(1/2)^{-}}=0$~GeV$^4$ and the values of $\Lambda_f$ and $\Lambda_f'$ from the entries for $r_0=0.5$~fm in Table~\ref{fits2p} and $r_0=0.3$~fm in Table~\ref{fits4p}. Although in some cases the potentials in Fig.~\ref{plot_pot} show significant variation depending on the parameter set used, we will show in the following section that this is not the case for the values of the hyperfine splittings.

\subsection{Doubly heavy baryon spectra}\label{sec:dhbs}

Now we compute the spectrum of double charm and bottom baryons including the hyperfine contributions using the interpolated potentials in Eqs.~\eqref{s1int}-\eqref{lint}. For the states associated to the $(1/2)_g$ static energy, we take values of the parameters from the entry $r_0=0.5$~fm in Table~\ref{fits2p} and for comparison the entry $r_0=0.3$~fm in Table~\ref{fits4p}. The results can be found in Tables~\ref{hft12gc} and \ref{hft12gb} for double charm and double bottom baryons, respectively. For the states associated to the $(1/2)_u'$ static energy we set $\Delta^{(0)}_{(1/2)^{-}}=0.075$~GeV$^2$ and $\Delta^{(1,0)}_{(1/2)^{-}}=\Delta^{(1,2)}_{(1/2)^{-}}=0$~GeV$^4$ and take the values of $\Lambda_f$ and $\Lambda_f'$ from the $r_0=0.5$~fm entry in Table~\ref{fits2p} and for comparison the entry $r_0=0.3$~fm in Table~\ref{fits4p}. The results can be found in Tables~\ref{hft12upc} and \ref{hft12upb} for double charm and bottom baryons respectively. The results for the spectra for the two sets of parameters are very close.

\begin{table}
\begin{tabular}{||c|c|c|c|c|c|c|c||} \hline \hline
\multirow{2}{*}{$l$}    & \multirow{2}{*}{$n$} & \multirow{2}{*}{$M^{(0)}$} & \multirow{2}{*}{$j$} & \multicolumn{2}{c|}{$r_0=0.5$~fm\,Table~\ref{fits2p}} & \multicolumn{2}{c||}{$r_0=0.3$~fm\,Table~\ref{fits4p}}  \\ \cline{5-8}
                        &                      &                          &       & $M^{(1)}$    & $M$     & $M^{(1)}$    & $M$     \\ \hline 
\multirow{7}{*}{$0$}    & \multirow{2}{*}{$1$} & \multirow{2}{*}{$3.712$} & $1/2$ & $-0.059(2)$  & $3.653$ & $-0.058(5)$  & $3.654$ \\ \cline{4-8}
                        &                      &                          & $3/2$ & $0.029(1)$   & $3.741$ & $0.029(2)$   & $3.741$ \\ \cline{2-8}
                        & \multirow{2}{*}{$2$} & \multirow{2}{*}{$4.286$} & $1/2$ & $-0.020(2)$  & $4.266$ & $-0.029(3)$  & $4.257$ \\ \cline{4-8}
                        &                      &                          & $3/2$ & $0.010(1)$   & $4.296$ & $0.015(1)$   & $4.301$ \\ \cline{2-8}
                        & \multirow{2}{*}{$3$} & \multirow{2}{*}{$4.748$} & $1/2$ & $-0.013(2)$  & $4.735$ & $-0.020(2)$  & $4.728$ \\ \cline{4-8}
                        &                      &                          & $3/2$ & $0.007(1)$   & $4.755$ & $0.010(1)$   & $4.758$ \\ \hline
\multirow{4}{*}{$1$}  & \multirow{2}{*}{$1$} & \multirow{2}{*}{$4.062$}   & $1/2$ & $-0.034(4)$  & $4.028$ & $-0.035(8)$  & $4.027$ \\ \cline{4-8}
                        &                      &                          & $3/2$ & $0.017(2)$   & $4.079$ & $0.017(4)$   & $4.079$ \\ \cline{2-8}
                        & \multirow{2}{*}{$2$} & \multirow{2}{*}{$4.552$} & $1/2$ & $-0.024(3)$  & $4.528$ & $-0.026(6)$  & $4.526$ \\ \cline{4-8}
                        &                      &                          & $3/2$ & $0.012(1)$   & $4.564$ & $0.013(3)$   & $4.565$ \\ \hline               
\multirow{12}{*}{$2$} & \multirow{7}{*}{$1$} & \multirow{7}{*}{$4.353$}   & $1/2$ & $-0.020(7)$  & $4.333$ & $-0.009(8)$  & $4.344$ \\ \cline{4-8}
                        &                      &                          & $3/2$ & $-0.032(6)$  & $4.321$ & $-0.026(6)$  & $4.327$ \\ \cline{4-8}
											  &                      &                          & $3/2$ & $0.015(4)$   & $4.368$ & $0.009(5)$   & $4.362$ \\ \cline{4-8}
                        &                      &                          & $5/2$ & $-0.052(6)$  & $4.301$ & $-0.053(5)$  & $4.300$ \\ \cline{4-8}
                        &                      &                          & $5/2$ & $0.023(3)$   & $4.376$ & $0.020(4)$   & $4.373$ \\ \cline{4-8}
											  &                      &                          & $7/2$ & $0.035(4)$   & $4.388$ & $0.035(4)$   & $4.388$ \\ \cline{2-8}
                        & \multirow{7}{*}{$2$} & \multirow{7}{*}{$4.794$} & $1/2$ & $-0.017(5)$  & $4.777$ & $-0.008(8)$  & $4.786$ \\ \cline{4-8}
                        &                      &                          & $3/2$ & $-0.026(4)$  & $4.768$ & $-0.022(6)$  & $4.772$ \\ \cline{4-8}
											  &                      &                          & $3/2$ & $0.011(3)$   & $4.805$ & $0.007(4)$   & $4.801$ \\ \cline{4-8}
                        &                      &                          & $5/2$ & $-0.042(4)$  & $4.752$ & $-0.044(4)$  & $4.750$ \\ \cline{4-8}
                        &                      &                          & $5/2$ & $0.019(3)$   & $4.813$ & $0.017(4)$   & $4.811$ \\ \cline{4-8}
											  &                      &                          & $7/2$ & $0.029(3)$   & $4.823$ & $0.030(4)$   & $4.824$ \\ \hline
\multirow{2}{*}{$3$}    & \multirow{2}{*}{$1$} & \multirow{2}{*}{$4.612$} & $5/2$ & $-0.043(8)$  & $4.569$ & $-0.037(5)$  & $4.575$ \\ \cline{4-8}
                        &                      &                          & $7/2$ & $0.032(6)$   & $4.644$ & $0.028(4)$   & $4.640$ \\ \hline\hline
\end{tabular}
\caption{Hyperfine contributions to the double charm baryons for the $(1/2)_g$ static energy for two sets of parameters of the hyperfine potentials. All masses in GeV units.}
\label{hft12gc}
\end{table}

\begin{table}
\begin{tabular}{||c|c|c|c|c|c|c|c||} \hline \hline
\multirow{2}{*}{$l$}    & \multirow{2}{*}{$n$} & \multirow{2}{*}{$M^{(0)}$} & \multirow{2}{*}{$j$} & \multicolumn{2}{c|}{$r_0=0.5$~fm\,Table~\ref{fits2p}} & \multicolumn{2}{c||}{$r_0=0.3$~fm\,Table~\ref{fits4p}}  \\ \cline{5-8}
                        &                      &                            &       & $M^{(1)}$ & $M$      & $M^{(1)}$ & $M$ \\ \hline 
\multirow{8}{*}{$0$}  & \multirow{2}{*}{$1$} & \multirow{2}{*}{$10.140$}    & $1/2$ & $-0.020(1)$  & $10.120$ & $-0.023(1)$  & $10.117$ \\ \cline{4-8}
                        &                      &                            & $3/2$ & $0.010(0)$   & $10.150$ & $0.011(1)$   & $10.151$ \\ \cline{2-8}
                        & \multirow{2}{*}{$2$} & \multirow{2}{*}{$10.542$}  & $1/2$ & $-0.009(1)$  & $10.533$ & $-0.009(1)$  & $10.533$ \\ \cline{4-8}
                        &                      &                            & $3/2$ & $0.004(0)$   & $10.546$ & $0.005(0)$   & $10.547$ \\ \cline{2-8}
                        & \multirow{2}{*}{$3$} & \multirow{2}{*}{$10.856$}  & $1/2$ & $-0.006(1)$  & $10.850$ & $-0.006(0)$  & $10.850$ \\ \cline{4-8}
                        &                      &                            & $3/2$ & $0.003(0)$   & $10.859$ & $0.003(0)$   & $10.859$ \\ \cline{2-8}
										    & \multirow{2}{*}{$4$} & \multirow{2}{*}{$11.131$}  & $1/2$ & $-0.004(0)$  & $11.127$ & $-0.005(1)$  & $11.126$ \\ \cline{4-8}
                        &                      &                            & $3/2$ & $0.002(0)$   & $11.133$ & $0.003(0)$   & $11.134$ \\ \hline
\multirow{6}{*}{$1$}  & \multirow{2}{*}{$1$} & \multirow{2}{*}{$10.398$}    & $1/2$ & $-0.012(1)$  & $10.386$ & $-0.016(5)$  & $10.382$ \\ \cline{4-8}
                        &                      &                            & $3/2$ & $0.006(0)$   & $10.404$ & $0.008(3)$   & $10.406$ \\ \cline{2-8}
                        & \multirow{2}{*}{$2$} & \multirow{2}{*}{$10.731$}  & $1/2$ & $-0.010(1)$  & $10.721$ & $-0.011(3)$  & $10.720$ \\ \cline{4-8}
                        &                      &                            & $3/2$ & $0.005(1)$   & $10.736$ & $0.006(1)$   & $10.737$ \\ \cline{2-8}      
										    & \multirow{2}{*}{$3$} & \multirow{2}{*}{$11.016$}  & $1/2$ & $-0.008(1)$  & $11.008$ & $-0.009(2)$  & $11.007$ \\ \cline{4-8}
                        &                      &                            & $3/2$ & $0.004(0)$   & $11.020$ & $0.004(1)$   & $11.020$ \\ \hline      
\multirow{18}{*}{$2$} & \multirow{7}{*}{$1$} & \multirow{7}{*}{$10.600$}    & $1/2$ & $-0.009(2)$  & $10.591$ & $-0.007(7)$  & $10.593$ \\ \cline{4-8}
                        &                      &                            & $3/2$ & $-0.015(2)$  & $10.585$ & $-0.014(5)$  & $10.586$ \\ \cline{4-8}
											  &                      &                            & $3/2$ & $0.005(1)$   & $10.605$ & $0.004(4)$   & $10.604$ \\ \cline{4-8}
                        &                      &                            & $5/2$ & $-0.023(2)$  & $10.577$ & $-0.026(4)$  & $10.574$ \\ \cline{4-8}
                        &                      &                            & $5/2$ & $0.010(1)$   & $10.610$ & $0.010(3)$   & $10.610$ \\ \cline{4-8}
											  &                      &                            & $7/2$ & $0.017(1)$   & $10.617$ & $0.018(3)$   & $10.618$ \\ \cline{2-8}
                        & \multirow{7}{*}{$2$} & \multirow{7}{*}{$10.897$}  & $1/2$ & $-0.007(2)$  & $10.890$ & $-0.006(5)$  & $10.891$ \\ \cline{4-8}
                        &                      &                            & $3/2$ & $-0.011(1)$  & $10.886$ & $-0.011(4)$  & $10.886$ \\ \cline{4-8}
											  &                      &                            & $3/2$ & $0.004(1)$   & $10.901$ & $0.003(3)$   & $10.900$ \\ \cline{4-8}
                        &                      &                            & $5/2$ & $-0.017(1)$  & $10.880$ & $-0.020(3)$  & $10.877$ \\ \cline{4-8}
                        &                      &                            & $5/2$ & $0.008(1)$   & $10.905$ & $0.008(2)$   & $10.905$ \\ \cline{4-8}
											  &                      &                            & $7/2$ & $0.012(1)$   & $10.909$ & $0.015(3)$   & $10.912$ \\ \cline{2-8}
											  & \multirow{7}{*}{$3$} & \multirow{7}{*}{$11.162$}  & $1/2$ & $-0.006(1)$  & $11.156$ & $-0.005(4)$  & $11.157$ \\ \cline{4-8}
                        &                      &                            & $3/2$ & $-0.009(1)$  & $11.153$ & $-0.010(3)$  & $11.152$ \\ \cline{4-8}
											  &                      &                            & $3/2$ & $0.004(1)$   & $11.166$ & $0.003(2)$   & $11.165$ \\ \cline{4-8}
                        &                      &                            & $5/2$ & $-0.014(1)$  & $11.148$ & $-0.017(2)$  & $11.145$ \\ \cline{4-8}
                        &                      &                            & $5/2$ & $0.007(1)$   & $11.169$ & $0.007(2)$   & $11.169$ \\ \cline{4-8}
											  &                      &                            & $7/2$ & $0.010(1)$   & $11.172$ & $0.012(2)$   & $11.174$ \\ \hline
\multirow{4}{*}{$3$}  & \multirow{2}{*}{$1$} & \multirow{2}{*}{$10.777$}    & $5/2$ & $-0.020(3)$  & $10.757$ & $-0.019(4)$  & $10.758$ \\ \cline{4-8}
                        &                      &                            & $7/2$ & $0.015(2)$   & $10.792$ & $0.014(3)$   & $10.791$ \\ \cline{2-8}
										    & \multirow{2}{*}{$2$} & \multirow{2}{*}{$11.051$}  & $5/2$ & $-0.016(2)$  & $11.035$ & $-0.017(3)$  & $11.034$ \\ \cline{4-8}
                        &                      &                            & $7/2$ & $0.012(2)$   & $11.063$ & $0.012(2)$   & $11.063$ \\ \hline\hline
\end{tabular}
\caption{Hyperfine contributions to the double bottom baryons for the $(1/2)_g$ static energy for two sets of parameters of the hyperfine potentials. All masses in GeV units.}
\label{hft12gb}
\end{table}

\begin{table}
\begin{tabular}{||c|c|c|c|c|c|c|c||} \hline \hline
\multirow{2}{*}{$l$}   & \multirow{2}{*}{$n$} & \multirow{2}{*}{$M^{(0)}$} & \multirow{2}{*}{$j$} & \multicolumn{2}{c|}{$r_0=0.5$~fm\,table~\ref{fits2p}} &  \multicolumn{2}{c||}{$r_0=0.3$~fm\,table~\ref{fits4p}}  \\ \cline{5-8}
                       &                      &                            &       & $M^{(1)}$ & $M$     & $M^{(1)}$ & $M$     \\ \hline 
  \multirow{4}{*}{$0$} & \multirow{2}{*}{$1$} & \multirow{2}{*}{$4.095$}   & $1/2$ & $-0.033(9)$  & $4.062$ & $-0.025(6)$  & $4.070$ \\ \cline{4-8}
                       &                      &                            & $3/2$ & $0.016(4)$   & $4.111$ & $0.012(3)$   & $4.107$ \\ \cline{2-8}
                       & \multirow{2}{*}{$2$} & \multirow{2}{*}{$4.667$}   & $1/2$ & $-0.07(5)$   & $4.660$ & $-0.015(4)$  & $4.652$ \\ \cline{4-8}
                       &                      &                            & $3/2$ & $0.003(2)$   & $4.670$ & $0.008(2)$   & $4.675$ \\ \hline
\multirow{2}{*}{$1$}   & \multirow{2}{*}{$1$} & \multirow{2}{*}{$4.443$}   & $1/2$ & $-0.033(5)$  & $4.410$ & $-0.033(4)$  & $4.410$ \\ \cline{4-8}
                       &                      &                            & $3/2$ & $0.016(3)$   & $4.459$ & $0.016(2)$   & $4.459$ \\ \hline               
\multirow{7}{*}{$2$}   & \multirow{7}{*}{$1$} & \multirow{7}{*}{$4.732$}   & $1/2$ & $-0.017(9)$  & $4.715$ & $0.000(6)$   & $4.732$ \\ \cline{4-8}
                       &                      &                            & $3/2$ & $-0.035(7)$  & $4.697$ & $-0.025(4)$  & $4.707$ \\ \cline{4-8}
											 &                      &                            & $3/2$ & $ 0.022(5)$  & $4.754$ & $0.008(3)$   & $4.740$ \\ \cline{4-8}
                       &                      &                            & $5/2$ & $-0.063(7)$  & $4.669$ & $-0.066(5)$  & $4.666$ \\ \cline{4-8}
                       &                      &                            & $5/2$ & $0.029(4)$   & $4.761$ & $0.022(3)$   & $4.754$ \\ \cline{4-8}
											 &                      &                            & $7/2$ & $0.037(6)$   & $4.769$ & $0.041(4)$   & $4.773$ \\ \hline\hline
\end{tabular}
\caption{Hyperfine contributions to the double charm baryons for the $(1/2)_u'$ static energy for two sets of parameters $\Lambda_f$, $\Lambda_f'$ of the hyperfine potentials ($\Delta^{(0)}_{(1/2)^{-}}=0.075$~GeV$^2$, $\Delta^{(1,0)}_{(1/2)^{-}}=\Delta^{(1,2)}_{(1/2)^{-}}=0$~GeV$^4$). All masses in GeV units.}
\label{hft12upc}
\end{table}

\begin{table}
\begin{tabular}{||c|c|c|c|c|c|c|c||} \hline \hline
\multirow{2}{*}{$l$}   & \multirow{2}{*}{$n$} & \multirow{2}{*}{$M^{(0)}$} & \multirow{2}{*}{$j$} & \multicolumn{2}{c|}{$r_0=0.5$~fm\,Table~\ref{fits2p}} &  \multicolumn{2}{c||}{$r_0=0.3$~fm\,Table~\ref{fits4p}}  \\ \cline{5-8}
                       &                      &                            &       & $M^{(1)}$ & $M$      & $M^{(1)}$ & $M$      \\ \hline 
  \multirow{4}{*}{$0$} & \multirow{2}{*}{$1$} & \multirow{2}{*}{$10.527$}  & $1/2$ & $-0.012(2)$  & $10.515$ & $-0.010(2)$  & $10.517$ \\ \cline{4-8}
                       &                      &                            & $3/2$ & $0.006(1)$   & $10.533$ & $0.005(1)$   & $10.532$ \\ \cline{2-8}
                       & \multirow{2}{*}{$2$} & \multirow{2}{*}{$10.924$}  & $1/2$ & $-0.004(1)$  & $10.920$ & $-0.005(1)$  & $10.919$ \\ \cline{4-8}
                       &                      &                            & $3/2$ & $0.002(1)$   & $10.926$ & $0.002(1)$   & $10.926$ \\ \hline
  \multirow{4}{*}{$1$} & \multirow{2}{*}{$1$} & \multirow{2}{*}{$10.781$}  & $1/2$ & $-0.010(1)$  & $10.771$ & $-0.012(1)$  & $10.769$ \\ \cline{4-8}
                       &                      &                            & $3/2$ & $0.005(1)$   & $10.786$ & $0.006(1)$   & $10.787$ \\ \cline{2-8}
                       & \multirow{2}{*}{$2$} & \multirow{2}{*}{$11.112$}  & $1/2$ & $-0.009(1)$  & $11.103$ & $-0.009(1)$  & $11.103$ \\ \cline{4-8}
                       &                      &                            & $3/2$ & $0.005(1)$   & $11.117$ & $0.005(0)$   & $11.117$ \\ \hline      
\multirow{7}{*}{$2$}   & \multirow{7}{*}{$1$} & \multirow{7}{*}{$10.981$}  & $1/2$ & $-0.008(2)$  & $10.973$ & $-0.004(2)$  & $10.977$ \\ \cline{4-8}
                       &                      &                            & $3/2$ & $-0.012(2)$  & $10.969$ & $-0.011(2)$  & $10.970$ \\ \cline{4-8}
											 &                      &                            & $3/2$ & $0.005(1)$   & $10.986$ & $0.004(1)$   & $10.985$ \\ \cline{4-8}
                       &                      &                            & $5/2$ & $-0.020(2)$  & $10.961$ & $-0.024(2)$  & $10.957$ \\ \cline{4-8}
                       &                      &                            & $5/2$ & $0.009(1)$   & $10.990$ & $0.009(1)$   & $10.990$ \\ \cline{4-8}
											 &                      &                            & $7/2$ & $0.014(2)$   & $10.995$ & $0.016(1)$   & $10.997$ \\ \hline
\multirow{2}{*}{$3$}   & \multirow{2}{*}{$1$} & \multirow{2}{*}{$11.157$}  & $5/2$ & $-0.020(3)$  & $11.137$ & $-0.019(2)$  & $11.138$ \\ \cline{4-8}
                       &                      &                            & $7/2$ & $0.015(2)$   & $11.172$ & $0.014(2)$   & $11.171$ \\ \hline\hline
\end{tabular}
\caption{Hyperfine contributions to the double bottom baryons for the $(1/2)_u'$ static energy for two sets of parameters $\Lambda_f$, $\Lambda_f'$ of the hyperfine potentials ($\Delta^{(0)}_{(1/2)^{-}}=0.075$~GeV$^2$, $\Delta^{(1,0)}_{(1/2)^{-}}=\Delta^{(1,2)}_{(1/2)^{-}}=0$~GeV$^4$). All masses in GeV units.}
\label{hft12upb}
\end{table}

\begin{figure}[ht!]
   \centerline{\includegraphics[width=.6\textwidth]{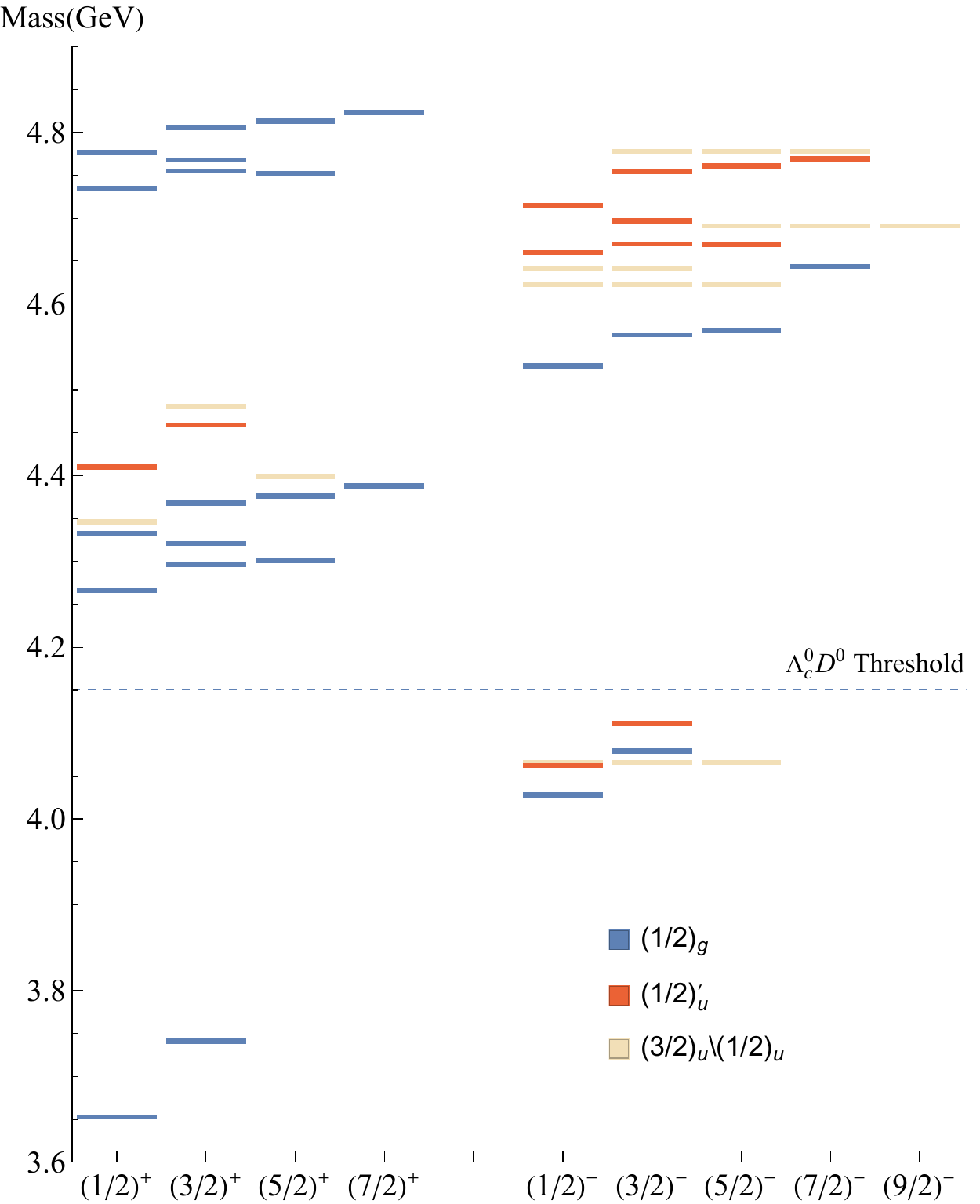}}
	\caption{Spectrum of double charm baryons in terms of $j^{\eta_P}$ states. Each line represents a state. The spectrum corresponds to the results of Tables~\ref{hft12gc} and \ref{hft12upc} for states associated to the $(1/2)_g$ and $(1/2)_u'$ static energies and the results for Ref.~\cite{Soto:2020pfa} for the mixed $(3/2)_u\backslash(1/2)_u$ static energies, which do not include hyperfine contributions. The color indicates the static energies that generate each state.}
	\label{ccplot}
 \end{figure}

\begin{figure}[ht!]
   \centerline{\includegraphics[width=.6\textwidth]{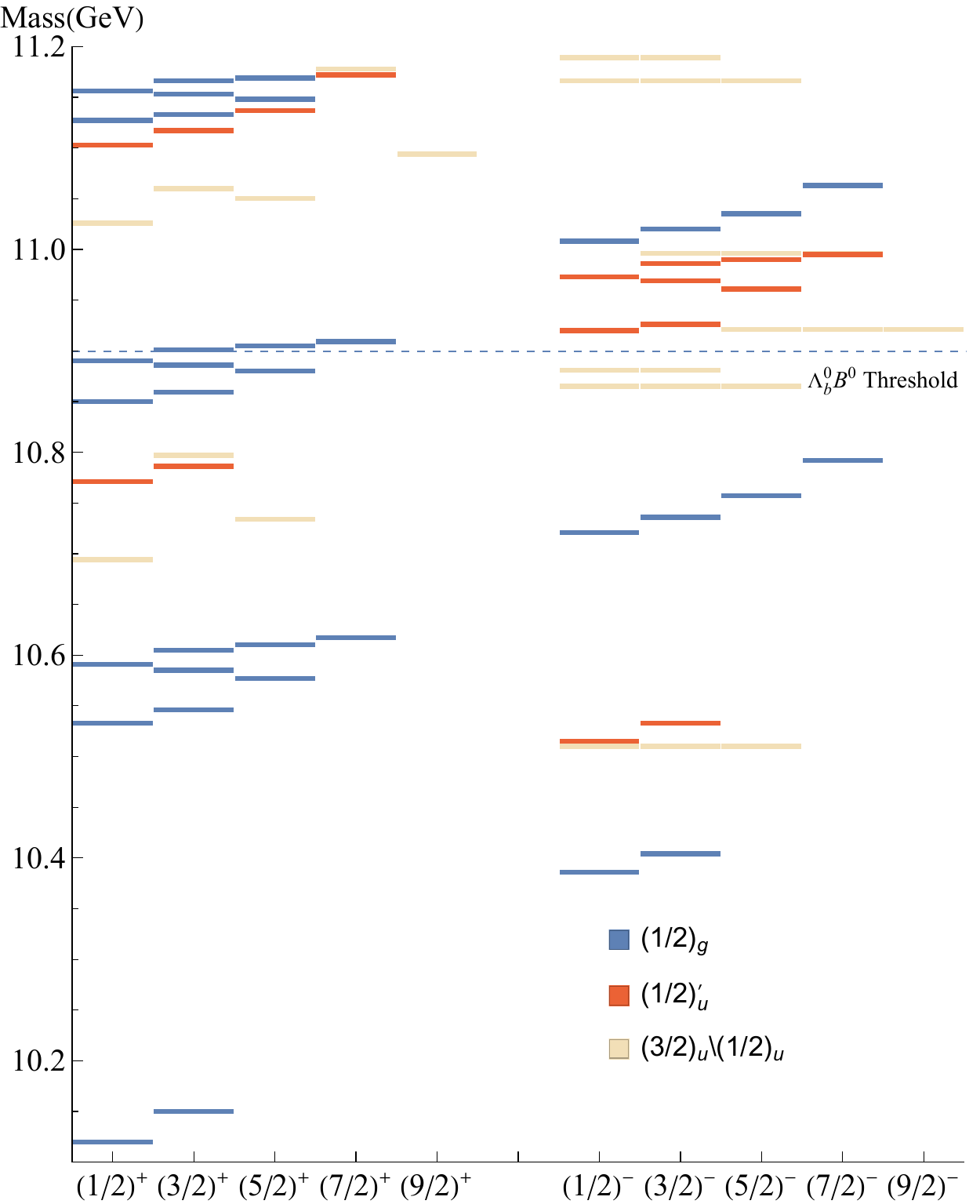}}
	\caption{Spectrum of double bottom baryons in terms of $j^{\eta_P}$ states. Each line represents a state. The spectrum corresponds to the results of Tables~\ref{hft12gb} and \ref{hft12upb} for states associated to the $(1/2)_g$ and $(1/2)_u'$ static energies and the results for Ref.~\cite{Soto:2020pfa} for the mixed $(3/2)_u\backslash(1/2)_u$ static energies, which do not include hyperfine contributions. The color indicates the static energies that generate each state.}
	\label{bbplot}
 \end{figure}

We plot the spectra for double charm and bottom baryons in Figs.~\ref{ccplot} and \ref{bbplot}, respectively, for the parameters of the entry $r_0=0.5$~fm in Table~\ref{fits2p}.

Let us discuss the uncertainties of our results. The leading order masses, $M^{(0)}$, have uncertainties associated to the values of the heavy-quark masses and $\overline{\Lambda}_{(1/2)^+}$, in Eqs.~\eqref{mc}-\eqref{lbar12}, as well as the uncertainty in the parametrization of the static potentials which was estimated as $10$~MeV in Ref.~\cite{Soto:2020pfa}.  Adding these uncertainties in quadrature we obtain $\delta M_{ccq}^{(0)}=39$~MeV and $\delta M_{bbq}^{(0)}=48$~MeV. Furthermore, there is in principle an uncertainty related to the use of an unphysical light-quark mass in the lattice determinations of the static potentials of Refs.~\cite{Najjar:2009da,Najjarthesis} that we used to obtain $M^{(0)}$ in Ref.~\cite{Soto:2020pfa}. We expect the contribution due to the unphysical light-quark mass to be almost independent of $r$. This is supported by the calculations of the charmonium spectrum (with respect to the $\eta_c$ mass) at $m_\pi\sim 400$ MeV \cite{HadronSpectrum:2012gic} and $m_\pi\sim 240$ MeV \cite{Cheung:2016bym}, in which almost no difference is observed for the masses of the states below threshold.\footnote{An increase of the charmonium masses when the light-quark mass  decreases is observed for states about $1$~GeV higher than the $\eta_c$ mass or beyond. If this is interpreted as due to an increase of the string tension with decreasing light-quark masses, then it is consistent with our findings in Appendix  \ref{app:ce}, provided the mass of our fermion on the string is an increasing function of the light-quark mass.} Hence, it will just produce an overall shift to the static energies computed on the lattice. However, in the computation of Ref.~\cite{Soto:2020pfa} the static energies were rescaled in order for the ground state static energy to be given in the short distance by the expression in Eq.~\eqref{sdstp}. Therefore any additive constant contribution to the static energies produces no change in our results.

The hyperfine contribution, $M^{(1)}$, has uncertainties associated to the statistical errors of the values of the parameters and interpolation of the potentials. The former ones are displayed in parentheses in Tables~\ref{hft12gc}-\ref{hft12upb} and are about a few MeV for most cases, although in some instances larger values up to $9$~MeV can also be found. To assess the uncertainty associated to the choice of interpolation of the potentials in Eqs.~\eqref{s1int}-\eqref{lint} we take the difference of the hyperfine contributions computed with the parameter sets for $r_0=0.5$~fm of Table~\ref{fits2p} and $r_0=0.3$~fm of Table~\ref{fits4p}. This uncertainty of the hyperfine contribution amounts to $1$~MeV-$6$~MeV for double charm baryons and $1$~MeV-$4$~MeV for double bottom baryons except for a few cases in Tables~\ref{hft12gc} and \ref{hft12upc} for double charm states where the difference is larger. Finally, one should consider the size of higher-order contributions to the doubly heavy baryon masses. The most important is the contribution form heavy-quark-spin and angular-momentum independent $1/m_Q$ suppressed potential of ${\cal O}(\Lambda^2_{\rm QCD}/m_Q)$, which we estimate as $\sim 64$~MeV and $\sim 19$~MeV for double charm and double bottom baryons, respectively. However, in the case of the hyperfine splittings the previous contribution cancels out and the higher order corrections correspond to the $1/m^2_Q$ suppressed potentials of ${\cal O}(\Lambda^3_{\rm QCD}/m^2_Q)$, which we take as $\sim 14$~MeV and $\sim 1$~MeV for double charm and double bottom baryons, respectively.

As an example, in the following we show the value of the masses for the double charm ground state doublet, often refereed as $\Xi_{cc}[(1/2)^+]$ and $\Xi^{*}_{cc}[(3/2)^+]$, adding the different uncertainties in quadrature:
\begin{align}
&m_{\Xi_{cc}}=3.653(75)~{\rm GeV}\,,\\
&m_{\Xi^{*}_{cc}}=3.741(75)~{\rm GeV}\,,
\end{align}
and for the double bottom ground state doublet:
\begin{align}
&m_{\Xi_{bb}}=10.120(52)~{\rm GeV}\,,\\
&m_{\Xi^{*}_{bb}}=10.150(52)~{\rm GeV}\,.
\end{align}
In the hyperfine splittings most of the uncertainties cancel out and hence our results have higher precision
\begin{align}
m_{\Xi^{*}_{cc}}-m_{\Xi_{cc}}=88(14)~{\rm MeV}\,,\label{hfsccn}\\
m_{\Xi^{*}_{bb}}-m_{\Xi_{bb}}=30(5)~{\rm MeV}\,.
\end{align}
The figures above are compatible with all lattice determinations we are aware of, see Table~VI of Ref.~\cite{Soto:2020pfa} and Table~\ref{hfsbb} for doubly charmed and doubly bottom baryons respectively.

\begin{table}[ht!]
\begin{tabular}{||cc||}\hline\hline
Ref. & $\delta_{hf}~[{\rm MeV}]$     \\ \hline
Our value              & $30(5)$     \\
\cite{Brown:2014ena}   & $34.6(7.8)$ \\
\cite{Lewis:2008fu}    & $26(8)$     \\
\cite{Mohanta:2019mxo} & $32(5)$     \\ \hline\hline
\end{tabular}
\caption{Lattice results for the hyperfine splitting $\delta_{hf}=M_{\Xi^*_{bb}}-M_{\Xi_{bb}}$.}
\label{hfsbb}
\end{table}

Let us finally note that the hyperfine splittings of the $(1/2)_u'$ states are entirely predicted from the long-distance parameters $\Lambda_f$ and $\Lambda_f'$, obtained from fits to the hyperfine splittings of the $(1/2)_g$ states, and the only short-distance parameter, $\Delta^{(0)}_{(1/2)^-}$, obtained from the $D$-meson spectrum. It is then interesting to compare them with the lattice results of Ref.~\cite{Padmanath:2015jea}. For the $(1/2)_u'$ ground state doublet $(1/2^-,3/2^-)$, we obtain from Table~\ref{hft12upc} $49(21)$~MeV for the hyperfine splitting, which agrees well with the $41(21)$~MeV of \cite{Padmanath:2015jea}. This is a nontrivial test of the EST we use, since the $(1/2)^+$ potentials differ from the $(1/2)^-$ at long distances in a very particular way [see \eqref{ldps1}-\eqref{ldpl}]. For the  $(1/2^+,3/2^+)$ first angular excitation, we obtain $49(15)$~MeV, whereas there are two possible values from \cite{Padmanath:2015jea} depending on how the state identifications are made, $25(58)$~MeV or $85(35)$~MeV, both of them compatible with our number within errors. State identification is plagued with ambiguities for higher excitations, which prevent us from making further comparisons. 

\section{Comparison with models}\label{sec:models}

There is a substantial amount of literature regarding doubly heavy baryons in different approaches; various quark models~\cite{DeRujula:1975qlm,Ebert:1996ec,Gerasyuta:1999pc,Itoh:2000um,Gershtein:2000nx,Ebert:2002ig,Albertus:2006ya,Roberts:2007ni,Martynenko:2007je,Yoshida:2015tia,Kiselev:2017eic,Shah:2017liu,Weng:2018mmf,Lu:2017meb}, Bethe-Salpeter equations~\cite{Migura:2006ep,Weng:2010rb,Li:2019ekr}, Born-Oppenheimer approximation with model potential~\cite{Fleck:1989mb,Maiani:2019lpu}, semiempirical mass formulas~\cite{Roncaglia:1995az,Karliner:2014gca,Lichtenberg:1995kg}, QCD sum rules~\cite{Zhang:2008rt,Wang:2018lhz}, Faddeev equations~\cite{Silvestre-Brac:1996tmn}, and bag models~\cite{He:2004px}. In this section we compare our results with a selected set of model computations and other approaches (see \cite{Shah:2017liu,Chen:2016spr,Mohajery:2018qhz} for further comparisons). In Table~\ref{tab:mod:ccq} we have collected the masses of the ground state doublet in the double charm  baryon sector from different approaches. The values of the $\Xi_{cc}$ mass are in good agreement for about $3/4$ of the references, including our own value. Considering the uncertainties only a few works show very significant differences. The values for $\Xi^{*}_{cc}$ show more dispersion with only half of the references being compatible with our own value. On the other hand, the splitting between the two masses is compatible with our value for only $1/4$ of the references. This is in contrast with lattice QCD calculations, which are compatible with our current result for the hyperfine splitting (\ref{hfsccn})  (see Table VI of ref. \cite{Soto:2020pfa}).

\begin{table}[ht!]
\begin{tabular}{||c|c|c||} \hline\hline
Ref.                          & $\Xi_{cc}[(1/2)^+]$ & $\Xi^{*}_{cc}[(3/2)^+]$ \\ \hline 
Our results                   & $3.653(75)$     & $3.741(75)$     \\ 
\cite{DeRujula:1975qlm}       & $3.550-3.760$   & $3.620-3.830$   \\ 
\cite{Fleck:1989mb}           & $3.613$         & $3.741$         \\ 
\cite{Roncaglia:1995az}       & $3.66(7)$       & $3.74(7)$       \\ 
\cite{Lichtenberg:1995kg}     & $3.676$         & $3.746$         \\ 
\cite{Ebert:1996ec}           & $3.660$         & $3.810$         \\ 
\cite{Silvestre-Brac:1996tmn} & $3.608$         & $3.701$         \\ 
\cite{Gerasyuta:1999pc}       & $3.527$         & $3.597$         \\ 
\cite{Itoh:2000um}            & $3.649(10)$     & $3.734(10)$     \\ 
\cite{Ebert:2002ig}           & $3.620$         & $3.727$         \\ 
\cite{He:2004px}              & $3.550$         & $3.590$         \\ 
\cite{Migura:2006ep}          & $3.642$         & $3.723$         \\ 
\cite{Albertus:2006ya}        & $3.612^{+(17)}$ & $3.706^{+(23)}$ \\ 
\cite{Roberts:2007ni}         & $3.676$         & $4.029$         \\ 
\cite{Martynenko:2007je}      & $3.510$         & $3.548$         \\ 
\cite{Zhang:2008rt}           & $4.26(19)$      & $3.9(1)$        \\ 
\cite{Karliner:2014gca}       & $3.627(12)$     & $3.690(12)$     \\ 
\cite{Yoshida:2015tia}        & $3.685$         & $3.754$         \\ 
\cite{Kiselev:2017eic}        & $3.615(55)$     & $3.747(55)$     \\ 
\cite{Shah:2017liu}           & $3.511$         & $3.687$         \\ 
\cite{Lu:2017meb}							& $3.606$         & $3.675$          \\
\cite{Weng:2018mmf}						& $3.633$         & $3.696$           \\
\cite{Wang:2018lhz}           & $3.630^{+(80)}_{-(70)}$ & $3.750(70)$ \\
\cite{Maiani:2019lpu}         & $3.621^{+(17)}_{-(7)}$ & -        \\ 
\cite{Li:2019ekr}             & $3.601$         & $3.703$          \\ \hline
Exp. \cite{LHCb:2019epo}      & $3.6216(4)$   & - \\
\hline\hline 
\end{tabular}
\caption{Masses of double charm baryons from model computations in GeV units.}
\label{tab:mod:ccq}
\end{table}

The masses of the ground state doublet in the double bottom baryon sector are shown in Table~\ref{tab:mod:bbq}. In this case the  differences are a lot more significant. For both the $\Xi_{bb}$ and $\Xi^{*}_{bb}$ only Refs.~\cite{He:2004px,Martynenko:2007je,Karliner:2014gca,Gershtein:2000nx} are compatible with our results and in general there is more dispersion among the values of the different model approaches. Although the values of the hyperfine splittings present less variation in absolute values, in relative terms the variation is also larger than in the double charm baryon sector. Moreover, very few values are compatible with ours. This is due to our small uncertainty for the splitting produced by the cancellation of uncertainties associated to various parameters remaining only the uncertainty on higher order contributions, which are small for double bottom baryons. We note that no reference has compatible results with ours for both for the $\Xi_{bb}$ and $\Xi^{*}_{bb}$ masses and the hyperfine splitting. This is in contrast with the good agreement we get with the available lattice results (see Table \ref{hfsbb}).

\begin{table}[ht!]
\begin{tabular}{||c|c|c||} \hline\hline
Ref.                          & $\Xi_{bb}[(1/2)^+]$      & $\Xi^{*}_{bb}[(3/2)^+]$ \\ \hline
Our results                   & $10.120(52)$             & $10.150(52)$            \\ 
\cite{Roncaglia:1995az}       & $10.34(10)$              & $10.37(10)$             \\ 
\cite{Ebert:1996ec}           & $10.23$                  & $10.28$                 \\ 
\cite{Silvestre-Brac:1996tmn} & $10.198$                 & $10.236$                \\ 
\cite{Gershtein:2000nx}       & $10.093$                 & $10.133$                \\ 
\cite{Ebert:2002ig}           & $10.202$                 & $10.237$                \\ 
\cite{He:2004px}              & $10.10$                  & $10.11$                 \\ 
\cite{Albertus:2006ya}        & $10.197^{+(10)}_{-(17)}$ & $10.236^{+(9)}_{-(17)}$ \\ 
\cite{Roberts:2007ni}         & $10.340$                 & $10.367$                \\ 
\cite{Martynenko:2007je}      & $10.130$                 & $10.144$                \\ 
\cite{Zhang:2008rt}           & $9.78(7)$                & $10.28(5)$              \\ 
\cite{Karliner:2014gca}       & $10.162(12)$             & $10.184(12)$            \\ 
\cite{Yoshida:2015tia}        & $10.314$                 & $10.339$                \\ 
\cite{Shah:2017liu}           & $10.312$                 & $10.335$                \\ 
\cite{Li:2019ekr}             & $10.182$                 & $10.214$                \\
\cite{Weng:2018mmf}						& $10.169$                 & $10.189$                \\
\cite{Lu:2017meb}							& $10.138$                 & $10.169$                \\
\cite{Wang:2018lhz}           & $10.220(70)$             & $10.270(70)$            \\ 
\hline\hline  
\end{tabular}
\caption{Masses of double bottom baryons from model computations in GeV units.}
\label{tab:mod:bbq}
\end{table}

\begin{figure}[ht!]
\begin{tabular}{cc}
\includegraphics[width=.45\textwidth]{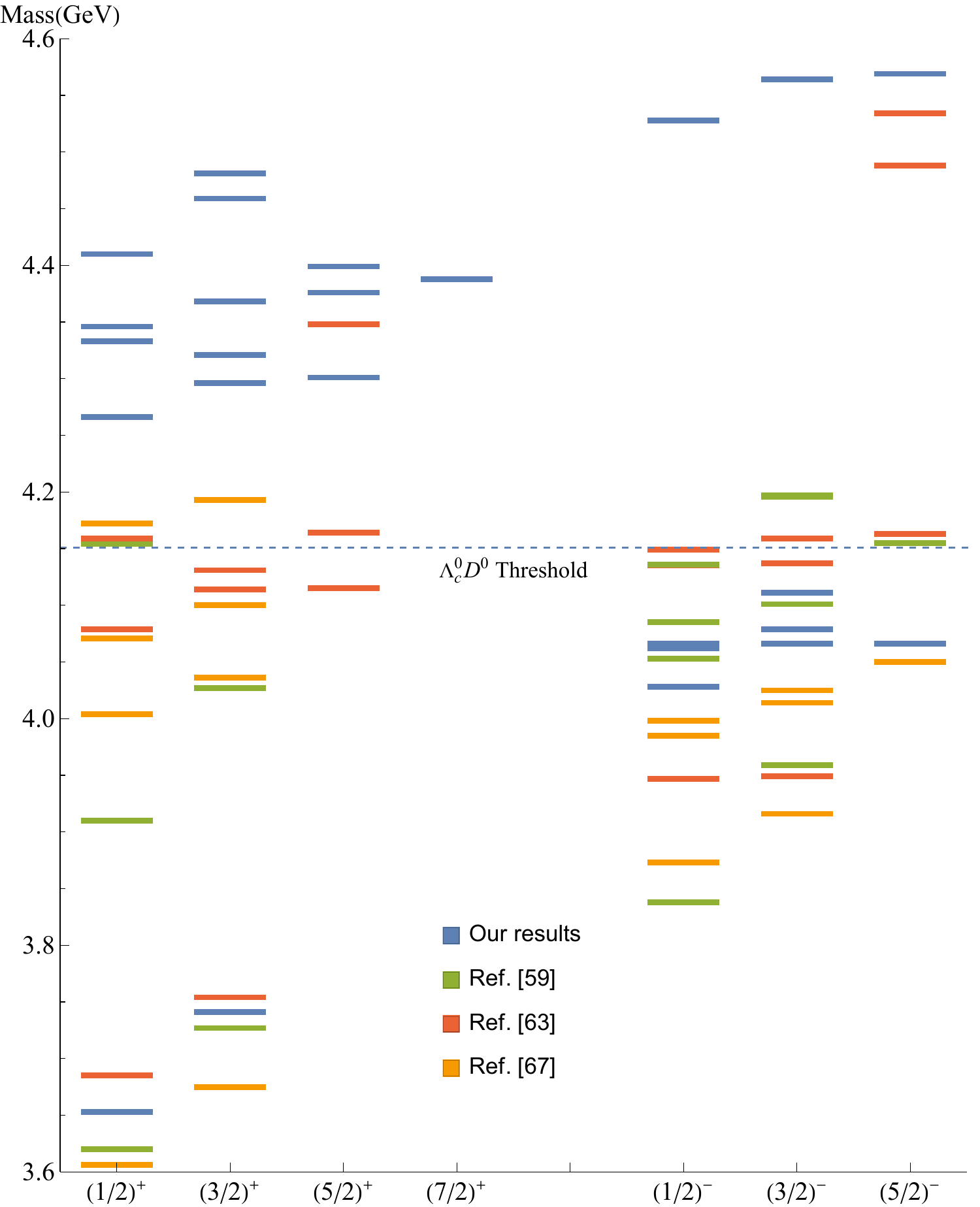} & \includegraphics[width=.45\textwidth]{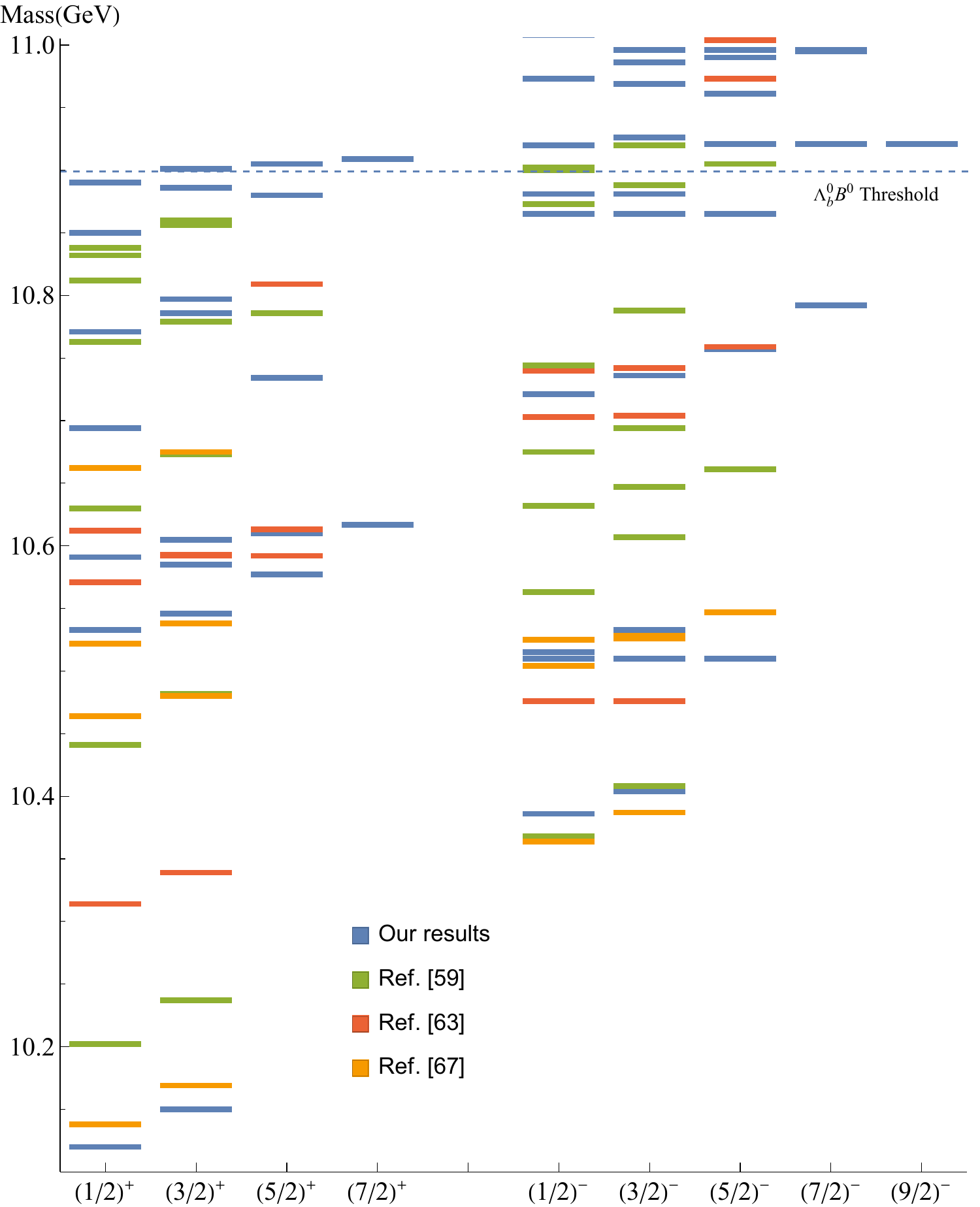}\\ 
(a) & (b) \\
\end{tabular} 
\caption{Comparison of our results with those of Refs.~\cite{Ebert:2002ig,Yoshida:2015tia,Lu:2017meb} (green, red, orange) for double charm and bottom baryons in (a) and (b), respectively. Our results (blue) correspond to the entries $r_0=0.5$~fm in Tables~\ref{hft12gc} - \ref{hft12upb}.}
\label{comp_all}
\end{figure}

The spectrum of doubly heavy baryons beyond the ground state doublet has also been studied in Refs.~\cite{Gershtein:2000nx,Ebert:2002ig,Yoshida:2015tia,Shah:2017liu,Lu:2017meb,Li:2019ekr}. In Fig.~\ref{comp_all}  we compare our spectra with the ones in Ref.~\cite{Ebert:2002ig,Yoshida:2015tia,Lu:2017meb} obtained with a quark model with a relativistic light quark, a nonrelativistic quark model, and a relativistic quark model with a diquark core respectively. The spectra of Ref.~\cite{Gershtein:2000nx} is derived from a similar quark model as in Ref.~\cite{Ebert:2002ig}, but the values are shifted down by about a $100$~MeV. Ref.~\cite{Li:2019ekr} uses the Bethe-Salpeter equation in a diquark picture and presents a limited number of states in the spin-symmetry limit. The results of Ref.~\cite{Shah:2017liu} do not include the Pauli principle for the heavy-quark wave functions and we do not consider it beyond the ground state. From Fig.~\ref{comp_all} (a) we can see that for double charm baryons the pattern of states beyond the ground state doublet does not agree with ours in none of the cases or among the quark model approaches themselves. For all displayed model spectra the excited states lie (much) lower than ours. This is in contrast to the overall agreement found with lattice calculations in~\cite{Soto:2020xpm}. For double bottom  baryons [see Fig.~\ref{comp_all} (b)] the discrepancies reach the ground state doublet, as the results of Ref.~\cite{Ebert:2002ig}, and to a lesser extend the ones of Ref.~\cite{Yoshida:2015tia}, lie higher than ours. However, there is agreement for the first excited (odd-parity) doublet, except for Ref.~\cite{Ebert:2002ig}. For higher states the discrepancies persist, except for the odd-parity states of Ref.~\cite{Lu:2017meb}, which are compatible with ours. 

\section{Conclusions}\label{sec:con}

An EFT describing doubly heavy hadrons was put forward in Ref.~\cite{Soto:2020xpm}. It is built upon the nonrelativistic expansion of the heavy quarks and the adiabatic expansion between the dynamics of the heavy quarks and the light degrees of freedom corresponding to the gluons and light quarks. The EFT was constructed in the single hadron sector up to the heavy-quark spin and angular momentum terms suppressed by $1/m_Q$. Expressions of the potentials as operator insertions in the Wilson loop were obtained by matching the EFT to NRQCD. The computation of the Wilson loop with operator insertions cannot be done using perturbative techniques and should be carried out (ideally) in lattice QCD or other nonperturbative approaches (see for instance \cite{Andreev:2020xor} for an AdS/CFT inspired proposal).

In Ref.~\cite{Soto:2020pfa} this EFT framework was applied to doubly heavy baryons. Using the lattice data of Refs.~\cite{Najjar:2009da,Najjarthesis} for the static energies the leading-order spectrum of doubly charm and bottom baryons was computed for the four lowest lying static states. However, since there are no available lattice determinations of the potentials of the heavy-quark spin and angular-momentum operators, the computation of the hyperfine contributions to the doubly heavy baryon masses was not possible.

In this paper we presented a parametrization of the $1/m_Q$ suppressed heavy-quark spin and angular-momentum operators with a minimal amount of modeling, the general idea of which can be extended to other potentials for doubly heavy hadrons, such as double charm tetraquark, $T^+_{cc}$, recently discovered by the LHCb Collaboration \cite{LHCb:2021vvq}. This parametrization of the potentials is based in their description in short- and long-distance regimes. In the short-distance regime, defined as $r\ll 1/\Lambda_{\rm QCD}$, the Wilson-loop expressions of the potentials can be expanded in the multipole expansion. This can be done using weakly-coupled pNRQCD, which is the EFT that incorporates the multipole expansion systematically, for two heavy quarks~\cite{Brambilla:2005yk}. This produces short-distance expressions of the potentials as an expansion in powers of $r^2$ with coefficients that encode the nonperturbative dynamics of the light degrees of freedom,\footnote{In general the potentials can have a nonanalytical term in $r$ originating from the perturbative integration of the heavy-quark momentum, i.e when the potential can be generated without interacting with the light degrees of freedom. However, this is not the case for the potentials of the $1/m_Q$ suppressed heavy-quark spin and angular momentum operators.} which we show in Sec.~\ref{sec:sdp} and Appendix~\ref{app:sdec}. At leading order in the multipole expansion only one coefficient is necessary and it can be determined in a model-independent way using the heavy quark-diquark duality from the heavy-mesons masses.

The long-distance regime is characterized by $r\gg 1/\Lambda_{\rm QCD}$. In the case of a heavy-quark-antiquark pair it is known from lattice QCD  that in this regime a gluonic flux tube connecting the two heavy quarks emerges. It is well-known that an Effective String Theory (EST)~\cite{Nambu:1978bd,Luscher:1980fr,Luscher:2002qv} reproduces accurately the lattice determinations~\cite{PerezNadal:2008vm,Hwang:2018rju}. In Sec.~\ref{sec:est} we propose an extension of the EST to include the presence of a fermion constrained to move on the string. We obtain a mapping of the NRQCD operators inserted in the Wilson loop to operators in the EST based on imposing the same transformation properties under $D_{\infty h}$ and flavor. Using this mapping we can translate the Wilson-loop expressions for the potentials to EST correlators and evaluate them. This procedure yields long-distance expressions of the potentials depending on two unknown coefficients of the EST. Additionally, we compute the vacuum  energy in the EST with fermions in Appendix~\ref{app:ce} and show that (i) the string tension runs with the square of the mass of the fermion and (ii) the sign of the L\"uscher term changes. These features can in principle be checked by lattice calculations of the ground state energy of two static quarks separated at a large distance with an additional light quark.\footnote{Beyond the string breaking scale, it would also require the calculation of excitations with the ground state quantum numbers.}

The final parametrization of the potentials is obtained by interpolating between the short- and long-distance descriptions. We choose the most simple interpolation that ensures that the correct short- and long-distance behavior are recovered in the corresponding limits. Nevertheless, an extra parameter is introduced in the definition of the interpolation. The hyperfine contributions to doubly heavy baryons can be computed using these parametrizations of the heavy-quark spin and angular-momentum dependent potentials. The values of the remaining unknown parameters are determined by fitting the hyperfine splittings obtained in lattice QCD in Refs.~\cite{Briceno:2012wt,Namekawa:2013vu,Brown:2014ena,Alexandrou:2014sha,Bali:2015lka,Padmanath:2015jea,Alexandrou:2017xwd,Lewis:2008fu,Brown:2014ena,Mohanta:2019mxo} for several $S$-, $P$- and , $D$-wave multiplets. This guarantees that all our inputs are from QCD, and the modeling is reduced to the choice of interpolation, provided that the EST we use is the correct effective theory at long distances. Using the parameters thus determined, we make predictions for the spectrum of double charm and bottom baryons including the hyperfine contributions. Our results are summarized in Tables~\ref{hft12gc}-\ref{hft12upb} and in Figs.~\ref{ccplot} and \ref{bbplot}. 

Finally in Sec.~\ref{sec:models} we compared our results with previous model approaches and sum rules determinations. We observe a huge dispersion of results. In the absence of lattice calculations for many states, specially for double bottom baryons, our EFT approach offers a framework in which modeling is minimal and errors can be reliably quantified, unlike in most models. Since lattice determinations of the potentials for doubly heavy hadrons are difficult, in particular when unquenched simulations are required, the procedure outlined in the paper and in Ref.~\cite{Oncala:2017hop}, to obtain reliable parametrizations of the potentials can be of significant utility in future studies of doubly heavy hadrons. In turn, this motivates further development of the EST to cases with multiple light quarks.

\section*{Acknowledgments}

J.S. acknowledges financial support from Grant No.~2017-SGR-929 from the Generalitat de Catalunya and from projects No.~PID2019-
105614GB-C21, No.~PID2019-110165GB-I00 and No.~CEX2019-000918-M from Ministerio de Ciencia, Innovaci\'on y Universidades.  J.T.C acknowledges the financial support from the European Union's Horizon 2020 research and innovation program under the Marie Sk\l{}odowska--Curie Grant Agreement No. 665919. He has also been supported in part by the Spanish Grants No.~FPA2017-86989-P and No.~SEV-2016-0588 from the Ministerio de Ciencia, Innovaci\'on y Universidades, and Grants No.~2017-SGR-1069 from the Generalitat de Catalunya. This research was supported by the Munich Institute for Astro and Particle Physics (MIAPP) which is funded by the Deutsche Forschungsgemeinschaft (DFG, German Research Foundation) under Germany's Excellence Strategy No.~EXC-2094–390783311.

\appendix

\section{The Casimir energy}\label{app:ce}

The Hamiltonian associated to the leading-order string action with fermionic degrees of freedom in Eqs.~\eqref{ngaplqexp} and \eqref{gstex} is
\begin{align}
H=\int^{r/2}_{-r/2}dz \left[\frac{\sigma}{2}\left(\pa_0\xi^l\pa_0\xi^l+\pa_3\xi^l\pa_3\xi^l\right)+i\psi^\dagger\pda_0\psi\right]\,,\label{ham}
\end{align}
where we have used the equation of motion of the fermion field to simplify the fermionic term.

Now, let us compute the expected value of the Hamiltonian in the vacuum 
\begin{align}
\langle 0|H|0\rangle=\frac{\pi}{r}\zeta(-1)-2n_f\sum^{\infty}_{n=1}\sqrt{\left(\frac{n\pi}{r}\right)^2+m_{\rm l.q.}^2}\,.
\end{align}
We have obtained the standard L\"uscher term for the bosonic string plus a new contribution from the fermion field. The new contribution is ill defined in an analogous way as the L\"uscher term is. The latter can easily be defined through the analytic continuation of the Riemann zeta function but the nonvanishing $m_{\rm l.q.}$ in the former requires extra care. 

In the nonrelativistic case (which implies a cutoff in the sum so that $\frac{n\pi}{r}\ll m_{\rm l.q.}$) there are no quantum fluctuation contributions of the fermion to the vacuum energy, 
\begin{align}
-\frac{1}{2}\langle 0|H|0\rangle_{\rm fermion}&\equiv\sum^{\infty}_{n=1}\sqrt{\left(\frac{n\pi}{r}\right)^2+m_{\rm l.q.}^2}=m_{\rm l.q.}\left(\zeta(0)+\frac{1}{2}\left(\frac{\pi}{m_{\rm l.q.}r}\right)^2\zeta(-2)-\frac{1}{8}\left(\frac{\pi}{m_{\rm l.q.}r}\right)^4\zeta(-4)+\dots\right)\nn\\
&=-\frac{1}{2}m_{\rm l.q.}\,.
\end{align}

In the general case, we will use dimensional regularization. Let us first write
\begin{align}
-\frac{1}{2}\langle 0|H|0\rangle_{\rm fermion}&=
-\frac{1}{2}\langle 0|H|0\rangle_{\rm fermion}^{\rm div.}-\frac{1}{2}\langle 0|H|0\rangle_{\rm fermion}^{\rm finite}\,,\\
-\frac{1}{2}\langle 0|H|0\rangle_{\rm fermion}^{\rm div.}&=\sum^{\infty}_{n=1}\left(\sqrt{\left(\frac{n\pi}{r}\right)^2}+\frac{m_{\rm l.q.}^2}{2\sqrt{\left(\frac{n\pi}{r}\right)^2}}\right)\,,\\
-\frac{1}{2}\langle 0|H|0\rangle_{\rm fermion}^{\rm finite}&=-\sum^{\infty}_{n=1}\frac{m_{\rm l.q.}^2r}{2n\pi}\frac{1}{\left(\sqrt{1+\left(\frac{n\pi}{m_{\rm l.q.}r}\right)^2}+\frac{n\pi}{m_{\rm l.q.}r}\right)^2}\,.\label{fin}
\end{align}
We regulate the divergent terms in dimensional regularization as follows: ($d=1+2\epsilon$, $\epsilon \to 0$)
\begin{align}\hspace{-0.5cm}
-\frac{1}{2}\langle 0|H|0\rangle_{\rm fermion}^{\rm div.}\to& \mu^{1-d}\int\frac{d^{d-1}\mathbf{k}}{(2\pi)^{d-1}}\sum^{\infty}_{n=1}\left(\sqrt{\mathbf{k}^2+\left(\frac{n\pi}{r}\right)^2}+\frac{m_{\rm l.q.}^2}{2\sqrt{\mathbf{k}^2+\left(\frac{n\pi}{r}\right)^2}}\right)\nn\\
=&\frac{2}{(4\pi\mu^2)^\epsilon\Gamma (\epsilon)}\int_0^\infty dk k^{-1+2\epsilon} \sum^{\infty}_{n=1}\left(\sqrt{k^2+\left(\frac{n\pi}{r}\right)^2}+\frac{m_{\rm l.q.}^2}{2\sqrt{k^2+\left(\frac{n\pi}{r}\right)^2}}\right)\nn\\
=& \frac{1}{\sqrt{\pi}}\left(\frac{\pi}{4\mu^2 r^2}\right)^\epsilon\left(-\frac{\pi}{2r}\zeta (-1-2\epsilon)\Gamma\left(-\frac{1}{2}-\epsilon\right)+\frac{m_{\rm l.q.}^2 r}{2\pi}\zeta(1-2\epsilon)\Gamma\left(\frac{1}{2}-\epsilon\right)\right)\nn\\
=& -\frac{\pi}{12r}+\frac{m_{\rm l.q.}^2 r}{4\pi}\left(-\frac{1}{\epsilon}+\log\frac{(\mu r)^2}{\pi}+\gamma_E\right)\,.\label{div}
\end{align}

Note that the $1/\epsilon$ pole can be absorbed in a redefinition of the string tension
\begin{align}
\sigma\to \sigma (\mu) -\frac{n_f}{2\pi\epsilon}m_{\rm l.q.}^2\,.\label{ren}
\end{align}
$\mu$-independence leads to
\begin{align}
\mu\frac{d \sigma(\mu)}{d\mu}=\frac{n_f}{\pi} m_{\rm l.q.}^2\,,
\end{align}
which implies that the string tension runs in such a way that it decreases at large distances. For this to be consistent within an EFT framework, we need $m_{\rm l.q.}^2 \ll\sigma (\mu)$, so that the replacement in Eq.~\eqref{ren} generates higher-order terms elsewhere. Since $\sigma (\mu)\sim \lQ^2$, we need $m_{\rm l.q.} \ll \lQ$, which may be achieved by implementing chiral symmetry linearly in the fermion fields on the world sheet. Numerically, we find $m_{\rm l.q.}^2\sim 0.051$ GeV$^2$ whereas $\sigma (\mu)\sim 0.21$ GeV$^2$.

Note that the first term in Eq.~\eqref{div} is the L\"uscher term. When the fermionic contribution is added to the bosonic one, the sign of the (total) L\"uscher term is reversed with respect to the purely bosonic contribution.

The finite piece in Eq.~\eqref{fin} can be evaluated numerically. In the $m_{\rm l.q.}r\to 0$ limit it reads as
\begin{align}
-\frac{1}{2}\langle 0|H|0\rangle_{\rm fermion}^{\rm finite}&=-\frac{m_{\rm l.q.}}{4}\left(\frac{m_{\rm l.q.}r}{\pi}\right)^3\zeta(3)+\dots\,.\label{ltff}
\end{align}
 
In the large $m_{\rm l.q.}r$ limit the sum Eq.~\eqref{fin} tends to an integral:
\begin{align}
-\frac{1}{2}\langle 0|H|0\rangle_{\rm fermion}^{\rm finite}&=
-\sum^{\infty}_{n=1}\frac{m_{\rm l.q.}^2r}{2n\pi}\frac{1}{\left(\sqrt{1+\left(\frac{n\pi}{m_{\rm l.q.}r}\right)^2}+\frac{n\pi}{m_{\rm l.q.}r}\right)^2}\xrightarrow{m_{\rm l.q.}r\to\infty}
\, -\frac{m^2r}{2\pi}\int^{\infty}_{\frac{\pi}{m_{\rm l.q.}r}}dx\frac{1}{x}\frac{1}{\left(\sqrt{1+x^2}+x\right)^2}\nn\\
&=-\frac{m_{\rm l.q.}^2r}{2\pi}\left[-\frac{1}{2}+\log\left(\frac{m_{\rm l.q.}r}{2\pi}\right)+\frac{\pi}{m_{\rm l.q.}r}-\frac{1}{6}\left(\frac{\pi}{m_{\rm l.q.}r}\right)^3+\dots\right]\,.
\end{align}
Adding up all the contributions
\begin{align}
-\frac{1}{2}\langle 0|H|0\rangle_{\rm fermion}\xrightarrow{m_{\rm l.q.}r\to\infty}-\frac{1}{12}\frac{\pi}{r}-\frac{m_{\rm l.q.}}{2}+\frac{m_{\rm l.q.}^2r}{4\pi}\left(1+\gamma_E-\log\left(\frac{m_{\rm l.q}^2}{4\pi\mu^2}\right)\right)+\frac{1}{12}\frac{\pi^2}{m_{\rm l.q.}r^2}+\dots\,,
\end{align}
where we have minimally subtracted the $1/\epsilon$ pole. We see that in this limit we get additional contributions to the string tension, a constant term, and $1/r$ corrections to the L\"uscher term, the leading one being $\mathcal{O}(1/r^2)$ rather than $\mathcal{O}(1/r^3)$ like in the bosonic part. The additional contribution to the string tension is particularly interesting. If we identify the standard string tension $\sigma=\sigma(\mu\sim \lQ)$, then for $m_{\rm l.q.} <\lQ$, this additional contribution makes the string tension diminish with $m_{\rm l.q.}^2$.

\section{Short-distance expansion coefficients}\label{app:sdec}

In this appendix we provide the expressions of the short-distance expansion of the potentials in terms of weakly-coupled pNRQCD correlators. Let us define $\bar{3}$ and $6$ tensor invariants as in Ref.~\cite{Brambilla:2005yk}
\begin{align}
&\underline{T}^l_{ij}  = \frac{1}{\sqrt{2}} \epsilon_{lij},\quad i,\,j,\,l=1,2,3\,\label{Tabg}\\
& \nn\\
& \underline{\Sigma}^\sigma_{ij}\quad i,\,j=1,2,3\,\quad \sigma=1,..,6\nn\\
&\underline{\Sigma}^1_{11}=\underline{\Sigma}^4_{22}=\underline{\Sigma}^6_{33}=1,\nn\\
&\underline{\Sigma}^2_{12}=\underline{\Sigma}^2_{21}=\underline{\Sigma}^3_{13}=\underline{\Sigma}^3_{31}=\underline{\Sigma}^5_{23}=\underline{\Sigma}^5_{32} = \frac{1}{\sqrt{2}}, 
\end{align}
where all other entries are zero. Both $\underline{T}^l_{ij}$ and $\underline{\Sigma}^\sigma_{ij}$ are real; $\underline{T}^l_{ij}$ is totally antisymmetric and $\underline{\Sigma}^\sigma_{ij}$  symmetric in the $i$ and $j$ indices. They satisfy the orthogonality and normalization relations:
\begin{align}
\sum_{ij=1}^3 \underline{T}^{l}_{ij} \, \underline{T}^{l'}_{ij} =\delta^{ll'}\,, \qquad
\sum_{ij=1}^3 \underline{\Sigma}^{\sigma}_{ij} \, \underline{\Sigma}^{\sigma'}_{ij} =\delta^{\sigma\sigma'}\,, \qquad \sum_{ij=1}^3 \underline{T}^l_{ij} \, \underline{\Sigma}^\sigma_{ij} = 0\,.
\end{align}

The Wilson lines associated to the propagation of the heavy-quark pair in an antitriplet and sextet color states are
\begin{align}
&\phi^{lk}_T(t_2,\,t_1)=e^{-ig\int^{t_2}_{t_1}dt'\,2{\rm Tr}\left[\underline{T}^{l}A_0(t',\,\bm{R})\underline{T}^{k}\right]}\,,\\
&\phi^{\sigma \sigma^{\prime}}_\Sigma(t_2,\,t_1)=e^{-ig\int^{t_2}_{t_1}dt'\,2{\rm Tr}\left[\underline{\Sigma}^{\sigma}A_0(t',\,\bm{R})\underline{\Sigma}^{\sigma'}\right]}\,.
\end{align}
Notice that the antitriplet and sextet representation generators can be written as $T^a_{\bar{3}}=-2{\rm Tr}\left[\underline{T}\,T^a\underline{T}\right]=(T^a)^*$ and $T^a_6=2{\rm Tr}\left[\underline{\Sigma}\, T^a\underline{\Sigma}\right]$. Let us also define
\begin{align}
&\bm{X}^{k\sigma}_{T\Sigma}(t)=g{\rm Tr}\left[\underline{T}^k \bm{X}(t)\underline{\Sigma}^{\sigma}\right]\,,\\
&\bm{X}^{\sigma k}_{\Sigma T}(t)=g{\rm Tr}\left[\underline{\Sigma}^{\sigma}\bm{X}(t)\underline{T}^k \right]\,,\\
&\bm{X}^{lk}_T(t)=g{\rm Tr}\left[\underline{T}^{l} \bm{X}(t)\underline{T}^{k}\right]\,,\\
&\bm{X}^{\sigma\sigma'}_\Sigma(t)=g{\rm Tr}\left[\underline{\Sigma}^{\sigma} \bm{X}(t)\underline{\Sigma}^{\sigma'}\right]\,,
\end{align}
with $\bm{X}=\bm{E},\,\bm{B}$ and 
\begin{align}
{\cal\underline{Q}}^l_{(1/2)^\pm}(t,\bm{x})&={\rm Tr}\left[\underline{T}^{l}{\cal Q}_{(1/2)^\pm}(t,\bm{x})\right]\,.
\end{align}

The nonperturbative constants up to next-to-leading order in the multipole expansion for the static potential are
\begin{align}
\overline{\Lambda}_{(1/2)^{\pm}}=&\lim_{t \to \infty}\frac{i}{t}\log{\rm Tr}\left[\langle 0|{\cal\underline{Q}}_{(1/2)^\pm}(t/2)\phi_T\,{\cal\underline{Q}}_{(1/2)^\pm}^\dagger(-t/2)|0\rangle\right]\,,\label{app:sdec:e1}\\
\overline{\Lambda}^{(1)}_{(1/2)^{\pm}}=&\lim_{t \to \infty}\frac{e^{i\overline{\Lambda}_{(1/2)^{\pm}}t}}{6it}{\rm Tr}\left[\langle 0|{\cal\underline{Q}}_{(1/2)^\pm}(t/2)\Biggl\{\int^{t/2}_{-t/2}dt_1 \phi_T\left(-\frac{1}{4}\right)\bm{D}\cdot\bm{E}_{\Sigma}(t_1)\phi_T+\int^{t/2}_{-t/2}dt_2\int^{t_2}_{-t/2}dt_1 \phi_T\right.\nn\\
&\left.\bm{E}_{T\Sigma}(t_2)\phi_\Sigma\,\bm{E}_{\Sigma T}(t_1)\phi_T\Biggr\}{\cal\underline{Q}}_{(1/2)^\pm}^\dagger(-t/2)|0\rangle\right]\,.\label{app:sdec:e2}
\end{align}
The Wilson lines should be understood as starting on the time of the operator immediately on the right and ending on the time of the operator immediately on the left.

For the heavy-quark spin or angular-momentum dependent potentials the nonperturbative constants correspond to the following correlators
\begin{align}
\Delta^{(0)}_{(1/2)^{\pm}}=&\lim_{t \to \infty}\frac{2e^{i\overline{\Lambda}_{(1/2)^{\pm}}t}}{3it}\int^{t/2}_{-t/2}dt' 
{\rm Tr}\left[\bm{S}_{1/2}\cdot\langle 0|{\cal\underline{Q}}_{(1/2)^\pm}(t/2)\phi_T\bm{B}_T(t')\phi_T{\cal\underline{Q}}_{(1/2)^\pm}^\dagger(-t/2)|0\rangle\right]\,,\label{app:sdec:e3}\\
\Delta^{(1,0)}_{(1/2)^{\pm}}=&\lim_{t \to \infty}\frac{2ie^{i\overline{\Lambda}_{(1/2)^{\pm}}t}}{3t}{\rm Tr}\left[\bm{S}_{1/2}\cdot \underline{\Delta}\right]\,,\label{app:sdec:e4}\\
\Delta^{(1,2)}_{(1/2)^{\pm}}=&\lim_{t \to \infty}\frac{3ie^{i\overline{\Lambda}_{(1/2)^{\pm}}t}}{t} {\rm Tr}\left[\left(\bm{S}_{1/2}\cdot\bm{{\cal T}}_2\right)\cdot \underline{\Delta}\right]\,,\label{app:sdec:e5}
\end{align}
with
\begin{align}
\underline{\Delta}^i=&\langle 0|{\cal\underline{Q}}_{(1/2)^\pm}(t/2)\Biggl\{\nn\\
&\int^{t/2}_{-t/2}dt_3\int^{t_3}_{-t/2}dt_2\int^{t_2}_{-t/2}dt_1 \phi_T\left[\bm{B}^i_T(t_3)\phi_T \hat{\bm{r}}\cdot\bm{E}_{T\Sigma}(t_2)\phi_\Sigma \hat{\bm{r}}\cdot\bm{E}_{\Sigma T}(t_1)+\hat{\bm{r}}\cdot\bm{E}_{T\Sigma}(t_3)\phi_\Sigma  \hat{\bm{r}}\cdot\bm{E}_{\Sigma T}(t_2)\phi_T\bm{B}^i_T(t_1)\right.\nn\\
&\left.+\hat{\bm{r}}\cdot\bm{E}_{T\Sigma}(t_3)\phi_\Sigma \bm{B}^i_\Sigma(t_2) \phi_\Sigma  \hat{\bm{r}}\cdot\bm{E}_{\Sigma T}(t_1)\right]\phi_T+\int^{t/2}_{-t/2}dt_2\int^{t_2}_{-t/2}dt_1 \phi_T\left[\frac{1}{4}\left( \hat{\bm{r}}^j\hat{\bm{r}}^k\bm{D}^j\bm{E}^k_{T}(t_2)\phi_T\bm{B}^i_T(t_1)+\bm{B}^i_T(t_2)\phi_T\right.\right. \nn\\
&\left.\left.\hat{\bm{r}}^j\hat{\bm{r}}^k\bm{D}^j\bm{E}^k_{T}(t_1)\right)-\frac{1}{2}\left(\hat{\bm{r}}\cdot\bm{D}\bm{B}^i_{T\Sigma}(t_2)\phi_\Sigma \hat{\bm{r}}\cdot\bm{E}_{\Sigma T}(t_1)+
 \hat{\bm{r}}\cdot\bm{E}_{T\Sigma }(t_2)\phi_\Sigma \hat{\bm{r}}\cdot\bm{D}\bm{B}^i_{\Sigma T}(t_1)\right)\right]\phi_T+\int^{t/2}_{-t/2}dt_1 \phi_T\left[-\frac{1}{8}\hat{\bm{r}}^j\hat{\bm{r}}^k\right.\nn\\
&\left.\bm{D}^j\bm{D}^k\bm{B}^i_{T}(t_1)\right] \phi_T\Biggr\} {\cal\underline{Q}}_{(1/2)^\pm}^\dagger(-t/2)|0\rangle\,.
\end{align}

Notice that we have omitted the dependence of all the operators on $\bm{R}$ and that the traces in Eqs.~\eqref{app:sdec:e1}-\eqref{app:sdec:e5} act both on the light-quark spin and color indices.

\bibliographystyle{apsrev4-1}
\bibliography{est_dhbbiblio}

\begin{thebibliography}{84}%
\makeatletter
\providecommand \@ifxundefined [1]{%
 \@ifx{#1\undefined}
}%
\providecommand \@ifnum [1]{%
 \ifnum #1\expandafter \@firstoftwo
 \else \expandafter \@secondoftwo
 \fi
}%
\providecommand \@ifx [1]{%
 \ifx #1\expandafter \@firstoftwo
 \else \expandafter \@secondoftwo
 \fi
}%
\providecommand \natexlab [1]{#1}%
\providecommand \enquote  [1]{``#1''}%
\providecommand \bibnamefont  [1]{#1}%
\providecommand \bibfnamefont [1]{#1}%
\providecommand \citenamefont [1]{#1}%
\providecommand \href@noop [0]{\@secondoftwo}%
\providecommand \href [0]{\begingroup \@sanitize@url \@href}%
\providecommand \@href[1]{\@@startlink{#1}\@@href}%
\providecommand \@@href[1]{\endgroup#1\@@endlink}%
\providecommand \@sanitize@url [0]{\catcode `\\12\catcode `\$12\catcode
  `\&12\catcode `\#12\catcode `\^12\catcode `\_12\catcode `\%12\relax}%
\providecommand \@@startlink[1]{}%
\providecommand \@@endlink[0]{}%
\providecommand \url  [0]{\begingroup\@sanitize@url \@url }%
\providecommand \@url [1]{\endgroup\@href {#1}{\urlprefix }}%
\providecommand \urlprefix  [0]{URL }%
\providecommand \Eprint [0]{\href }%
\providecommand \doibase [0]{http://dx.doi.org/}%
\providecommand \selectlanguage [0]{\@gobble}%
\providecommand \bibinfo  [0]{\@secondoftwo}%
\providecommand \bibfield  [0]{\@secondoftwo}%
\providecommand \translation [1]{[#1]}%
\providecommand \BibitemOpen [0]{}%
\providecommand \bibitemStop [0]{}%
\providecommand \bibitemNoStop [0]{.\EOS\space}%
\providecommand \EOS [0]{\spacefactor3000\relax}%
\providecommand \BibitemShut  [1]{\csname bibitem#1\endcsname}%
\let\auto@bib@innerbib\@empty
\bibitem [{\citenamefont {Soto}\ and\ \citenamefont
  {Tarr\'us~Castell\`a}(2020{\natexlab{a}})}]{Soto:2020xpm}%
  \BibitemOpen
  \bibfield  {author} {\bibinfo {author} {\bibfnamefont {J.}~\bibnamefont
  {Soto}}\ and\ \bibinfo {author} {\bibfnamefont {J.}~\bibnamefont
  {Tarr\'us~Castell\`a}},\ }\href {\doibase 10.1103/PhysRevD.102.014012}
  {\bibfield  {journal} {\bibinfo  {journal} {Phys. Rev. D}\ }\textbf {\bibinfo
  {volume} {102}},\ \bibinfo {pages} {014012} (\bibinfo {year}
  {2020}{\natexlab{a}})},\ \Eprint {http://arxiv.org/abs/2005.00552}
  {arXiv:2005.00552 [hep-ph]} \BibitemShut {NoStop}%
\bibitem [{\citenamefont {Brambilla}\ \emph {et~al.}(2001)\citenamefont
  {Brambilla}, \citenamefont {Pineda}, \citenamefont {Soto},\ and\
  \citenamefont {Vairo}}]{Brambilla:2000gk}%
  \BibitemOpen
  \bibfield  {author} {\bibinfo {author} {\bibfnamefont {N.}~\bibnamefont
  {Brambilla}}, \bibinfo {author} {\bibfnamefont {A.}~\bibnamefont {Pineda}},
  \bibinfo {author} {\bibfnamefont {J.}~\bibnamefont {Soto}}, \ and\ \bibinfo
  {author} {\bibfnamefont {A.}~\bibnamefont {Vairo}},\ }\href {\doibase
  10.1103/PhysRevD.63.014023} {\bibfield  {journal} {\bibinfo  {journal} {Phys.
  Rev. D}\ }\textbf {\bibinfo {volume} {63}},\ \bibinfo {pages} {014023}
  (\bibinfo {year} {2001})},\ \Eprint {http://arxiv.org/abs/hep-ph/0002250}
  {arXiv:hep-ph/0002250} \BibitemShut {NoStop}%
\bibitem [{\citenamefont {Pineda}\ and\ \citenamefont
  {Vairo}(2001)}]{Pineda:2000sz}%
  \BibitemOpen
  \bibfield  {author} {\bibinfo {author} {\bibfnamefont {A.}~\bibnamefont
  {Pineda}}\ and\ \bibinfo {author} {\bibfnamefont {A.}~\bibnamefont {Vairo}},\
  }\href {\doibase 10.1103/PhysRevD.64.039902} {\bibfield  {journal} {\bibinfo
  {journal} {Phys. Rev. D}\ }\textbf {\bibinfo {volume} {63}},\ \bibinfo
  {pages} {054007} (\bibinfo {year} {2001})},\ \bibinfo {note} {[Erratum:
  Phys.Rev.D 64, 039902 (2001)]},\ \Eprint
  {http://arxiv.org/abs/hep-ph/0009145} {arXiv:hep-ph/0009145} \BibitemShut
  {NoStop}%
\bibitem [{\citenamefont {Caswell}\ and\ \citenamefont
  {Lepage}(1986)}]{Caswell:1985ui}%
  \BibitemOpen
  \bibfield  {author} {\bibinfo {author} {\bibfnamefont {W.~E.}\ \bibnamefont
  {Caswell}}\ and\ \bibinfo {author} {\bibfnamefont {G.~P.}\ \bibnamefont
  {Lepage}},\ }\href {\doibase 10.1016/0370-2693(86)91297-9} {\bibfield
  {journal} {\bibinfo  {journal} {Phys. Lett. B}\ }\textbf {\bibinfo {volume}
  {167}},\ \bibinfo {pages} {437} (\bibinfo {year} {1986})}\BibitemShut
  {NoStop}%
\bibitem [{\citenamefont {Bodwin}\ \emph {et~al.}(1995)\citenamefont {Bodwin},
  \citenamefont {Braaten},\ and\ \citenamefont {Lepage}}]{Bodwin:1994jh}%
  \BibitemOpen
  \bibfield  {author} {\bibinfo {author} {\bibfnamefont {G.~T.}\ \bibnamefont
  {Bodwin}}, \bibinfo {author} {\bibfnamefont {E.}~\bibnamefont {Braaten}}, \
  and\ \bibinfo {author} {\bibfnamefont {G.~P.}\ \bibnamefont {Lepage}},\
  }\href {\doibase 10.1103/PhysRevD.55.5853} {\bibfield  {journal} {\bibinfo
  {journal} {Phys. Rev. D}\ }\textbf {\bibinfo {volume} {51}},\ \bibinfo
  {pages} {1125} (\bibinfo {year} {1995})},\ \bibinfo {note} {[Erratum:
  Phys.Rev.D 55, 5853 (1997)]},\ \Eprint {http://arxiv.org/abs/hep-ph/9407339}
  {arXiv:hep-ph/9407339} \BibitemShut {NoStop}%
\bibitem [{\citenamefont {Manohar}(1997)}]{Manohar:1997qy}%
  \BibitemOpen
  \bibfield  {author} {\bibinfo {author} {\bibfnamefont {A.~V.}\ \bibnamefont
  {Manohar}},\ }\href {\doibase 10.1103/PhysRevD.56.230} {\bibfield  {journal}
  {\bibinfo  {journal} {Phys. Rev. D}\ }\textbf {\bibinfo {volume} {56}},\
  \bibinfo {pages} {230} (\bibinfo {year} {1997})},\ \Eprint
  {http://arxiv.org/abs/hep-ph/9701294} {arXiv:hep-ph/9701294} \BibitemShut
  {NoStop}%
\bibitem [{\citenamefont {Soto}\ and\ \citenamefont
  {Tarr\'us~Castell\`a}(2020{\natexlab{b}})}]{Soto:2020pfa}%
  \BibitemOpen
  \bibfield  {author} {\bibinfo {author} {\bibfnamefont {J.}~\bibnamefont
  {Soto}}\ and\ \bibinfo {author} {\bibfnamefont {J.}~\bibnamefont
  {Tarr\'us~Castell\`a}},\ }\href {\doibase 10.1103/PhysRevD.102.014013}
  {\bibfield  {journal} {\bibinfo  {journal} {Phys. Rev. D}\ }\textbf {\bibinfo
  {volume} {102}},\ \bibinfo {pages} {014013} (\bibinfo {year}
  {2020}{\natexlab{b}})},\ \Eprint {http://arxiv.org/abs/2005.00551}
  {arXiv:2005.00551 [hep-ph]} \BibitemShut {NoStop}%
\bibitem [{\citenamefont {Najjar}\ and\ \citenamefont
  {Bali}(2009)}]{Najjar:2009da}%
  \BibitemOpen
  \bibfield  {author} {\bibinfo {author} {\bibfnamefont {J.}~\bibnamefont
  {Najjar}}\ and\ \bibinfo {author} {\bibfnamefont {G.}~\bibnamefont {Bali}},\
  }\bibfield  {booktitle} {\emph {\bibinfo {booktitle} {{Proceedings, 27th
  International Symposium on Lattice field theory (Lattice 2009): Beijing, P.R.
  China, July 26-31, 2009}}},\ }\href {\doibase 10.22323/1.091.0089} {\bibfield
   {journal} {\bibinfo  {journal} {PoS}\ }\textbf {\bibinfo {volume}
  {LAT2009}},\ \bibinfo {pages} {089} (\bibinfo {year} {2009})},\ \Eprint
  {http://arxiv.org/abs/0910.2824} {arXiv:0910.2824 [hep-lat]} \BibitemShut
  {NoStop}%
\bibitem [{\citenamefont {Najjar}(2009)}]{Najjarthesis}%
  \BibitemOpen
  \bibfield  {author} {\bibinfo {author} {\bibfnamefont {J.}~\bibnamefont
  {Najjar}},\ }\emph {\bibinfo {title} {Static-static-light Baryonic potentials
  from lattice QCD}},\ \href@noop {} {\bibinfo {type} {Diploma thesis}},\
  \bibinfo  {school} {Universit\"at Regensburg} (\bibinfo {year}
  {2009})\BibitemShut {NoStop}%
\bibitem [{\citenamefont {Pineda}\ and\ \citenamefont
  {Soto}(1998)}]{Pineda:1997bj}%
  \BibitemOpen
  \bibfield  {author} {\bibinfo {author} {\bibfnamefont {A.}~\bibnamefont
  {Pineda}}\ and\ \bibinfo {author} {\bibfnamefont {J.}~\bibnamefont {Soto}},\
  }\href {\doibase 10.1016/S0920-5632(97)01102-X} {\bibfield  {journal}
  {\bibinfo  {journal} {Nucl. Phys. B Proc. Suppl.}\ }\textbf {\bibinfo
  {volume} {64}},\ \bibinfo {pages} {428} (\bibinfo {year} {1998})},\ \Eprint
  {http://arxiv.org/abs/hep-ph/9707481} {arXiv:hep-ph/9707481} \BibitemShut
  {NoStop}%
\bibitem [{\citenamefont {Brambilla}\ \emph {et~al.}(2000)\citenamefont
  {Brambilla}, \citenamefont {Pineda}, \citenamefont {Soto},\ and\
  \citenamefont {Vairo}}]{Brambilla:1999xf}%
  \BibitemOpen
  \bibfield  {author} {\bibinfo {author} {\bibfnamefont {N.}~\bibnamefont
  {Brambilla}}, \bibinfo {author} {\bibfnamefont {A.}~\bibnamefont {Pineda}},
  \bibinfo {author} {\bibfnamefont {J.}~\bibnamefont {Soto}}, \ and\ \bibinfo
  {author} {\bibfnamefont {A.}~\bibnamefont {Vairo}},\ }\href {\doibase
  10.1016/S0550-3213(99)00693-8} {\bibfield  {journal} {\bibinfo  {journal}
  {Nucl. Phys. B}\ }\textbf {\bibinfo {volume} {566}},\ \bibinfo {pages} {275}
  (\bibinfo {year} {2000})},\ \Eprint {http://arxiv.org/abs/hep-ph/9907240}
  {arXiv:hep-ph/9907240} \BibitemShut {NoStop}%
\bibitem [{\citenamefont {Brambilla}\ \emph {et~al.}(2005)\citenamefont
  {Brambilla}, \citenamefont {Vairo},\ and\ \citenamefont
  {Rosch}}]{Brambilla:2005yk}%
  \BibitemOpen
  \bibfield  {author} {\bibinfo {author} {\bibfnamefont {N.}~\bibnamefont
  {Brambilla}}, \bibinfo {author} {\bibfnamefont {A.}~\bibnamefont {Vairo}}, \
  and\ \bibinfo {author} {\bibfnamefont {T.}~\bibnamefont {Rosch}},\ }\href
  {\doibase 10.1103/PhysRevD.72.034021} {\bibfield  {journal} {\bibinfo
  {journal} {Phys. Rev. D}\ }\textbf {\bibinfo {volume} {72}},\ \bibinfo
  {pages} {034021} (\bibinfo {year} {2005})},\ \Eprint
  {http://arxiv.org/abs/hep-ph/0506065} {arXiv:hep-ph/0506065} \BibitemShut
  {NoStop}%
\bibitem [{\citenamefont {Brambilla}\ \emph {et~al.}(2019)\citenamefont
  {Brambilla}, \citenamefont {Lai}, \citenamefont {Segovia}, \citenamefont
  {Tarr\'us~Castell\`a},\ and\ \citenamefont {Vairo}}]{Brambilla:2018pyn}%
  \BibitemOpen
  \bibfield  {author} {\bibinfo {author} {\bibfnamefont {N.}~\bibnamefont
  {Brambilla}}, \bibinfo {author} {\bibfnamefont {W.~K.}\ \bibnamefont {Lai}},
  \bibinfo {author} {\bibfnamefont {J.}~\bibnamefont {Segovia}}, \bibinfo
  {author} {\bibfnamefont {J.}~\bibnamefont {Tarr\'us~Castell\`a}}, \ and\
  \bibinfo {author} {\bibfnamefont {A.}~\bibnamefont {Vairo}},\ }\href
  {\doibase 10.1103/PhysRevD.99.014017} {\bibfield  {journal} {\bibinfo
  {journal} {Phys. Rev. D}\ }\textbf {\bibinfo {volume} {99}},\ \bibinfo
  {pages} {014017} (\bibinfo {year} {2019})},\ \bibinfo {note} {[Erratum:
  Phys.Rev.D 101, 099902 (2020)]},\ \Eprint {http://arxiv.org/abs/1805.07713}
  {arXiv:1805.07713 [hep-ph]} \BibitemShut {NoStop}%
\bibitem [{\citenamefont {Brambilla}\ \emph {et~al.}(2020)\citenamefont
  {Brambilla}, \citenamefont {Lai}, \citenamefont {Segovia},\ and\
  \citenamefont {Tarr\'us~Castell\`a}}]{Brambilla:2019jfi}%
  \BibitemOpen
  \bibfield  {author} {\bibinfo {author} {\bibfnamefont {N.}~\bibnamefont
  {Brambilla}}, \bibinfo {author} {\bibfnamefont {W.~K.}\ \bibnamefont {Lai}},
  \bibinfo {author} {\bibfnamefont {J.}~\bibnamefont {Segovia}}, \ and\
  \bibinfo {author} {\bibfnamefont {J.}~\bibnamefont {Tarr\'us~Castell\`a}},\
  }\href {\doibase 10.1103/PhysRevD.101.054040} {\bibfield  {journal} {\bibinfo
   {journal} {Phys. Rev. D}\ }\textbf {\bibinfo {volume} {101}},\ \bibinfo
  {pages} {054040} (\bibinfo {year} {2020})},\ \Eprint
  {http://arxiv.org/abs/1908.11699} {arXiv:1908.11699 [hep-ph]} \BibitemShut
  {NoStop}%
\bibitem [{\citenamefont {Pineda}\ and\ \citenamefont
  {Tarr\'us~Castell\`a}(2019)}]{Pineda:2019mhw}%
  \BibitemOpen
  \bibfield  {author} {\bibinfo {author} {\bibfnamefont {A.}~\bibnamefont
  {Pineda}}\ and\ \bibinfo {author} {\bibfnamefont {J.}~\bibnamefont
  {Tarr\'us~Castell\`a}},\ }\href {\doibase 10.1103/PhysRevD.100.054021}
  {\bibfield  {journal} {\bibinfo  {journal} {Phys. Rev. D}\ }\textbf {\bibinfo
  {volume} {100}},\ \bibinfo {pages} {054021} (\bibinfo {year} {2019})},\
  \Eprint {http://arxiv.org/abs/1905.03794} {arXiv:1905.03794 [hep-ph]}
  \BibitemShut {NoStop}%
\bibitem [{\citenamefont {Tarr\'us~Castell\`a}\ and\ \citenamefont
  {Passemar}(2021)}]{Castella:2021hul}%
  \BibitemOpen
  \bibfield  {author} {\bibinfo {author} {\bibfnamefont {J.}~\bibnamefont
  {Tarr\'us~Castell\`a}}\ and\ \bibinfo {author} {\bibfnamefont
  {E.}~\bibnamefont {Passemar}},\ }\href {\doibase 10.1103/PhysRevD.104.034019}
  {\bibfield  {journal} {\bibinfo  {journal} {Phys. Rev. D}\ }\textbf {\bibinfo
  {volume} {104}},\ \bibinfo {pages} {034019} (\bibinfo {year} {2021})},\
  \Eprint {http://arxiv.org/abs/2104.03975} {arXiv:2104.03975 [hep-ph]}
  \BibitemShut {NoStop}%
\bibitem [{\citenamefont {Nambu}(1979)}]{Nambu:1978bd}%
  \BibitemOpen
  \bibfield  {author} {\bibinfo {author} {\bibfnamefont {Y.}~\bibnamefont
  {Nambu}},\ }\href {\doibase 10.1016/0370-2693(79)91193-6} {\bibfield
  {journal} {\bibinfo  {journal} {Phys. Lett. B}\ }\textbf {\bibinfo {volume}
  {80}},\ \bibinfo {pages} {372} (\bibinfo {year} {1979})}\BibitemShut
  {NoStop}%
\bibitem [{\citenamefont {Luscher}\ \emph {et~al.}(1980)\citenamefont
  {Luscher}, \citenamefont {Symanzik},\ and\ \citenamefont
  {Weisz}}]{Luscher:1980fr}%
  \BibitemOpen
  \bibfield  {author} {\bibinfo {author} {\bibfnamefont {M.}~\bibnamefont
  {Luscher}}, \bibinfo {author} {\bibfnamefont {K.}~\bibnamefont {Symanzik}}, \
  and\ \bibinfo {author} {\bibfnamefont {P.}~\bibnamefont {Weisz}},\ }\href
  {\doibase 10.1016/0550-3213(80)90009-7} {\bibfield  {journal} {\bibinfo
  {journal} {Nucl. Phys. B}\ }\textbf {\bibinfo {volume} {173}},\ \bibinfo
  {pages} {365} (\bibinfo {year} {1980})}\BibitemShut {NoStop}%
\bibitem [{\citenamefont {Luscher}\ and\ \citenamefont
  {Weisz}(2002)}]{Luscher:2002qv}%
  \BibitemOpen
  \bibfield  {author} {\bibinfo {author} {\bibfnamefont {M.}~\bibnamefont
  {Luscher}}\ and\ \bibinfo {author} {\bibfnamefont {P.}~\bibnamefont
  {Weisz}},\ }\href {\doibase 10.1088/1126-6708/2002/07/049} {\bibfield
  {journal} {\bibinfo  {journal} {JHEP}\ }\textbf {\bibinfo {volume} {07}},\
  \bibinfo {pages} {049} (\bibinfo {year} {2002})},\ \Eprint
  {http://arxiv.org/abs/hep-lat/0207003} {arXiv:hep-lat/0207003} \BibitemShut
  {NoStop}%
\bibitem [{\citenamefont {Polchinski}\ and\ \citenamefont
  {Strominger}(1991)}]{Polchinski:1991ax}%
  \BibitemOpen
  \bibfield  {author} {\bibinfo {author} {\bibfnamefont {J.}~\bibnamefont
  {Polchinski}}\ and\ \bibinfo {author} {\bibfnamefont {A.}~\bibnamefont
  {Strominger}},\ }\href {\doibase 10.1103/PhysRevLett.67.1681} {\bibfield
  {journal} {\bibinfo  {journal} {Phys. Rev. Lett.}\ }\textbf {\bibinfo
  {volume} {67}},\ \bibinfo {pages} {1681} (\bibinfo {year}
  {1991})}\BibitemShut {NoStop}%
\bibitem [{\citenamefont {Aharony}\ and\ \citenamefont
  {Komargodski}(2013)}]{Aharony:2013ipa}%
  \BibitemOpen
  \bibfield  {author} {\bibinfo {author} {\bibfnamefont {O.}~\bibnamefont
  {Aharony}}\ and\ \bibinfo {author} {\bibfnamefont {Z.}~\bibnamefont
  {Komargodski}},\ }\href {\doibase 10.1007/JHEP05(2013)118} {\bibfield
  {journal} {\bibinfo  {journal} {JHEP}\ }\textbf {\bibinfo {volume} {05}},\
  \bibinfo {pages} {118} (\bibinfo {year} {2013})},\ \Eprint
  {http://arxiv.org/abs/1302.6257} {arXiv:1302.6257 [hep-th]} \BibitemShut
  {NoStop}%
\bibitem [{\citenamefont {Juge}\ \emph {et~al.}(2003)\citenamefont {Juge},
  \citenamefont {Kuti},\ and\ \citenamefont {Morningstar}}]{Juge:2002br}%
  \BibitemOpen
  \bibfield  {author} {\bibinfo {author} {\bibfnamefont {K.~J.}\ \bibnamefont
  {Juge}}, \bibinfo {author} {\bibfnamefont {J.}~\bibnamefont {Kuti}}, \ and\
  \bibinfo {author} {\bibfnamefont {C.}~\bibnamefont {Morningstar}},\ }\href
  {\doibase 10.1103/PhysRevLett.90.161601} {\bibfield  {journal} {\bibinfo
  {journal} {Phys. Rev. Lett.}\ }\textbf {\bibinfo {volume} {90}},\ \bibinfo
  {pages} {161601} (\bibinfo {year} {2003})},\ \Eprint
  {http://arxiv.org/abs/hep-lat/0207004} {arXiv:hep-lat/0207004} \BibitemShut
  {NoStop}%
\bibitem [{\citenamefont {Brandt}(2017)}]{Brandt:2017yzw}%
  \BibitemOpen
  \bibfield  {author} {\bibinfo {author} {\bibfnamefont {B.~B.}\ \bibnamefont
  {Brandt}},\ }\href {\doibase 10.1007/JHEP07(2017)008} {\bibfield  {journal}
  {\bibinfo  {journal} {JHEP}\ }\textbf {\bibinfo {volume} {07}},\ \bibinfo
  {pages} {008} (\bibinfo {year} {2017})},\ \Eprint
  {http://arxiv.org/abs/1705.03828} {arXiv:1705.03828 [hep-lat]} \BibitemShut
  {NoStop}%
\bibitem [{\citenamefont {Perez-Nadal}\ and\ \citenamefont
  {Soto}(2009)}]{PerezNadal:2008vm}%
  \BibitemOpen
  \bibfield  {author} {\bibinfo {author} {\bibfnamefont {G.}~\bibnamefont
  {Perez-Nadal}}\ and\ \bibinfo {author} {\bibfnamefont {J.}~\bibnamefont
  {Soto}},\ }\href {\doibase 10.1103/PhysRevD.79.114002} {\bibfield  {journal}
  {\bibinfo  {journal} {Phys. Rev. D}\ }\textbf {\bibinfo {volume} {79}},\
  \bibinfo {pages} {114002} (\bibinfo {year} {2009})},\ \Eprint
  {http://arxiv.org/abs/0811.2762} {arXiv:0811.2762 [hep-ph]} \BibitemShut
  {NoStop}%
\bibitem [{\citenamefont {Brambilla}\ \emph {et~al.}(2014)\citenamefont
  {Brambilla}, \citenamefont {Groher}, \citenamefont {Martinez},\ and\
  \citenamefont {Vairo}}]{Brambilla:2014eaa}%
  \BibitemOpen
  \bibfield  {author} {\bibinfo {author} {\bibfnamefont {N.}~\bibnamefont
  {Brambilla}}, \bibinfo {author} {\bibfnamefont {M.}~\bibnamefont {Groher}},
  \bibinfo {author} {\bibfnamefont {H.~E.}\ \bibnamefont {Martinez}}, \ and\
  \bibinfo {author} {\bibfnamefont {A.}~\bibnamefont {Vairo}},\ }\href
  {\doibase 10.1103/PhysRevD.90.114032} {\bibfield  {journal} {\bibinfo
  {journal} {Phys. Rev. D}\ }\textbf {\bibinfo {volume} {90}},\ \bibinfo
  {pages} {114032} (\bibinfo {year} {2014})},\ \Eprint
  {http://arxiv.org/abs/1407.7761} {arXiv:1407.7761 [hep-ph]} \BibitemShut
  {NoStop}%
\bibitem [{\citenamefont {Hwang}(2018)}]{Hwang:2018rju}%
  \BibitemOpen
  \bibfield  {author} {\bibinfo {author} {\bibfnamefont {S.}~\bibnamefont
  {Hwang}},\ }\emph {\bibinfo {title} {{Symmetries and determination of heavy
  quark potentials in effective string theory}}},\ \href@noop {} {Ph.D.
  thesis},\ \bibinfo  {school} {Munich, Tech. U.} (\bibinfo {year}
  {2018})\BibitemShut {NoStop}%
\bibitem [{\citenamefont {Koma}\ \emph {et~al.}(2006)\citenamefont {Koma},
  \citenamefont {Koma},\ and\ \citenamefont {Wittig}}]{Koma:2006si}%
  \BibitemOpen
  \bibfield  {author} {\bibinfo {author} {\bibfnamefont {Y.}~\bibnamefont
  {Koma}}, \bibinfo {author} {\bibfnamefont {M.}~\bibnamefont {Koma}}, \ and\
  \bibinfo {author} {\bibfnamefont {H.}~\bibnamefont {Wittig}},\ }\href
  {\doibase 10.1103/PhysRevLett.97.122003} {\bibfield  {journal} {\bibinfo
  {journal} {Phys. Rev. Lett.}\ }\textbf {\bibinfo {volume} {97}},\ \bibinfo
  {pages} {122003} (\bibinfo {year} {2006})},\ \Eprint
  {http://arxiv.org/abs/hep-lat/0607009} {arXiv:hep-lat/0607009} \BibitemShut
  {NoStop}%
\bibitem [{\citenamefont {Koma}\ and\ \citenamefont
  {Koma}(2007)}]{Koma:2006fw}%
  \BibitemOpen
  \bibfield  {author} {\bibinfo {author} {\bibfnamefont {Y.}~\bibnamefont
  {Koma}}\ and\ \bibinfo {author} {\bibfnamefont {M.}~\bibnamefont {Koma}},\
  }\href {\doibase 10.1016/j.nuclphysb.2007.01.033} {\bibfield  {journal}
  {\bibinfo  {journal} {Nucl. Phys. B}\ }\textbf {\bibinfo {volume} {769}},\
  \bibinfo {pages} {79} (\bibinfo {year} {2007})},\ \Eprint
  {http://arxiv.org/abs/hep-lat/0609078} {arXiv:hep-lat/0609078} \BibitemShut
  {NoStop}%
\bibitem [{\citenamefont {Oncala}\ and\ \citenamefont
  {Soto}(2017)}]{Oncala:2017hop}%
  \BibitemOpen
  \bibfield  {author} {\bibinfo {author} {\bibfnamefont {R.}~\bibnamefont
  {Oncala}}\ and\ \bibinfo {author} {\bibfnamefont {J.}~\bibnamefont {Soto}},\
  }\href {\doibase 10.1103/PhysRevD.96.014004} {\bibfield  {journal} {\bibinfo
  {journal} {Phys. Rev. D}\ }\textbf {\bibinfo {volume} {96}},\ \bibinfo
  {pages} {014004} (\bibinfo {year} {2017})},\ \Eprint
  {http://arxiv.org/abs/1702.03900} {arXiv:1702.03900 [hep-ph]} \BibitemShut
  {NoStop}%
\bibitem [{\citenamefont {Briceno}\ \emph {et~al.}(2012)\citenamefont
  {Briceno}, \citenamefont {Lin},\ and\ \citenamefont
  {Bolton}}]{Briceno:2012wt}%
  \BibitemOpen
  \bibfield  {author} {\bibinfo {author} {\bibfnamefont {R.~A.}\ \bibnamefont
  {Briceno}}, \bibinfo {author} {\bibfnamefont {H.-W.}\ \bibnamefont {Lin}}, \
  and\ \bibinfo {author} {\bibfnamefont {D.~R.}\ \bibnamefont {Bolton}},\
  }\href {\doibase 10.1103/PhysRevD.86.094504} {\bibfield  {journal} {\bibinfo
  {journal} {Phys. Rev. D}\ }\textbf {\bibinfo {volume} {86}},\ \bibinfo
  {pages} {094504} (\bibinfo {year} {2012})},\ \Eprint
  {http://arxiv.org/abs/1207.3536} {arXiv:1207.3536 [hep-lat]} \BibitemShut
  {NoStop}%
\bibitem [{\citenamefont {Namekawa}\ \emph {et~al.}(2013)\citenamefont
  {Namekawa} \emph {et~al.}}]{Namekawa:2013vu}%
  \BibitemOpen
  \bibfield  {author} {\bibinfo {author} {\bibfnamefont {Y.}~\bibnamefont
  {Namekawa}} \emph {et~al.} (\bibinfo {collaboration} {PACS-CS}),\ }\href
  {\doibase 10.1103/PhysRevD.87.094512} {\bibfield  {journal} {\bibinfo
  {journal} {Phys. Rev. D}\ }\textbf {\bibinfo {volume} {87}},\ \bibinfo
  {pages} {094512} (\bibinfo {year} {2013})},\ \Eprint
  {http://arxiv.org/abs/1301.4743} {arXiv:1301.4743 [hep-lat]} \BibitemShut
  {NoStop}%
\bibitem [{\citenamefont {Brown}\ \emph {et~al.}(2014)\citenamefont {Brown},
  \citenamefont {Detmold}, \citenamefont {Meinel},\ and\ \citenamefont
  {Orginos}}]{Brown:2014ena}%
  \BibitemOpen
  \bibfield  {author} {\bibinfo {author} {\bibfnamefont {Z.~S.}\ \bibnamefont
  {Brown}}, \bibinfo {author} {\bibfnamefont {W.}~\bibnamefont {Detmold}},
  \bibinfo {author} {\bibfnamefont {S.}~\bibnamefont {Meinel}}, \ and\ \bibinfo
  {author} {\bibfnamefont {K.}~\bibnamefont {Orginos}},\ }\href {\doibase
  10.1103/PhysRevD.90.094507} {\bibfield  {journal} {\bibinfo  {journal} {Phys.
  Rev. D}\ }\textbf {\bibinfo {volume} {90}},\ \bibinfo {pages} {094507}
  (\bibinfo {year} {2014})},\ \Eprint {http://arxiv.org/abs/1409.0497}
  {arXiv:1409.0497 [hep-lat]} \BibitemShut {NoStop}%
\bibitem [{\citenamefont {Alexandrou}\ \emph {et~al.}(2014)\citenamefont
  {Alexandrou}, \citenamefont {Drach}, \citenamefont {Jansen}, \citenamefont
  {Kallidonis},\ and\ \citenamefont {Koutsou}}]{Alexandrou:2014sha}%
  \BibitemOpen
  \bibfield  {author} {\bibinfo {author} {\bibfnamefont {C.}~\bibnamefont
  {Alexandrou}}, \bibinfo {author} {\bibfnamefont {V.}~\bibnamefont {Drach}},
  \bibinfo {author} {\bibfnamefont {K.}~\bibnamefont {Jansen}}, \bibinfo
  {author} {\bibfnamefont {C.}~\bibnamefont {Kallidonis}}, \ and\ \bibinfo
  {author} {\bibfnamefont {G.}~\bibnamefont {Koutsou}},\ }\href {\doibase
  10.1103/PhysRevD.90.074501} {\bibfield  {journal} {\bibinfo  {journal} {Phys.
  Rev. D}\ }\textbf {\bibinfo {volume} {90}},\ \bibinfo {pages} {074501}
  (\bibinfo {year} {2014})},\ \Eprint {http://arxiv.org/abs/1406.4310}
  {arXiv:1406.4310 [hep-lat]} \BibitemShut {NoStop}%
\bibitem [{\citenamefont {P\'erez-Rubio}\ \emph {et~al.}(2015)\citenamefont
  {P\'erez-Rubio}, \citenamefont {Collins},\ and\ \citenamefont
  {Bali}}]{Bali:2015lka}%
  \BibitemOpen
  \bibfield  {author} {\bibinfo {author} {\bibfnamefont {P.}~\bibnamefont
  {P\'erez-Rubio}}, \bibinfo {author} {\bibfnamefont {S.}~\bibnamefont
  {Collins}}, \ and\ \bibinfo {author} {\bibfnamefont {G.~S.}\ \bibnamefont
  {Bali}},\ }\href {\doibase 10.1103/PhysRevD.92.034504} {\bibfield  {journal}
  {\bibinfo  {journal} {Phys. Rev. D}\ }\textbf {\bibinfo {volume} {92}},\
  \bibinfo {pages} {034504} (\bibinfo {year} {2015})},\ \Eprint
  {http://arxiv.org/abs/1503.08440} {arXiv:1503.08440 [hep-lat]} \BibitemShut
  {NoStop}%
\bibitem [{\citenamefont {Padmanath}\ \emph {et~al.}(2015)\citenamefont
  {Padmanath}, \citenamefont {Edwards}, \citenamefont {Mathur},\ and\
  \citenamefont {Peardon}}]{Padmanath:2015jea}%
  \BibitemOpen
  \bibfield  {author} {\bibinfo {author} {\bibfnamefont {M.}~\bibnamefont
  {Padmanath}}, \bibinfo {author} {\bibfnamefont {R.~G.}\ \bibnamefont
  {Edwards}}, \bibinfo {author} {\bibfnamefont {N.}~\bibnamefont {Mathur}}, \
  and\ \bibinfo {author} {\bibfnamefont {M.}~\bibnamefont {Peardon}},\ }\href
  {\doibase 10.1103/PhysRevD.91.094502} {\bibfield  {journal} {\bibinfo
  {journal} {Phys. Rev. D}\ }\textbf {\bibinfo {volume} {91}},\ \bibinfo
  {pages} {094502} (\bibinfo {year} {2015})},\ \Eprint
  {http://arxiv.org/abs/1502.01845} {arXiv:1502.01845 [hep-lat]} \BibitemShut
  {NoStop}%
\bibitem [{\citenamefont {Alexandrou}\ and\ \citenamefont
  {Kallidonis}(2017)}]{Alexandrou:2017xwd}%
  \BibitemOpen
  \bibfield  {author} {\bibinfo {author} {\bibfnamefont {C.}~\bibnamefont
  {Alexandrou}}\ and\ \bibinfo {author} {\bibfnamefont {C.}~\bibnamefont
  {Kallidonis}},\ }\href {\doibase 10.1103/PhysRevD.96.034511} {\bibfield
  {journal} {\bibinfo  {journal} {Phys. Rev. D}\ }\textbf {\bibinfo {volume}
  {96}},\ \bibinfo {pages} {034511} (\bibinfo {year} {2017})},\ \Eprint
  {http://arxiv.org/abs/1704.02647} {arXiv:1704.02647 [hep-lat]} \BibitemShut
  {NoStop}%
\bibitem [{\citenamefont {Lewis}\ and\ \citenamefont
  {Woloshyn}(2009)}]{Lewis:2008fu}%
  \BibitemOpen
  \bibfield  {author} {\bibinfo {author} {\bibfnamefont {R.}~\bibnamefont
  {Lewis}}\ and\ \bibinfo {author} {\bibfnamefont {R.~M.}\ \bibnamefont
  {Woloshyn}},\ }\href {\doibase 10.1103/PhysRevD.79.014502} {\bibfield
  {journal} {\bibinfo  {journal} {Phys. Rev. D}\ }\textbf {\bibinfo {volume}
  {79}},\ \bibinfo {pages} {014502} (\bibinfo {year} {2009})},\ \Eprint
  {http://arxiv.org/abs/0806.4783} {arXiv:0806.4783 [hep-lat]} \BibitemShut
  {NoStop}%
\bibitem [{\citenamefont {Mohanta}\ and\ \citenamefont
  {Basak}(2020)}]{Mohanta:2019mxo}%
  \BibitemOpen
  \bibfield  {author} {\bibinfo {author} {\bibfnamefont {P.}~\bibnamefont
  {Mohanta}}\ and\ \bibinfo {author} {\bibfnamefont {S.}~\bibnamefont
  {Basak}},\ }\href {\doibase 10.1103/PhysRevD.101.094503} {\bibfield
  {journal} {\bibinfo  {journal} {Phys. Rev. D}\ }\textbf {\bibinfo {volume}
  {101}},\ \bibinfo {pages} {094503} (\bibinfo {year} {2020})},\ \Eprint
  {http://arxiv.org/abs/1911.03741} {arXiv:1911.03741 [hep-lat]} \BibitemShut
  {NoStop}%
\bibitem [{\citenamefont {Bahtiyar}\ \emph {et~al.}(2020)\citenamefont
  {Bahtiyar}, \citenamefont {Can}, \citenamefont {Erkol}, \citenamefont
  {Gubler}, \citenamefont {Oka},\ and\ \citenamefont
  {Takahashi}}]{Bahtiyar:2020uuj}%
  \BibitemOpen
  \bibfield  {author} {\bibinfo {author} {\bibfnamefont {H.}~\bibnamefont
  {Bahtiyar}}, \bibinfo {author} {\bibfnamefont {K.~U.}\ \bibnamefont {Can}},
  \bibinfo {author} {\bibfnamefont {G.}~\bibnamefont {Erkol}}, \bibinfo
  {author} {\bibfnamefont {P.}~\bibnamefont {Gubler}}, \bibinfo {author}
  {\bibfnamefont {M.}~\bibnamefont {Oka}}, \ and\ \bibinfo {author}
  {\bibfnamefont {T.~T.}\ \bibnamefont {Takahashi}},\ }\href {\doibase
  10.1103/PhysRevD.102.054513} {\bibfield  {journal} {\bibinfo  {journal}
  {Phys. Rev. D}\ }\textbf {\bibinfo {volume} {102}},\ \bibinfo {pages}
  {054513} (\bibinfo {year} {2020})},\ \Eprint
  {http://arxiv.org/abs/2004.08999} {arXiv:2004.08999 [hep-lat]} \BibitemShut
  {NoStop}%
\bibitem [{\citenamefont {Savage}\ and\ \citenamefont
  {Wise}(1990)}]{Savage:1990di}%
  \BibitemOpen
  \bibfield  {author} {\bibinfo {author} {\bibfnamefont {M.~J.}\ \bibnamefont
  {Savage}}\ and\ \bibinfo {author} {\bibfnamefont {M.~B.}\ \bibnamefont
  {Wise}},\ }\href {\doibase 10.1016/0370-2693(90)90035-5} {\bibfield
  {journal} {\bibinfo  {journal} {Phys. Lett. B}\ }\textbf {\bibinfo {volume}
  {248}},\ \bibinfo {pages} {177} (\bibinfo {year} {1990})}\BibitemShut
  {NoStop}%
\bibitem [{\citenamefont {Hu}\ and\ \citenamefont {Mehen}(2006)}]{Hu:2005gf}%
  \BibitemOpen
  \bibfield  {author} {\bibinfo {author} {\bibfnamefont {J.}~\bibnamefont
  {Hu}}\ and\ \bibinfo {author} {\bibfnamefont {T.}~\bibnamefont {Mehen}},\
  }\href {\doibase 10.1103/PhysRevD.73.054003} {\bibfield  {journal} {\bibinfo
  {journal} {Phys. Rev. D}\ }\textbf {\bibinfo {volume} {73}},\ \bibinfo
  {pages} {054003} (\bibinfo {year} {2006})},\ \Eprint
  {http://arxiv.org/abs/hep-ph/0511321} {arXiv:hep-ph/0511321} \BibitemShut
  {NoStop}%
\bibitem [{\citenamefont {Mehen}(2017)}]{Mehen:2017nrh}%
  \BibitemOpen
  \bibfield  {author} {\bibinfo {author} {\bibfnamefont {T.}~\bibnamefont
  {Mehen}},\ }\href {\doibase 10.1103/PhysRevD.96.094028} {\bibfield  {journal}
  {\bibinfo  {journal} {Phys. Rev. D}\ }\textbf {\bibinfo {volume} {96}},\
  \bibinfo {pages} {094028} (\bibinfo {year} {2017})},\ \Eprint
  {http://arxiv.org/abs/1708.05020} {arXiv:1708.05020 [hep-ph]} \BibitemShut
  {NoStop}%
\bibitem [{\citenamefont {Mehen}\ and\ \citenamefont
  {Mohapatra}(2019)}]{Mehen:2019cxn}%
  \BibitemOpen
  \bibfield  {author} {\bibinfo {author} {\bibfnamefont {T.~C.}\ \bibnamefont
  {Mehen}}\ and\ \bibinfo {author} {\bibfnamefont {A.}~\bibnamefont
  {Mohapatra}},\ }\href {\doibase 10.1103/PhysRevD.100.076014} {\bibfield
  {journal} {\bibinfo  {journal} {Phys. Rev. D}\ }\textbf {\bibinfo {volume}
  {100}},\ \bibinfo {pages} {076014} (\bibinfo {year} {2019})},\ \Eprint
  {http://arxiv.org/abs/1905.06965} {arXiv:1905.06965 [hep-ph]} \BibitemShut
  {NoStop}%
\bibitem [{\citenamefont {Bazavov}\ \emph {et~al.}(2018)\citenamefont {Bazavov}
  \emph {et~al.}}]{Bazavov:2018omf}%
  \BibitemOpen
  \bibfield  {author} {\bibinfo {author} {\bibfnamefont {A.}~\bibnamefont
  {Bazavov}} \emph {et~al.} (\bibinfo {collaboration} {Fermilab Lattice, MILC,
  TUMQCD}),\ }\href {\doibase 10.1103/PhysRevD.98.054517} {\bibfield  {journal}
  {\bibinfo  {journal} {Phys. Rev. D}\ }\textbf {\bibinfo {volume} {98}},\
  \bibinfo {pages} {054517} (\bibinfo {year} {2018})},\ \Eprint
  {http://arxiv.org/abs/1802.04248} {arXiv:1802.04248 [hep-lat]} \BibitemShut
  {NoStop}%
\bibitem [{\citenamefont {Ayala}\ \emph {et~al.}(2020)\citenamefont {Ayala},
  \citenamefont {Lobregat},\ and\ \citenamefont {Pineda}}]{Ayala:2019hkn}%
  \BibitemOpen
  \bibfield  {author} {\bibinfo {author} {\bibfnamefont {C.}~\bibnamefont
  {Ayala}}, \bibinfo {author} {\bibfnamefont {X.}~\bibnamefont {Lobregat}}, \
  and\ \bibinfo {author} {\bibfnamefont {A.}~\bibnamefont {Pineda}},\ }\href
  {\doibase 10.1103/PhysRevD.101.034002} {\bibfield  {journal} {\bibinfo
  {journal} {Phys. Rev. D}\ }\textbf {\bibinfo {volume} {101}},\ \bibinfo
  {pages} {034002} (\bibinfo {year} {2020})},\ \Eprint
  {http://arxiv.org/abs/1909.01370} {arXiv:1909.01370 [hep-ph]} \BibitemShut
  {NoStop}%
\bibitem [{\citenamefont {Zyla}\ \emph {et~al.}(2020)\citenamefont {Zyla} \emph
  {et~al.}}]{Zyla:2020zbs}%
  \BibitemOpen
  \bibfield  {author} {\bibinfo {author} {\bibfnamefont {P.~A.}\ \bibnamefont
  {Zyla}} \emph {et~al.} (\bibinfo {collaboration} {Particle Data Group}),\
  }\href {\doibase 10.1093/ptep/ptaa104} {\bibfield  {journal} {\bibinfo
  {journal} {PTEP}\ }\textbf {\bibinfo {volume} {2020}},\ \bibinfo {pages}
  {083C01} (\bibinfo {year} {2020})}\BibitemShut {NoStop}%
\bibitem [{\citenamefont {Bali}\ \emph {et~al.}(1995)\citenamefont {Bali},
  \citenamefont {Schilling},\ and\ \citenamefont {Schlichter}}]{Bali:1994de}%
  \BibitemOpen
  \bibfield  {author} {\bibinfo {author} {\bibfnamefont {G.~S.}\ \bibnamefont
  {Bali}}, \bibinfo {author} {\bibfnamefont {K.}~\bibnamefont {Schilling}}, \
  and\ \bibinfo {author} {\bibfnamefont {C.}~\bibnamefont {Schlichter}},\
  }\href {\doibase 10.1103/PhysRevD.51.5165} {\bibfield  {journal} {\bibinfo
  {journal} {Phys. Rev. D}\ }\textbf {\bibinfo {volume} {51}},\ \bibinfo
  {pages} {5165} (\bibinfo {year} {1995})},\ \Eprint
  {http://arxiv.org/abs/hep-lat/9409005} {arXiv:hep-lat/9409005} \BibitemShut
  {NoStop}%
\bibitem [{\citenamefont {Ichie}\ \emph {et~al.}(2003)\citenamefont {Ichie},
  \citenamefont {Bornyakov}, \citenamefont {Streuer},\ and\ \citenamefont
  {Schierholz}}]{Ichie:2002dy}%
  \BibitemOpen
  \bibfield  {author} {\bibinfo {author} {\bibfnamefont {H.}~\bibnamefont
  {Ichie}}, \bibinfo {author} {\bibfnamefont {V.}~\bibnamefont {Bornyakov}},
  \bibinfo {author} {\bibfnamefont {T.}~\bibnamefont {Streuer}}, \ and\
  \bibinfo {author} {\bibfnamefont {G.}~\bibnamefont {Schierholz}},\ }\href
  {\doibase 10.1016/S0375-9474(03)01238-7} {\bibfield  {journal} {\bibinfo
  {journal} {Nucl. Phys. A}\ }\textbf {\bibinfo {volume} {721}},\ \bibinfo
  {pages} {899} (\bibinfo {year} {2003})},\ \Eprint
  {http://arxiv.org/abs/hep-lat/0212036} {arXiv:hep-lat/0212036} \BibitemShut
  {NoStop}%
\bibitem [{\citenamefont {Bali}\ \emph {et~al.}(2005)\citenamefont {Bali},
  \citenamefont {Neff}, \citenamefont {Duessel}, \citenamefont {Lippert},\ and\
  \citenamefont {Schilling}}]{Bali:2005fu}%
  \BibitemOpen
  \bibfield  {author} {\bibinfo {author} {\bibfnamefont {G.~S.}\ \bibnamefont
  {Bali}}, \bibinfo {author} {\bibfnamefont {H.}~\bibnamefont {Neff}}, \bibinfo
  {author} {\bibfnamefont {T.}~\bibnamefont {Duessel}}, \bibinfo {author}
  {\bibfnamefont {T.}~\bibnamefont {Lippert}}, \ and\ \bibinfo {author}
  {\bibfnamefont {K.}~\bibnamefont {Schilling}} (\bibinfo {collaboration}
  {SESAM}),\ }\href {\doibase 10.1103/PhysRevD.71.114513} {\bibfield  {journal}
  {\bibinfo  {journal} {Phys. Rev. D}\ }\textbf {\bibinfo {volume} {71}},\
  \bibinfo {pages} {114513} (\bibinfo {year} {2005})},\ \Eprint
  {http://arxiv.org/abs/hep-lat/0505012} {arXiv:hep-lat/0505012} \BibitemShut
  {NoStop}%
\bibitem [{\citenamefont {Bulava}\ \emph {et~al.}(2019)\citenamefont {Bulava},
  \citenamefont {H\"orz}, \citenamefont {Knechtli}, \citenamefont {Koch},
  \citenamefont {Moir}, \citenamefont {Morningstar},\ and\ \citenamefont
  {Peardon}}]{Bulava:2019iut}%
  \BibitemOpen
  \bibfield  {author} {\bibinfo {author} {\bibfnamefont {J.}~\bibnamefont
  {Bulava}}, \bibinfo {author} {\bibfnamefont {B.}~\bibnamefont {H\"orz}},
  \bibinfo {author} {\bibfnamefont {F.}~\bibnamefont {Knechtli}}, \bibinfo
  {author} {\bibfnamefont {V.}~\bibnamefont {Koch}}, \bibinfo {author}
  {\bibfnamefont {G.}~\bibnamefont {Moir}}, \bibinfo {author} {\bibfnamefont
  {C.}~\bibnamefont {Morningstar}}, \ and\ \bibinfo {author} {\bibfnamefont
  {M.}~\bibnamefont {Peardon}},\ }\href {\doibase
  10.1016/j.physletb.2019.05.018} {\bibfield  {journal} {\bibinfo  {journal}
  {Phys. Lett. B}\ }\textbf {\bibinfo {volume} {793}},\ \bibinfo {pages} {493}
  (\bibinfo {year} {2019})},\ \Eprint {http://arxiv.org/abs/1902.04006}
  {arXiv:1902.04006 [hep-lat]} \BibitemShut {NoStop}%
\bibitem [{\citenamefont {Yamamoto}\ \emph {et~al.}(2008)\citenamefont
  {Yamamoto}, \citenamefont {Suganuma},\ and\ \citenamefont
  {Iida}}]{Yamamoto:2008jz}%
  \BibitemOpen
  \bibfield  {author} {\bibinfo {author} {\bibfnamefont {A.}~\bibnamefont
  {Yamamoto}}, \bibinfo {author} {\bibfnamefont {H.}~\bibnamefont {Suganuma}},
  \ and\ \bibinfo {author} {\bibfnamefont {H.}~\bibnamefont {Iida}},\ }\href
  {\doibase 10.1103/PhysRevD.78.014513} {\bibfield  {journal} {\bibinfo
  {journal} {Phys. Rev. D}\ }\textbf {\bibinfo {volume} {78}},\ \bibinfo
  {pages} {014513} (\bibinfo {year} {2008})},\ \Eprint
  {http://arxiv.org/abs/0806.3554} {arXiv:0806.3554 [hep-lat]} \BibitemShut
  {NoStop}%
\bibitem [{\citenamefont {Liu}\ \emph {et~al.}(2012)\citenamefont {Liu},
  \citenamefont {Moir}, \citenamefont {Peardon}, \citenamefont {Ryan},
  \citenamefont {Thomas}, \citenamefont {Vilaseca}, \citenamefont {Dudek},
  \citenamefont {Edwards}, \citenamefont {Joo},\ and\ \citenamefont
  {Richards}}]{HadronSpectrum:2012gic}%
  \BibitemOpen
  \bibfield  {author} {\bibinfo {author} {\bibfnamefont {L.}~\bibnamefont
  {Liu}}, \bibinfo {author} {\bibfnamefont {G.}~\bibnamefont {Moir}}, \bibinfo
  {author} {\bibfnamefont {M.}~\bibnamefont {Peardon}}, \bibinfo {author}
  {\bibfnamefont {S.~M.}\ \bibnamefont {Ryan}}, \bibinfo {author}
  {\bibfnamefont {C.~E.}\ \bibnamefont {Thomas}}, \bibinfo {author}
  {\bibfnamefont {P.}~\bibnamefont {Vilaseca}}, \bibinfo {author}
  {\bibfnamefont {J.~J.}\ \bibnamefont {Dudek}}, \bibinfo {author}
  {\bibfnamefont {R.~G.}\ \bibnamefont {Edwards}}, \bibinfo {author}
  {\bibfnamefont {B.}~\bibnamefont {Joo}}, \ and\ \bibinfo {author}
  {\bibfnamefont {D.~G.}\ \bibnamefont {Richards}} (\bibinfo {collaboration}
  {Hadron Spectrum}),\ }\href {\doibase 10.1007/JHEP07(2012)126} {\bibfield
  {journal} {\bibinfo  {journal} {JHEP}\ }\textbf {\bibinfo {volume} {07}},\
  \bibinfo {pages} {126} (\bibinfo {year} {2012})},\ \Eprint
  {http://arxiv.org/abs/1204.5425} {arXiv:1204.5425 [hep-ph]} \BibitemShut
  {NoStop}%
\bibitem [{\citenamefont {Cheung}\ \emph {et~al.}(2016)\citenamefont {Cheung},
  \citenamefont {O'Hara}, \citenamefont {Moir}, \citenamefont {Peardon},
  \citenamefont {Ryan}, \citenamefont {Thomas},\ and\ \citenamefont
  {Tims}}]{Cheung:2016bym}%
  \BibitemOpen
  \bibfield  {author} {\bibinfo {author} {\bibfnamefont {G.~K.~C.}\
  \bibnamefont {Cheung}}, \bibinfo {author} {\bibfnamefont {C.}~\bibnamefont
  {O'Hara}}, \bibinfo {author} {\bibfnamefont {G.}~\bibnamefont {Moir}},
  \bibinfo {author} {\bibfnamefont {M.}~\bibnamefont {Peardon}}, \bibinfo
  {author} {\bibfnamefont {S.~M.}\ \bibnamefont {Ryan}}, \bibinfo {author}
  {\bibfnamefont {C.~E.}\ \bibnamefont {Thomas}}, \ and\ \bibinfo {author}
  {\bibfnamefont {D.}~\bibnamefont {Tims}} (\bibinfo {collaboration} {Hadron
  Spectrum}),\ }\href {\doibase 10.1007/JHEP12(2016)089} {\bibfield  {journal}
  {\bibinfo  {journal} {JHEP}\ }\textbf {\bibinfo {volume} {12}},\ \bibinfo
  {pages} {089} (\bibinfo {year} {2016})},\ \Eprint
  {http://arxiv.org/abs/1610.01073} {arXiv:1610.01073 [hep-lat]} \BibitemShut
  {NoStop}%
\bibitem [{\citenamefont {De~Rujula}\ \emph {et~al.}(1975)\citenamefont
  {De~Rujula}, \citenamefont {Georgi},\ and\ \citenamefont
  {Glashow}}]{DeRujula:1975qlm}%
  \BibitemOpen
  \bibfield  {author} {\bibinfo {author} {\bibfnamefont {A.}~\bibnamefont
  {De~Rujula}}, \bibinfo {author} {\bibfnamefont {H.}~\bibnamefont {Georgi}}, \
  and\ \bibinfo {author} {\bibfnamefont {S.~L.}\ \bibnamefont {Glashow}},\
  }\href {\doibase 10.1103/PhysRevD.12.147} {\bibfield  {journal} {\bibinfo
  {journal} {Phys. Rev. D}\ }\textbf {\bibinfo {volume} {12}},\ \bibinfo
  {pages} {147} (\bibinfo {year} {1975})}\BibitemShut {NoStop}%
\bibitem [{\citenamefont {Ebert}\ \emph {et~al.}(1997)\citenamefont {Ebert},
  \citenamefont {Faustov}, \citenamefont {Galkin}, \citenamefont {Martynenko},\
  and\ \citenamefont {Saleev}}]{Ebert:1996ec}%
  \BibitemOpen
  \bibfield  {author} {\bibinfo {author} {\bibfnamefont {D.}~\bibnamefont
  {Ebert}}, \bibinfo {author} {\bibfnamefont {R.~N.}\ \bibnamefont {Faustov}},
  \bibinfo {author} {\bibfnamefont {V.~O.}\ \bibnamefont {Galkin}}, \bibinfo
  {author} {\bibfnamefont {A.~P.}\ \bibnamefont {Martynenko}}, \ and\ \bibinfo
  {author} {\bibfnamefont {V.~A.}\ \bibnamefont {Saleev}},\ }\href {\doibase
  10.1007/s002880050534} {\bibfield  {journal} {\bibinfo  {journal} {Z. Phys.
  C}\ }\textbf {\bibinfo {volume} {76}},\ \bibinfo {pages} {111} (\bibinfo
  {year} {1997})},\ \Eprint {http://arxiv.org/abs/hep-ph/9607314}
  {arXiv:hep-ph/9607314} \BibitemShut {NoStop}%
\bibitem [{\citenamefont {Gerasyuta}\ and\ \citenamefont
  {Ivanov}(1999)}]{Gerasyuta:1999pc}%
  \BibitemOpen
  \bibfield  {author} {\bibinfo {author} {\bibfnamefont {S.~M.}\ \bibnamefont
  {Gerasyuta}}\ and\ \bibinfo {author} {\bibfnamefont {D.~V.}\ \bibnamefont
  {Ivanov}},\ }\href {\doibase 10.1007/BF03035848} {\bibfield  {journal}
  {\bibinfo  {journal} {Nuovo Cim. A}\ }\textbf {\bibinfo {volume} {112}},\
  \bibinfo {pages} {261} (\bibinfo {year} {1999})},\ \Eprint
  {http://arxiv.org/abs/hep-ph/0101310} {arXiv:hep-ph/0101310} \BibitemShut
  {NoStop}%
\bibitem [{\citenamefont {Itoh}\ \emph {et~al.}(2000)\citenamefont {Itoh},
  \citenamefont {Minamikawa}, \citenamefont {Miura},\ and\ \citenamefont
  {Watanabe}}]{Itoh:2000um}%
  \BibitemOpen
  \bibfield  {author} {\bibinfo {author} {\bibfnamefont {C.}~\bibnamefont
  {Itoh}}, \bibinfo {author} {\bibfnamefont {T.}~\bibnamefont {Minamikawa}},
  \bibinfo {author} {\bibfnamefont {K.}~\bibnamefont {Miura}}, \ and\ \bibinfo
  {author} {\bibfnamefont {T.}~\bibnamefont {Watanabe}},\ }\href {\doibase
  10.1103/PhysRevD.61.057502} {\bibfield  {journal} {\bibinfo  {journal} {Phys.
  Rev. D}\ }\textbf {\bibinfo {volume} {61}},\ \bibinfo {pages} {057502}
  (\bibinfo {year} {2000})}\BibitemShut {NoStop}%
\bibitem [{\citenamefont {Gershtein}\ \emph {et~al.}(2000)\citenamefont
  {Gershtein}, \citenamefont {Kiselev}, \citenamefont {Likhoded},\ and\
  \citenamefont {Onishchenko}}]{Gershtein:2000nx}%
  \BibitemOpen
  \bibfield  {author} {\bibinfo {author} {\bibfnamefont {S.~S.}\ \bibnamefont
  {Gershtein}}, \bibinfo {author} {\bibfnamefont {V.~V.}\ \bibnamefont
  {Kiselev}}, \bibinfo {author} {\bibfnamefont {A.~K.}\ \bibnamefont
  {Likhoded}}, \ and\ \bibinfo {author} {\bibfnamefont {A.~I.}\ \bibnamefont
  {Onishchenko}},\ }\href {\doibase 10.1103/PhysRevD.62.054021} {\bibfield
  {journal} {\bibinfo  {journal} {Phys. Rev. D}\ }\textbf {\bibinfo {volume}
  {62}},\ \bibinfo {pages} {054021} (\bibinfo {year} {2000})}\BibitemShut
  {NoStop}%
\bibitem [{\citenamefont {Ebert}\ \emph {et~al.}(2002)\citenamefont {Ebert},
  \citenamefont {Faustov}, \citenamefont {Galkin},\ and\ \citenamefont
  {Martynenko}}]{Ebert:2002ig}%
  \BibitemOpen
  \bibfield  {author} {\bibinfo {author} {\bibfnamefont {D.}~\bibnamefont
  {Ebert}}, \bibinfo {author} {\bibfnamefont {R.~N.}\ \bibnamefont {Faustov}},
  \bibinfo {author} {\bibfnamefont {V.~O.}\ \bibnamefont {Galkin}}, \ and\
  \bibinfo {author} {\bibfnamefont {A.~P.}\ \bibnamefont {Martynenko}},\ }\href
  {\doibase 10.1103/PhysRevD.66.014008} {\bibfield  {journal} {\bibinfo
  {journal} {Phys. Rev. D}\ }\textbf {\bibinfo {volume} {66}},\ \bibinfo
  {pages} {014008} (\bibinfo {year} {2002})},\ \Eprint
  {http://arxiv.org/abs/hep-ph/0201217} {arXiv:hep-ph/0201217} \BibitemShut
  {NoStop}%
\bibitem [{\citenamefont {Albertus}\ \emph {et~al.}(2007)\citenamefont
  {Albertus}, \citenamefont {Hernandez}, \citenamefont {Nieves},\ and\
  \citenamefont {Verde-Velasco}}]{Albertus:2006ya}%
  \BibitemOpen
  \bibfield  {author} {\bibinfo {author} {\bibfnamefont {C.}~\bibnamefont
  {Albertus}}, \bibinfo {author} {\bibfnamefont {E.}~\bibnamefont {Hernandez}},
  \bibinfo {author} {\bibfnamefont {J.}~\bibnamefont {Nieves}}, \ and\ \bibinfo
  {author} {\bibfnamefont {J.~M.}\ \bibnamefont {Verde-Velasco}},\ }\href
  {\doibase 10.1140/epja/i2007-10364-y} {\bibfield  {journal} {\bibinfo
  {journal} {Eur. Phys. J. A}\ }\textbf {\bibinfo {volume} {32}},\ \bibinfo
  {pages} {183} (\bibinfo {year} {2007})},\ \bibinfo {note} {[Erratum:
  Eur.Phys.J.A 36, 119 (2008)]},\ \Eprint {http://arxiv.org/abs/hep-ph/0610030}
  {arXiv:hep-ph/0610030} \BibitemShut {NoStop}%
\bibitem [{\citenamefont {Roberts}\ and\ \citenamefont
  {Pervin}(2008)}]{Roberts:2007ni}%
  \BibitemOpen
  \bibfield  {author} {\bibinfo {author} {\bibfnamefont {W.}~\bibnamefont
  {Roberts}}\ and\ \bibinfo {author} {\bibfnamefont {M.}~\bibnamefont
  {Pervin}},\ }\href {\doibase 10.1142/S0217751X08041219} {\bibfield  {journal}
  {\bibinfo  {journal} {Int. J. Mod. Phys. A}\ }\textbf {\bibinfo {volume}
  {23}},\ \bibinfo {pages} {2817} (\bibinfo {year} {2008})},\ \Eprint
  {http://arxiv.org/abs/0711.2492} {arXiv:0711.2492 [nucl-th]} \BibitemShut
  {NoStop}%
\bibitem [{\citenamefont {Martynenko}(2008)}]{Martynenko:2007je}%
  \BibitemOpen
  \bibfield  {author} {\bibinfo {author} {\bibfnamefont {A.~P.}\ \bibnamefont
  {Martynenko}},\ }\href {\doibase 10.1016/j.physletb.2008.04.030} {\bibfield
  {journal} {\bibinfo  {journal} {Phys. Lett. B}\ }\textbf {\bibinfo {volume}
  {663}},\ \bibinfo {pages} {317} (\bibinfo {year} {2008})},\ \Eprint
  {http://arxiv.org/abs/0708.2033} {arXiv:0708.2033 [hep-ph]} \BibitemShut
  {NoStop}%
\bibitem [{\citenamefont {Yoshida}\ \emph {et~al.}(2015)\citenamefont
  {Yoshida}, \citenamefont {Hiyama}, \citenamefont {Hosaka}, \citenamefont
  {Oka},\ and\ \citenamefont {Sadato}}]{Yoshida:2015tia}%
  \BibitemOpen
  \bibfield  {author} {\bibinfo {author} {\bibfnamefont {T.}~\bibnamefont
  {Yoshida}}, \bibinfo {author} {\bibfnamefont {E.}~\bibnamefont {Hiyama}},
  \bibinfo {author} {\bibfnamefont {A.}~\bibnamefont {Hosaka}}, \bibinfo
  {author} {\bibfnamefont {M.}~\bibnamefont {Oka}}, \ and\ \bibinfo {author}
  {\bibfnamefont {K.}~\bibnamefont {Sadato}},\ }\href {\doibase
  10.1103/PhysRevD.92.114029} {\bibfield  {journal} {\bibinfo  {journal} {Phys.
  Rev. D}\ }\textbf {\bibinfo {volume} {92}},\ \bibinfo {pages} {114029}
  (\bibinfo {year} {2015})},\ \Eprint {http://arxiv.org/abs/1510.01067}
  {arXiv:1510.01067 [hep-ph]} \BibitemShut {NoStop}%
\bibitem [{\citenamefont {Kiselev}\ \emph {et~al.}(2018)\citenamefont
  {Kiselev}, \citenamefont {Berezhnoy},\ and\ \citenamefont
  {Likhoded}}]{Kiselev:2017eic}%
  \BibitemOpen
  \bibfield  {author} {\bibinfo {author} {\bibfnamefont {V.~V.}\ \bibnamefont
  {Kiselev}}, \bibinfo {author} {\bibfnamefont {A.~V.}\ \bibnamefont
  {Berezhnoy}}, \ and\ \bibinfo {author} {\bibfnamefont {A.~K.}\ \bibnamefont
  {Likhoded}},\ }\href {\doibase 10.1134/S1063778818030134} {\bibfield
  {journal} {\bibinfo  {journal} {Phys. Atom. Nucl.}\ }\textbf {\bibinfo
  {volume} {81}},\ \bibinfo {pages} {369} (\bibinfo {year} {2018})},\ \Eprint
  {http://arxiv.org/abs/1706.09181} {arXiv:1706.09181 [hep-ph]} \BibitemShut
  {NoStop}%
\bibitem [{\citenamefont {Shah}\ and\ \citenamefont
  {Rai}(2017)}]{Shah:2017liu}%
  \BibitemOpen
  \bibfield  {author} {\bibinfo {author} {\bibfnamefont {Z.}~\bibnamefont
  {Shah}}\ and\ \bibinfo {author} {\bibfnamefont {A.~K.}\ \bibnamefont {Rai}},\
  }\href {\doibase 10.1140/epjc/s10052-017-4688-x} {\bibfield  {journal}
  {\bibinfo  {journal} {Eur. Phys. J. C}\ }\textbf {\bibinfo {volume} {77}},\
  \bibinfo {pages} {129} (\bibinfo {year} {2017})},\ \Eprint
  {http://arxiv.org/abs/1702.02726} {arXiv:1702.02726 [hep-ph]} \BibitemShut
  {NoStop}%
\bibitem [{\citenamefont {Weng}\ \emph {et~al.}(2018)\citenamefont {Weng},
  \citenamefont {Chen},\ and\ \citenamefont {Deng}}]{Weng:2018mmf}%
  \BibitemOpen
  \bibfield  {author} {\bibinfo {author} {\bibfnamefont {X.-Z.}\ \bibnamefont
  {Weng}}, \bibinfo {author} {\bibfnamefont {X.-L.}\ \bibnamefont {Chen}}, \
  and\ \bibinfo {author} {\bibfnamefont {W.-Z.}\ \bibnamefont {Deng}},\ }\href
  {\doibase 10.1103/PhysRevD.97.054008} {\bibfield  {journal} {\bibinfo
  {journal} {Phys. Rev. D}\ }\textbf {\bibinfo {volume} {97}},\ \bibinfo
  {pages} {054008} (\bibinfo {year} {2018})},\ \Eprint
  {http://arxiv.org/abs/1801.08644} {arXiv:1801.08644 [hep-ph]} \BibitemShut
  {NoStop}%
\bibitem [{\citenamefont {L\"u}\ \emph {et~al.}(2017)\citenamefont {L\"u},
  \citenamefont {Wang}, \citenamefont {Xiao},\ and\ \citenamefont
  {Zhong}}]{Lu:2017meb}%
  \BibitemOpen
  \bibfield  {author} {\bibinfo {author} {\bibfnamefont {Q.-F.}\ \bibnamefont
  {L\"u}}, \bibinfo {author} {\bibfnamefont {K.-L.}\ \bibnamefont {Wang}},
  \bibinfo {author} {\bibfnamefont {L.-Y.}\ \bibnamefont {Xiao}}, \ and\
  \bibinfo {author} {\bibfnamefont {X.-H.}\ \bibnamefont {Zhong}},\ }\href
  {\doibase 10.1103/PhysRevD.96.114006} {\bibfield  {journal} {\bibinfo
  {journal} {Phys. Rev. D}\ }\textbf {\bibinfo {volume} {96}},\ \bibinfo
  {pages} {114006} (\bibinfo {year} {2017})},\ \Eprint
  {http://arxiv.org/abs/1708.04468} {arXiv:1708.04468 [hep-ph]} \BibitemShut
  {NoStop}%
\bibitem [{\citenamefont {Migura}\ \emph {et~al.}(2006)\citenamefont {Migura},
  \citenamefont {Merten}, \citenamefont {Metsch},\ and\ \citenamefont
  {Petry}}]{Migura:2006ep}%
  \BibitemOpen
  \bibfield  {author} {\bibinfo {author} {\bibfnamefont {S.}~\bibnamefont
  {Migura}}, \bibinfo {author} {\bibfnamefont {D.}~\bibnamefont {Merten}},
  \bibinfo {author} {\bibfnamefont {B.}~\bibnamefont {Metsch}}, \ and\ \bibinfo
  {author} {\bibfnamefont {H.-R.}\ \bibnamefont {Petry}},\ }\href {\doibase
  10.1140/epja/i2006-10017-9} {\bibfield  {journal} {\bibinfo  {journal} {Eur.
  Phys. J. A}\ }\textbf {\bibinfo {volume} {28}},\ \bibinfo {pages} {41}
  (\bibinfo {year} {2006})},\ \Eprint {http://arxiv.org/abs/hep-ph/0602153}
  {arXiv:hep-ph/0602153} \BibitemShut {NoStop}%
\bibitem [{\citenamefont {Weng}\ \emph {et~al.}(2011)\citenamefont {Weng},
  \citenamefont {Guo},\ and\ \citenamefont {Thomas}}]{Weng:2010rb}%
  \BibitemOpen
  \bibfield  {author} {\bibinfo {author} {\bibfnamefont {M.~H.}\ \bibnamefont
  {Weng}}, \bibinfo {author} {\bibfnamefont {X.~H.}\ \bibnamefont {Guo}}, \
  and\ \bibinfo {author} {\bibfnamefont {A.~W.}\ \bibnamefont {Thomas}},\
  }\href {\doibase 10.1103/PhysRevD.83.056006} {\bibfield  {journal} {\bibinfo
  {journal} {Phys. Rev. D}\ }\textbf {\bibinfo {volume} {83}},\ \bibinfo
  {pages} {056006} (\bibinfo {year} {2011})},\ \Eprint
  {http://arxiv.org/abs/1012.0082} {arXiv:1012.0082 [hep-ph]} \BibitemShut
  {NoStop}%
\bibitem [{\citenamefont {Li}\ \emph {et~al.}(2020)\citenamefont {Li},
  \citenamefont {Chang}, \citenamefont {Qin},\ and\ \citenamefont
  {Wang}}]{Li:2019ekr}%
  \BibitemOpen
  \bibfield  {author} {\bibinfo {author} {\bibfnamefont {Q.}~\bibnamefont
  {Li}}, \bibinfo {author} {\bibfnamefont {C.-H.}\ \bibnamefont {Chang}},
  \bibinfo {author} {\bibfnamefont {S.-X.}\ \bibnamefont {Qin}}, \ and\
  \bibinfo {author} {\bibfnamefont {G.-L.}\ \bibnamefont {Wang}},\ }\href
  {\doibase 10.1088/1674-1137/44/1/013102} {\bibfield  {journal} {\bibinfo
  {journal} {Chin. Phys. C}\ }\textbf {\bibinfo {volume} {44}},\ \bibinfo
  {pages} {013102} (\bibinfo {year} {2020})},\ \Eprint
  {http://arxiv.org/abs/1903.02282} {arXiv:1903.02282 [hep-ph]} \BibitemShut
  {NoStop}%
\bibitem [{\citenamefont {Fleck}\ and\ \citenamefont
  {Richard}(1989)}]{Fleck:1989mb}%
  \BibitemOpen
  \bibfield  {author} {\bibinfo {author} {\bibfnamefont {S.}~\bibnamefont
  {Fleck}}\ and\ \bibinfo {author} {\bibfnamefont {J.~M.}\ \bibnamefont
  {Richard}},\ }\href {\doibase 10.1143/PTP.82.760} {\bibfield  {journal}
  {\bibinfo  {journal} {Prog. Theor. Phys.}\ }\textbf {\bibinfo {volume}
  {82}},\ \bibinfo {pages} {760} (\bibinfo {year} {1989})}\BibitemShut
  {NoStop}%
\bibitem [{\citenamefont {Maiani}\ \emph {et~al.}(2019)\citenamefont {Maiani},
  \citenamefont {Polosa},\ and\ \citenamefont {Riquer}}]{Maiani:2019lpu}%
  \BibitemOpen
  \bibfield  {author} {\bibinfo {author} {\bibfnamefont {L.}~\bibnamefont
  {Maiani}}, \bibinfo {author} {\bibfnamefont {A.~D.}\ \bibnamefont {Polosa}},
  \ and\ \bibinfo {author} {\bibfnamefont {V.}~\bibnamefont {Riquer}},\ }\href
  {\doibase 10.1103/PhysRevD.100.074002} {\bibfield  {journal} {\bibinfo
  {journal} {Phys. Rev. D}\ }\textbf {\bibinfo {volume} {100}},\ \bibinfo
  {pages} {074002} (\bibinfo {year} {2019})},\ \Eprint
  {http://arxiv.org/abs/1908.03244} {arXiv:1908.03244 [hep-ph]} \BibitemShut
  {NoStop}%
\bibitem [{\citenamefont {Roncaglia}\ \emph {et~al.}(1995)\citenamefont
  {Roncaglia}, \citenamefont {Lichtenberg},\ and\ \citenamefont
  {Predazzi}}]{Roncaglia:1995az}%
  \BibitemOpen
  \bibfield  {author} {\bibinfo {author} {\bibfnamefont {R.}~\bibnamefont
  {Roncaglia}}, \bibinfo {author} {\bibfnamefont {D.~B.}\ \bibnamefont
  {Lichtenberg}}, \ and\ \bibinfo {author} {\bibfnamefont {E.}~\bibnamefont
  {Predazzi}},\ }\href {\doibase 10.1103/PhysRevD.52.1722} {\bibfield
  {journal} {\bibinfo  {journal} {Phys. Rev. D}\ }\textbf {\bibinfo {volume}
  {52}},\ \bibinfo {pages} {1722} (\bibinfo {year} {1995})},\ \Eprint
  {http://arxiv.org/abs/hep-ph/9502251} {arXiv:hep-ph/9502251} \BibitemShut
  {NoStop}%
\bibitem [{\citenamefont {Karliner}\ and\ \citenamefont
  {Rosner}(2014)}]{Karliner:2014gca}%
  \BibitemOpen
  \bibfield  {author} {\bibinfo {author} {\bibfnamefont {M.}~\bibnamefont
  {Karliner}}\ and\ \bibinfo {author} {\bibfnamefont {J.~L.}\ \bibnamefont
  {Rosner}},\ }\href {\doibase 10.1103/PhysRevD.90.094007} {\bibfield
  {journal} {\bibinfo  {journal} {Phys. Rev. D}\ }\textbf {\bibinfo {volume}
  {90}},\ \bibinfo {pages} {094007} (\bibinfo {year} {2014})},\ \Eprint
  {http://arxiv.org/abs/1408.5877} {arXiv:1408.5877 [hep-ph]} \BibitemShut
  {NoStop}%
\bibitem [{\citenamefont {Lichtenberg}\ \emph {et~al.}(1996)\citenamefont
  {Lichtenberg}, \citenamefont {Roncaglia},\ and\ \citenamefont
  {Predazzi}}]{Lichtenberg:1995kg}%
  \BibitemOpen
  \bibfield  {author} {\bibinfo {author} {\bibfnamefont {D.~B.}\ \bibnamefont
  {Lichtenberg}}, \bibinfo {author} {\bibfnamefont {R.}~\bibnamefont
  {Roncaglia}}, \ and\ \bibinfo {author} {\bibfnamefont {E.}~\bibnamefont
  {Predazzi}},\ }\href {\doibase 10.1103/PhysRevD.53.6678} {\bibfield
  {journal} {\bibinfo  {journal} {Phys. Rev. D}\ }\textbf {\bibinfo {volume}
  {53}},\ \bibinfo {pages} {6678} (\bibinfo {year} {1996})},\ \Eprint
  {http://arxiv.org/abs/hep-ph/9511461} {arXiv:hep-ph/9511461} \BibitemShut
  {NoStop}%
\bibitem [{\citenamefont {Zhang}\ and\ \citenamefont
  {Huang}(2008)}]{Zhang:2008rt}%
  \BibitemOpen
  \bibfield  {author} {\bibinfo {author} {\bibfnamefont {J.-R.}\ \bibnamefont
  {Zhang}}\ and\ \bibinfo {author} {\bibfnamefont {M.-Q.}\ \bibnamefont
  {Huang}},\ }\href {\doibase 10.1103/PhysRevD.78.094007} {\bibfield  {journal}
  {\bibinfo  {journal} {Phys. Rev. D}\ }\textbf {\bibinfo {volume} {78}},\
  \bibinfo {pages} {094007} (\bibinfo {year} {2008})},\ \Eprint
  {http://arxiv.org/abs/0810.5396} {arXiv:0810.5396 [hep-ph]} \BibitemShut
  {NoStop}%
\bibitem [{\citenamefont {Wang}(2018)}]{Wang:2018lhz}%
  \BibitemOpen
  \bibfield  {author} {\bibinfo {author} {\bibfnamefont {Z.-G.}\ \bibnamefont
  {Wang}},\ }\href {\doibase 10.1140/epjc/s10052-018-6300-4} {\bibfield
  {journal} {\bibinfo  {journal} {Eur. Phys. J. C}\ }\textbf {\bibinfo {volume}
  {78}},\ \bibinfo {pages} {826} (\bibinfo {year} {2018})},\ \Eprint
  {http://arxiv.org/abs/1808.09820} {arXiv:1808.09820 [hep-ph]} \BibitemShut
  {NoStop}%
\bibitem [{\citenamefont {Silvestre-Brac}(1996)}]{Silvestre-Brac:1996tmn}%
  \BibitemOpen
  \bibfield  {author} {\bibinfo {author} {\bibfnamefont {B.}~\bibnamefont
  {Silvestre-Brac}},\ }\href {\doibase 10.1016/0146-6410(96)00030-0} {\bibfield
   {journal} {\bibinfo  {journal} {Prog. Part. Nucl. Phys.}\ }\textbf {\bibinfo
  {volume} {36}},\ \bibinfo {pages} {263} (\bibinfo {year} {1996})}\BibitemShut
  {NoStop}%
\bibitem [{\citenamefont {He}\ \emph {et~al.}(2004)\citenamefont {He},
  \citenamefont {Qian}, \citenamefont {Ding}, \citenamefont {Li},\ and\
  \citenamefont {Shen}}]{He:2004px}%
  \BibitemOpen
  \bibfield  {author} {\bibinfo {author} {\bibfnamefont {D.-H.}\ \bibnamefont
  {He}}, \bibinfo {author} {\bibfnamefont {K.}~\bibnamefont {Qian}}, \bibinfo
  {author} {\bibfnamefont {Y.-B.}\ \bibnamefont {Ding}}, \bibinfo {author}
  {\bibfnamefont {X.-Q.}\ \bibnamefont {Li}}, \ and\ \bibinfo {author}
  {\bibfnamefont {P.-N.}\ \bibnamefont {Shen}},\ }\href {\doibase
  10.1103/PhysRevD.70.094004} {\bibfield  {journal} {\bibinfo  {journal} {Phys.
  Rev. D}\ }\textbf {\bibinfo {volume} {70}},\ \bibinfo {pages} {094004}
  (\bibinfo {year} {2004})},\ \Eprint {http://arxiv.org/abs/hep-ph/0403301}
  {arXiv:hep-ph/0403301} \BibitemShut {NoStop}%
\bibitem [{\citenamefont {Chen}\ \emph {et~al.}(2017)\citenamefont {Chen},
  \citenamefont {Chen}, \citenamefont {Liu}, \citenamefont {Liu},\ and\
  \citenamefont {Zhu}}]{Chen:2016spr}%
  \BibitemOpen
  \bibfield  {author} {\bibinfo {author} {\bibfnamefont {H.-X.}\ \bibnamefont
  {Chen}}, \bibinfo {author} {\bibfnamefont {W.}~\bibnamefont {Chen}}, \bibinfo
  {author} {\bibfnamefont {X.}~\bibnamefont {Liu}}, \bibinfo {author}
  {\bibfnamefont {Y.-R.}\ \bibnamefont {Liu}}, \ and\ \bibinfo {author}
  {\bibfnamefont {S.-L.}\ \bibnamefont {Zhu}},\ }\href {\doibase
  10.1088/1361-6633/aa6420} {\bibfield  {journal} {\bibinfo  {journal} {Rept.
  Prog. Phys.}\ }\textbf {\bibinfo {volume} {80}},\ \bibinfo {pages} {076201}
  (\bibinfo {year} {2017})},\ \Eprint {http://arxiv.org/abs/1609.08928}
  {arXiv:1609.08928 [hep-ph]} \BibitemShut {NoStop}%
\bibitem [{\citenamefont {Mohajery}\ \emph {et~al.}(2018)\citenamefont
  {Mohajery}, \citenamefont {Salehi},\ and\ \citenamefont
  {Hassanabadi}}]{Mohajery:2018qhz}%
  \BibitemOpen
  \bibfield  {author} {\bibinfo {author} {\bibfnamefont {N.}~\bibnamefont
  {Mohajery}}, \bibinfo {author} {\bibfnamefont {N.}~\bibnamefont {Salehi}}, \
  and\ \bibinfo {author} {\bibfnamefont {H.}~\bibnamefont {Hassanabadi}},\
  }\href {\doibase 10.1155/2018/1326438} {\bibfield  {journal} {\bibinfo
  {journal} {Adv. High Energy Phys.}\ }\textbf {\bibinfo {volume} {2018}},\
  \bibinfo {pages} {1326438} (\bibinfo {year} {2018})},\ \Eprint
  {http://arxiv.org/abs/1807.06800} {arXiv:1807.06800 [nucl-th]} \BibitemShut
  {NoStop}%
\bibitem [{\citenamefont {Aaij}\ \emph {et~al.}(2020)\citenamefont {Aaij} \emph
  {et~al.}}]{LHCb:2019epo}%
  \BibitemOpen
  \bibfield  {author} {\bibinfo {author} {\bibfnamefont {R.}~\bibnamefont
  {Aaij}} \emph {et~al.} (\bibinfo {collaboration} {LHCb}),\ }\href {\doibase
  10.1007/JHEP02(2020)049} {\bibfield  {journal} {\bibinfo  {journal} {JHEP}\
  }\textbf {\bibinfo {volume} {02}},\ \bibinfo {pages} {049} (\bibinfo {year}
  {2020})},\ \Eprint {http://arxiv.org/abs/1911.08594} {arXiv:1911.08594
  [hep-ex]} \BibitemShut {NoStop}%
\bibitem [{\citenamefont {Andreev}(2021)}]{Andreev:2020xor}%
  \BibitemOpen
  \bibfield  {author} {\bibinfo {author} {\bibfnamefont {O.}~\bibnamefont
  {Andreev}},\ }\href {\doibase 10.1007/JHEP05(2021)173} {\bibfield  {journal}
  {\bibinfo  {journal} {JHEP}\ }\textbf {\bibinfo {volume} {05}},\ \bibinfo
  {pages} {173} (\bibinfo {year} {2021})},\ \Eprint
  {http://arxiv.org/abs/2007.15466} {arXiv:2007.15466 [hep-ph]} \BibitemShut
  {NoStop}%
\bibitem [{\citenamefont {Aaij}\ \emph {et~al.}(2021)\citenamefont {Aaij} \emph
  {et~al.}}]{LHCb:2021vvq}%
  \BibitemOpen
  \bibfield  {author} {\bibinfo {author} {\bibfnamefont {R.}~\bibnamefont
  {Aaij}} \emph {et~al.} (\bibinfo {collaboration} {LHCb}),\ }\href@noop {} {\
  (\bibinfo {year} {2021})},\ \Eprint {http://arxiv.org/abs/2109.01038}
  {arXiv:2109.01038 [hep-ex]} \BibitemShut {NoStop}%
\end{thebibliography}%

\end{document}